\documentclass[prd,a4paper,nofootinbib,reprint]{revtex4}
\pdfoutput=1 

\usepackage{graphicx}
\usepackage{amsfonts}
\usepackage{contractions}
\usepackage{soul}
\usepackage{amssymb}
\usepackage{amsmath}
\usepackage{slashed} 
\usepackage{enumerate}
\usepackage{color}
\usepackage{comment}
\usepackage{bbold}
\usepackage{tikz}
\usepackage{multirow}
\usepackage{hyperref}
\usepackage[utf8]{inputenc}
\usepackage{adjustbox}
\usepackage{erewhon}
\definecolor{VioletRed4}{rgb}{0, 0, 0}

\def\eq#1{{eq.~(\ref{#1})}}

\definecolor{oucrimsonred}{rgb}{0.6, 0.0, 0.0}
\definecolor{persianblue}{rgb}{0.11, 0.22, 0.73}
\definecolor{forestgreen}{rgb}{0.13,0.35,0.13}
\definecolor{lightgray}{rgb}{0.83, 0.83, 0.83}
 \hypersetup{colorlinks, citecolor=oucrimsonred, linkcolor=persianblue, urlcolor=oucrimsonred}
\def\hhref#1{\href{http://arxiv.org/abs/#1}{#1}} 
\newcommand{\be}{\begin{equation}}
\newcommand{\ee}{\end{equation}}
\newcommand{\bea}{\begin{eqnarray}}
\newcommand{\eea}{\end{eqnarray}}

\definecolor{oucrimsonred}{rgb}{0.6, 0.0, 0.0}
\newcommand{\fd}[2]{\parbox{#1}{\includegraphics[width=#1]{figs/#2}}}

\definecolor{verdechiaro}{rgb}{0.6,1,0.6}
\definecolor{giallochiaro}{rgb}{1,1,0.6}
\definecolor{bluscuro}{rgb}{0.15, 0.2, 0.9}
\definecolor{verdes}{rgb}{0.1, 0.5, 0.1}
\definecolor{tangerineyellow}{rgb}{1.0, 0.8, 0.0}
\definecolor{mtcolor}{rgb}{.8,.3,.1}
\definecolor{violachiaro}{rgb}{1,0.6,1}
\definecolor{jrcolor}{rgb}{.0,.0,.55}
\definecolor{gbcolor}{rgb}{.9,.22,.12}

\definecolor{gbcolor2}{rgb}{.9,.2,.6}
\definecolor{gbcolor3}{rgb}{.3,.2,.6}

\begin{document}

\begin{flushright}
\footnotesize
{IFT-UAM/CSIC-20-91}
\end{flushright}

\title[]{Stochastic inflationary dynamics beyond slow-roll and\\ consequences for primordial black hole formation}

\date{\today}
\author{Guillermo Ballesteros$^{a,b}$}
\author{Juli\'an Rey$^{a,b}$}
\author{Marco Taoso$^{c}$}
\author{Alfredo Urbano$^{d,e}$}
\affiliation{$^a$ Instituto de F\'isica Te\'orica UAM/CSIC, Calle Nicolás Cabrera 13--15 Cantoblanco E-28049 Madrid, Spain}
\affiliation{$^b$ Departamento de F\'isica Te\'orica, Universidad Aut\'onoma de Madrid (UAM) Campus de Cantoblanco, E-28049 Madrid, Spain}
\affiliation{$^c$I.N.F.N. sezione di Torino, via P. Giuria 1, I-10125 Torino, Italy}
\affiliation{$^d$I.N.F.N. sezione di Trieste, SISSA, via Bonomea 265, I-34132 Trieste, Italy}
\affiliation{$^e$I.F.P.U., Institute  for  Fundamental Physics  of  the  Universe, via  Beirut  2, I-34014 Trieste, Italy.}

\begin{abstract}
\noindent  
We consider the impact of quantum diffusion on inflationary dynamics during an ultra-slow-roll phase, which can be of particular significance for the formation of primordial black holes. We show, by means of a fully analytical approach, that the power spectrum of comoving curvature perturbations computed in stochastic inflation matches precisely, at the linear level, the result obtained by solving the Mukhanov-Sasaki equation, even in the presence of an ultra-slow-roll phase. We confirm this result numerically in a model in which the inflaton has a polynomial potential and is coupled quadratically to the Ricci scalar. En route, we assess the role that quantum noise plays in the presence of an ultra-slow-roll phase, and clarify the issue of the quantum-to-classical transition in this scenario.
\end{abstract}

\maketitle
 
\section{Motivation and main results}\label{sec:Mot}
In the standard description of single-field inflation, the homogeneous part of the inflaton field $\phi$ is usually assumed to behave classically, while small deviations from homogeneity and isotropy over the classical background are treated quantum mechanically. 
In the slow-roll regime,
the kinetic energy $\dot\phi^2/2$ and the acceleration $\ddot\phi$ are neglected, respectively, with respect to the potential energy $V(\phi)$ and the Hubble friction $3H\dot\phi$. 
This treatment leads to the usual $\mathcal{P_R}\sim H^4/\dot H$  dependence of the primordial power spectrum (with $H$ being the Hubble function and $\dot H \ll H^2$). This is known to account well for the temperature anisotropies in the Cosmic Microwave Background (CMB), provided that a suitable $V(\phi)$ is assumed. The formation of the primordial power spectrum in the slow-roll regime is usually pictured as follows: comoving curvature perturbations, starting their evolution in the Bunch-Davies vacuum in the far past, are said to be stretched out of the horizon (of size $H^{-1}$) by inflation, in such a way that fluctuations over a comoving extension  $k^{-1}$ become constant shortly after the condition $k<a\,H=\dot a$ is met (where $a$ is the scale factor of the Universe), determining $\mathcal{P_R}$ and becoming classical.

This description  leaves out of the frame the possibility that quantum fluctuations could back-react on the classical trajectory of the inflaton, modifying the way in which the background evolves, and potentially altering the inflationary predictions. This back-reaction can be described in the framework of stochastic inflation \cite{Starobinsky:1986fx}, which aims to understand not only how quantum effects can affect the background dynamics, but also the process through which fluctuations become classical during inflation. In this framework, the long wavelength fluctuations (small $k$) are sourced by the small wavelength ones (larger $k$) through a coarse-graining procedure in which the latter behave as classical (stochastic) noise for the former according to Langevin equations. The divide between short and long wavelength fluctuations is defined by considering a (non-comoving) length scale sufficiently larger than the horizon scale during inflation. The system of Langevin equations can be solved perturbatively at first order in the fluctuations. It is known that in the slow-roll regime this procedure leads to a $\mathcal{P_R}$ in agreement with the result of the more commonly used standard perturbation theory. 

Whether or not an analogous result holds away from the slow-roll regime may have important implications, especially if inflation undergoes a phase of ultra-slow-roll (USR) \cite{Kinney:2005vj}, in which the acceleration of the inflaton $\ddot \phi$ is not negligible with respect to Hubble friction and instead $V'(\phi)$ is subdominant. A temporary USR phase is the key ingredient of the vast majority of single-field inflation models that have been proposed in recent years with the aim of providing an explanation for a putative population of primordial black holes (PBHs) that could account for the totality of the dark matter. In these models, the USR phase arises from an approximate inflection point (typically a shallow minimum) in $V(\phi)$, which produces a rapid deceleration of the inflaton. In the standard approach to inflationary perturbations, this kind of dynamics leads to a localized enhancement of $\mathcal{P_R}$ for modes exiting the horizon around the USR phase. If the primordial spectrum is large enough, PBHs will form abundantly when those modes become sub-horizon after inflation ends. In the simplest scenario, in which PBH formation takes place during radiation domination, the (narrow) PBH mass distribution is determined by the location of the spectral peak in $k$-space and the PBH abundance depends exponentially on $\mathcal{P_R}$. Within this scenario, the value of $\mathcal{P_R}$ required to account for all the dark matter can be estimated to be several ($\sim 7$) orders of magnitude larger than the one inferred  at CMB scales. The local minimum of $V(\phi)$ has to be carefully crafted to generate this enhancement of $\mathcal{P_R}$, as small variations of its properties can lead to exponentially large effects on the PBH abundance. It is then natural to ask whether the back-reaction of small scale fluctuations changes this picture substantially. Given an USR phase, the back-reaction could modify the peak height of $\mathcal{P_R}$ or the location of the peak itself. It was also speculated in \cite{Ballesteros:2017fsr} that, for instance, a broadening of the PBH mass function might occur. 

The question was addressed quantitatively in reference \cite{Biagetti:2018pjj}, where it was argued that the usual estimate of the PBH abundance and distribution (based on standard perturbation theory) could be significantly altered once the effects of stochastic dynamics are properly taken into account. The matter was also studied in \cite{Ezquiaga:2018gbw}, which concluded that the back-reaction of short wavelength modes generates an enhancement of three to four orders of magnitude on the peak height of $\mathcal{P_R}$ for a concrete (but representative) example of PBH formation from USR in single-field inflation. Given the exponential dependence of the PBH abundance on $\mathcal{P_R}$, this would be a result with significant consequences extending into the model-building details of single-field inflation for PBH formation.\footnote{In particular, it could allow to reduce the tuning in the shape of $V(\phi)$ required to account for all dark matter.}  However, the results of \cite{Ezquiaga:2018gbw} where disputed in \cite{Cruces:2018cvq}, which instead concludes that in an USR phase the primordial power spectrum computed with traditional perturbation theory agrees with the one computed in the stochastic approach at zeroth order in the slow-roll expansion.\footnote{For other studies of stochastic effects during USR see\,\cite{Firouzjahi_2019,Pattison:2019hef}.}

Considering the relevance of these issues for PBH dark matter, we reassess the computation of the primordial power spectrum (in the presence of an USR phase) using the framework of stochastic inflation, both analytically and numerically. We construct an analytical toy model which implements the main features of the necessary transition from slow-roll to ultra-slow-roll (see \cite{Ballesteros:2020qam}) and allows us to calculate $\mathcal{P_R}$ in the stochastic approach by taking into account the stochastic noise in the comoving curvature perturbation and its conjugate momentum. We find that at linear order in fluctuations (and at all orders in the slow-roll expansion) the primordial spectra calculated in the stochastic framework and in the usual perturbative approach are the same. We confirm this result numerically using a simple model of polynomial inflation capable of accounting for all dark matter.\footnote{This model also addresses the issue of the smallness of the scalar spectral index in PBH formation from single-field inflation, see \cite{Ballesteros:2020qam} and also \cite{Ballesteros:2017fsr}.} With our analysis we help to clarify the role of the coarse-graining (distance) scale that separates long and short wavelength modes in defining the stochastic approach and characterizes the quantum-to-classical transition of the Fourier modes of the comoving curvature perturbation during inflation. We show that an inadequate choice of this scale can lead to spurious results for $\mathcal{P_R}$ and thus for the PBH abundance and mass. Our findings ensure that the standard computation of the primordial power spectrum in inflationary models of PBH formation featuring a transient USR period is robust under the inclusion of the leading order fluctuations from stochastic dynamics. However, let us mention that non-linear corrections, which we have not studied here, might be relevant, and induce sizeable non-Gaussianities, as those presented in\,\cite{Ezquiaga:2019ftu} for the case of slow-roll inflation.

In the next section we present our toy-model of inflation for PBH formation with an USR phase and obtain (analytically) the spectrum of comoving curvature fluctuations with the usual splitting into background and perturbations. In section \ref{sec:Stoch} we discuss the stochastic approach, emphasizing the role of the coarse-graining scale. In section \ref{sec:Res} we present our results. We discuss the quantum to classical transition for the analytical toy model and compute the stochastic noise and its primordial power spectrum. We compare the results with a numerical analysis for the model introduced in \cite{Ballesteros:2020qam}. We close the section with some relevant comments about non-Gaussianities. In section \ref{sec:Conc} we present our conclusions. In this paper we use natural units ($\hbar = c = 1$) and set the reduced Planck mass {$(8\pi G)^{-1/2}$} to 1.

\section{The classical dynamics and the standard perturbative description}\label{sec:Mod}

The key ingredient in the scenario we are focusing on is the presence of an ultra-slow-roll (USR) phase during inflation.
This part of the inflationary dynamics is defined by the condition $\eta > 3/2$ on the Hubble parameter $\eta$:
 \begin{align}\label{eq:HubbleEps}
\epsilon \equiv - \frac{\dot{H}}{H^2}  = \frac{1}{2}\left(\frac{d\phi}{dN_e}\right)^2\,,~~~~\eta \equiv -\frac{\ddot{H}}{2H\dot{H}} = \epsilon - \frac{1}{2}\frac{d\log \epsilon}{dN_e}\,,
\end{align}
where $\phi$ is the classical solution for the evolution of the inflaton, $N_e$ is the number of $e$-folds and $\dot H$ is the cosmic time derivative of the Hubble function.
As shown in~\cite{Ballesteros:2020qam}, the evolution of the comoving curvature perturbation during inflation is described by the differential equation of a damped harmonic oscillator. During the USR phase the friction term changes sign and becomes a driving force, exponentially enhancing the amplitude of the comoving curvature perturbations. 
These dynamics lead to an enhancement of the power spectrum of curvature perturbations at small scales, and in particular at those scales relevant for the formation of PBHs.

\begin{figure}[t]
\begin{center}
$$\includegraphics[width=.48\textwidth]{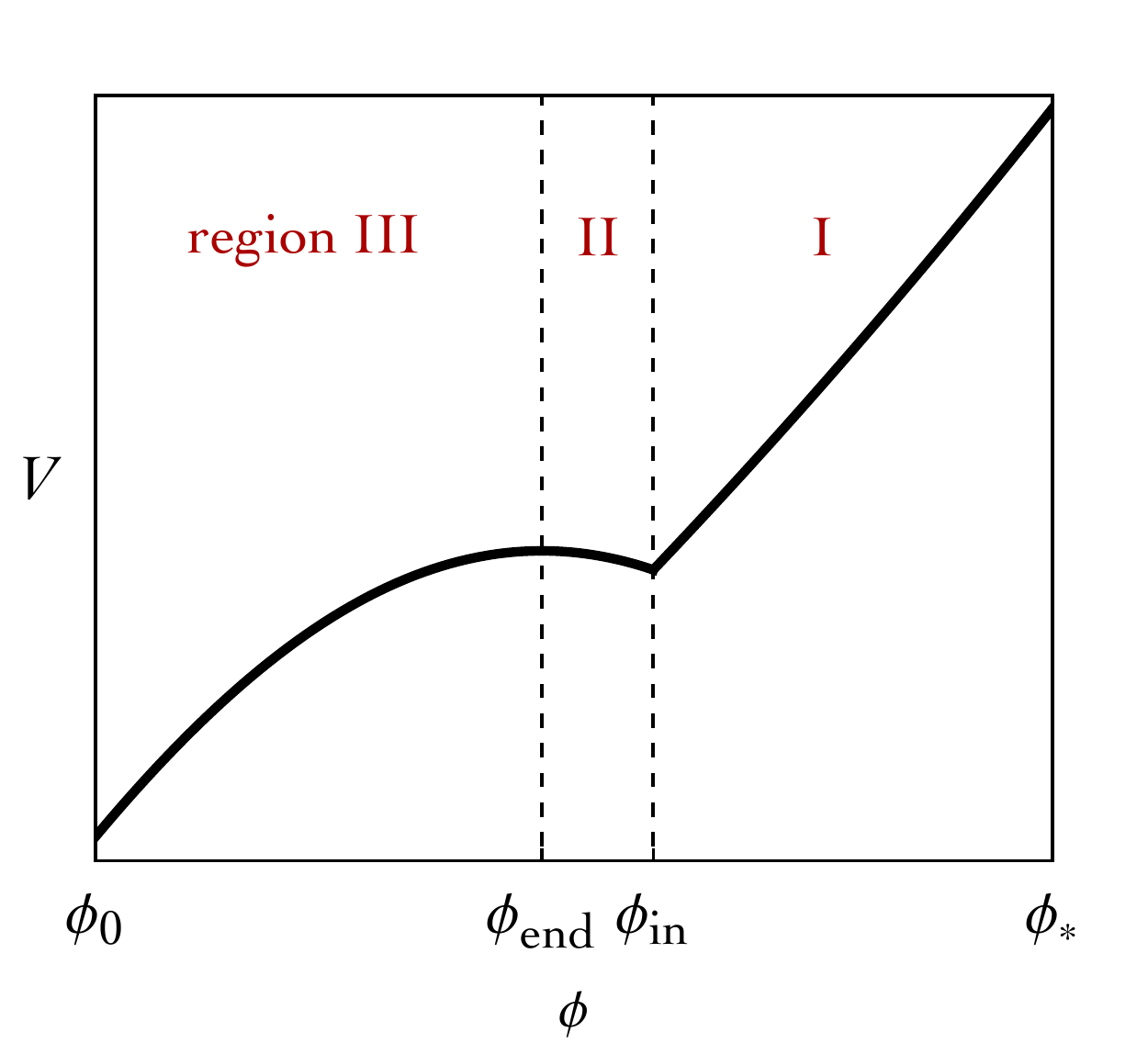}
\qquad\includegraphics[width=.48\textwidth]{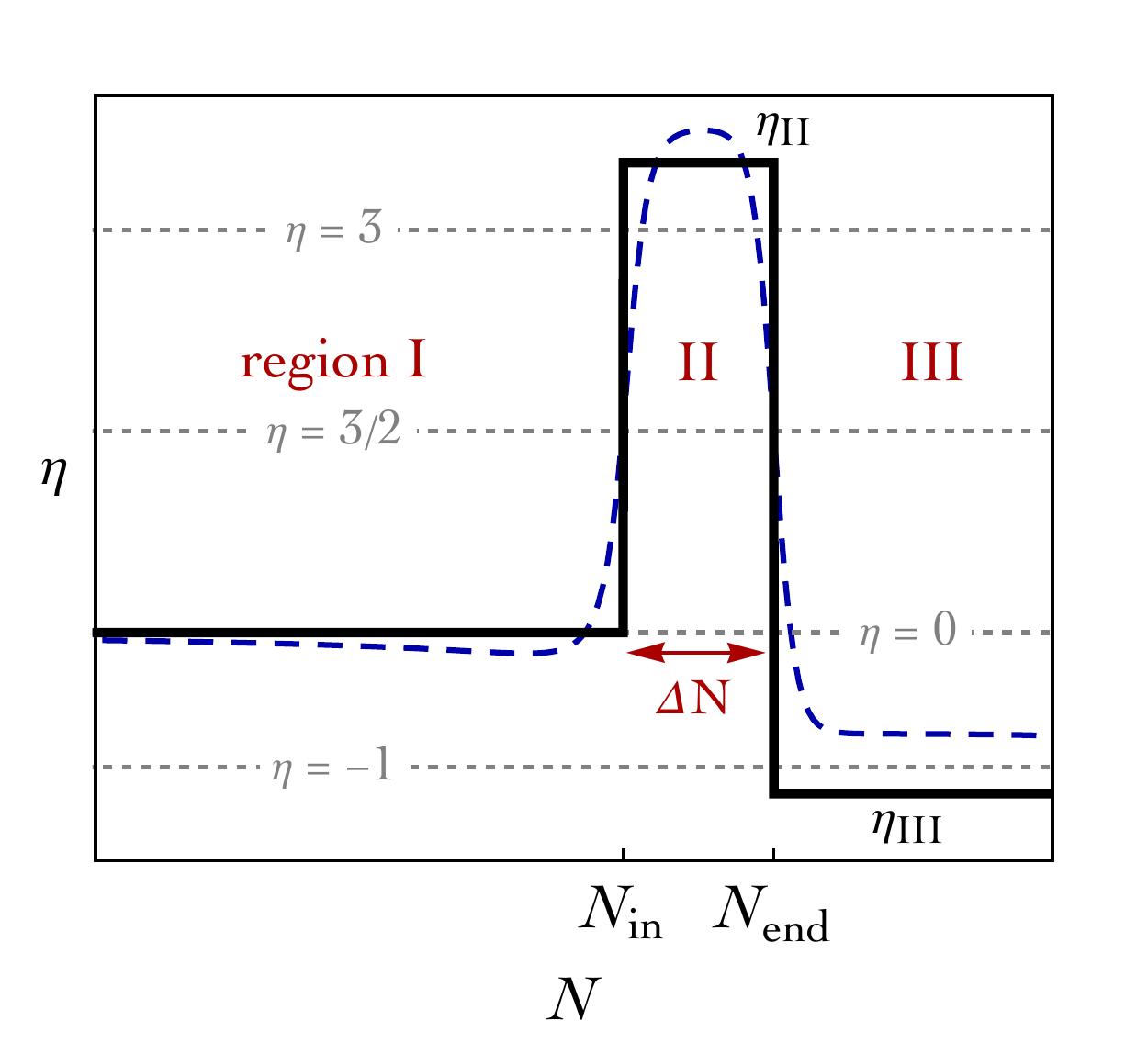}$$
\caption{\em \label{fig:Pot} 
Left panel. Profile of the potential as function of inflaton field values across the three regions discussed in our toy model. We take $\epsilon_{\rm I} = 0.1$, $\eta_{\rm II} = 4$, $\eta_{\rm III} = -1$ and $\Delta N = 3$.
Right panel. Piecewise approximation for $\eta$ as function of the number of $e$-folds (solid black line) across the three regions discussed in our toy model. 
We superimpose (dashed blue line) the exact value of $\eta$
obtained numerically in the context of a model in which one can generate 100\% of dark matter in terms of a population of PBHs (see ref.\,\cite{Ballesteros:2020qam} for details). Horizontal dotted gray lines mark 
 reference values of $\eta$. 
The exact numerical result  features $\eta_{\rm II} > 3$ and $\eta_{\rm III} < 0$;
 for the potential in the left panel and the computation of the power spectrum in fig.\,\ref{fig:Ka2}, we used $\eta_{\rm II} = 4$ and $\eta_{\rm III} = -1$.
 }
\end{center}
\end{figure}

In what follows we develop a simple, yet remarkably powerful, analytical model dividing the classical dynamics in three regions, each of them characterized by a constant value {of $\eta$.} Using the classical equation of motion of the inflaton we can also express the evolution of the inflaton itself and the potential in terms of Hubble parameters:\footnote{See \cite{Byrnes:2018txb} for another application of this strategy. Notice that the expression for $V(N)$ given there assumes $\epsilon \ll 3$.}
\begin{align}
V(N) &= V(\tilde{N})\exp\left\{
-2\int_{\tilde{N}}^{N}dN^{\prime}\left[\frac{\epsilon(3-\eta)}{3-\epsilon}\right]
\right\}\,,\label{eq:Master1}\\
\phi(N) &= \phi(\tilde{N}) \pm \int_{\tilde{N}}^N dN^{\prime}\sqrt{2\epsilon}\,,\label{eq:Master2}
\end{align}
where $\tilde{N}$ is some reference value for the $e$-fold time $N$.  We remark that no approximation is assumed in eqs.\,(\ref{eq:Master1},\,\ref{eq:Master2}). 
For a given time evolution of the Hubble parameters $\epsilon = \epsilon(N)$ and $\eta = \eta(N)$, one can solve --at worst numerically-- eqs.\,(\ref{eq:Master1},\,\ref{eq:Master2})  and reverse-engineer 
the potential and field profile. 
The three regions of the analytical model are defined as follows:
\begin{itemize}
\item [$\circ$] \textbf{Region\,I}.
The first phase of the classical dynamics extends from some initial ($e$-fold) time $N_*$ to some final time $N_{\rm in}$. During this phase, we assume $\eta = 0$ and constant $\epsilon = \epsilon_{\rm I}\ll 1$.
From eqs.\,(\ref{eq:Master1},\,\ref{eq:Master2}) we find
\begin{align}
V_{\rm I}(\phi_{\rm I}) &= V_*\exp\left[\frac{3\sqrt{2\epsilon_{\rm I}}}{3-\epsilon_{\rm I}}(\phi_{\rm I} - \phi_*)
\right]
\simeq V_*\left[
1 + \sqrt{2\epsilon_{\rm I}}(\phi_{\rm I} - \phi_*)
\right]
\,,\\
\phi_{\rm I}(N) &= \phi_* -  \sqrt{2\epsilon_{\rm I}}(N-N_*)\,.\label{eq:FieldSol1}
\end{align}
At $N=N_{\rm in}$, we have $\phi_{\rm in}\equiv \phi_{\rm I}(N_{\rm in})$ and $V_{\rm in}\equiv V_{\rm I}(\phi_{\rm in})$.
In eq.\,(\ref{eq:Master2}), we take the negative branch of the solution since we consider (without loss of generality) a large field inflation model in which the field rolls down the potential 
from large to small values.

\item [$\circ$] \textbf{Region\,II}. The second phase of the classical dynamics extends from $N_{\rm in}$ to some final time $N_{\rm end}$. 
We define $\Delta N \equiv N_{\rm end} - N_{\rm in}$. During this phase, we assume $\eta = \eta_{\rm II} \geqslant 3$. 
The Hubble parameter $\epsilon$ is not constant but it evolves in time according to 
\begin{align}
\frac{d\epsilon}{dN} = 2\epsilon(\epsilon-\eta) \simeq - 2\epsilon\eta~~~\Longrightarrow~~~ \epsilon_{\rm II}(N) = \epsilon_{\rm I}e^{-2\eta_{\rm II}(N-N_{\rm in})}\,.
\end{align}
As {it is} evident form this equation, $\epsilon_{\rm II}$ drops down to exponentially small values.
From eqs.\,(\ref{eq:Master1},\,\ref{eq:Master2}) we find
\begin{align}
V_{\rm II}(\phi_{\rm II}) &= V_{\rm in}\exp\left\{
\frac{(3-\eta_{\rm II})}{6}\left[
2\sqrt{2\epsilon_{\rm I}} + \eta_{\rm II}(\phi_{\rm II} - \phi_{\rm in})
\right](\phi_{\rm II} - \phi_{\rm in})
\right\}
\,,\label{eq:VregII}\\
\phi_{\rm II}(N) &= \phi_{\rm in} + \frac{\sqrt{2\epsilon_{\rm I}}}{\eta_{\rm II}}\left[
-1 + e^{-\eta_{\rm II}(N - N_{\rm in})}
\right]
\,.\label{eq:FieldSol2}
\end{align}
At $N=N_{\rm end}$, we have $\phi_{\rm end}\equiv \phi_{\rm II}(N_{\rm end})$ and $V_{\rm end}\equiv V_{\rm II}(\phi_{\rm end})$.

\item [$\circ$] \textbf{Region\,III}. The third phase of the classical dynamics extends from $N_{\rm end}$ to some final time $N_{0}$. 
During this phase we assume $\eta = \eta_{\rm III} < 0$. 
Consequently, the Hubble parameter $\epsilon$ evolves in time according to 
\begin{align}\label{eq:EpsilonRegionIII}
\epsilon_{\rm III}(N) = \epsilon_{\rm I}e^{-2\eta_{\rm II}\Delta N}e^{-2\eta_{\rm III}(N-N_{\rm end})}\,.
\end{align}
For negative $\eta_{\rm III}$, $\epsilon_{\rm III}$ increases exponentially starting from the small value reached at the end of region\,II. 
Thus, eventually 
the condition
$\epsilon_{\rm III}= 1$ will be satisfied and inflation will end. We shall consider values of $N_0$ not too far from $N_{\rm end}$, such that 
we can still use the approximation in which $\epsilon_{\rm III} \ll{\rm min}\{|\eta_{\rm III}|,1\}$. From eqs.\,(\ref{eq:Master1},\,\ref{eq:Master2}) we find
\begin{align}
V_{\rm III}(\phi_{\rm III}) &= V_{\rm end}\exp\left\{
\frac{(3-\eta_{\rm III})}{6}\left[
2\sqrt{2\epsilon_{\rm I}}e^{-\eta_{\rm II}\Delta N} + \eta_{\rm III}(\phi_{\rm III} - \phi_{\rm end})
\right](\phi_{\rm III} - \phi_{\rm end})
\right\}
\,,\label{eq:PotIII}\\
\phi_{\rm III}(N) &= \phi_{\rm end} + \frac{\sqrt{2\epsilon_{\rm I}}e^{-\eta_{\rm II}\Delta N}}{\eta_{\rm III}}\left[
-1 + e^{-\eta_{\rm III}(N - N_{\rm end})}
\right]
\,.\label{eq:FieldSol3}
\end{align}
\end{itemize}
This simple analytical model captures well the dynamics relevant for the production of PBHs, as discussed in~\cite{Ballesteros:2020qam}. Assuming a non-negligible $\eta$ in Region I is not essential to describe the relevant physics and only complicates the previous expressions and subsequent ones unnecessarily.
In the right panel of fig.\,\ref{fig:Pot} we show our piecewise approximation for $\eta$ as a function of the number of $e$-folds across the three regions discussed before.
We also superimpose the exact evolution of $\eta$ found solving numerically the inflaton dynamics in a model where PBHs can account for all the DM content of the Universe, see~\cite{Ballesteros:2020qam}.
The piecewise analytical model is designed to reproduce the behaviour of the exact solution. As we will see in a moment, it also performs well in describing the shape of the power spectrum of curvature perturbations.
We stress that the qualitative behaviour of the numerical result shown in the right panel of fig.\,\ref{fig:Pot} is by no means unique of \cite{Ballesteros:2020qam}. In the context of PBH production, an USR phase has been widely considered in inflation models capable of featuring an approximate stationary inflection point with a local minimum (see e.g.\ \cite{Garcia-Bellido:2017mdw,Ballesteros:2017fsr,Hertzberg:2017dkh,Cicoli:2018asa,Ozsoy:2018flq,Dalianis:2018frf,Bhaumik:2019tvl,Ballesteros:2019hus}, and \cite{Starobinsky:1992ts,Ivanov:1994pa} for earlier, related works). In these scenarios a rapid growth and subsequent drop of $\eta$ can happen when the inflaton approaches the local minimum and then slowly overcomes the barrier. Therefore, choosing appropriately the parameters $\eta_{\rm II}$, $\eta_{\rm III}$ and the duration of the USR phase $\Delta N,$ the analytical model allows to describe locally any such model.\footnote{The (rather generic) failure of the slow-roll approximation in inflationary PBH formation models with an approximate inflection point was pointed out in \cite{Germani:2017bcs,Motohashi:2017kbs,Ballesteros:2017fsr}. The power spectrum, $\mathcal{P_R}$, for such models was first correctly computed  using standard perturbation theory in \cite{Ballesteros:2017fsr}. An earlier analysis which assumed a quickly varying epsilon (also in the context of PBH formation) can be found in \cite{Chongchitnan:2006wx}. The deviation from slow-roll in the growth of $\mathcal{P_R}$ from a sudden change in the slope of the potential was studied before in \cite{Leach:2000yw,Leach:2001zf}, the latter of which considered the model of \cite{Starobinsky:1992ts}.}

For illustrative purposes, we show in the left panel of fig.\,\ref{fig:Pot} the typical shape of the potential we consider in this paper as a function 
of the inflaton field 
across the three regions discussed before. 
If $\eta_{\rm II} = 3$, the potential in region\,II is flat (see eq.\ (\ref{eq:VregII})) while for $\eta_{\rm II} > 3$ the potential 
increases for decreasing field values (in the left panel of fig.\,\ref{fig:Pot} we consider the case $\eta_{\rm II} = 4$), thus reproducing the typical situation, relevant for PBH production, that arises in the presence of an approximate stationary inflection point with a local minimum followed by a local maximum.

In this model it is possible to compute analytically the power spectrum of comoving curvature perturbations.
\begin{figure}[t]
\begin{center}
$$\includegraphics[width=.48\textwidth]{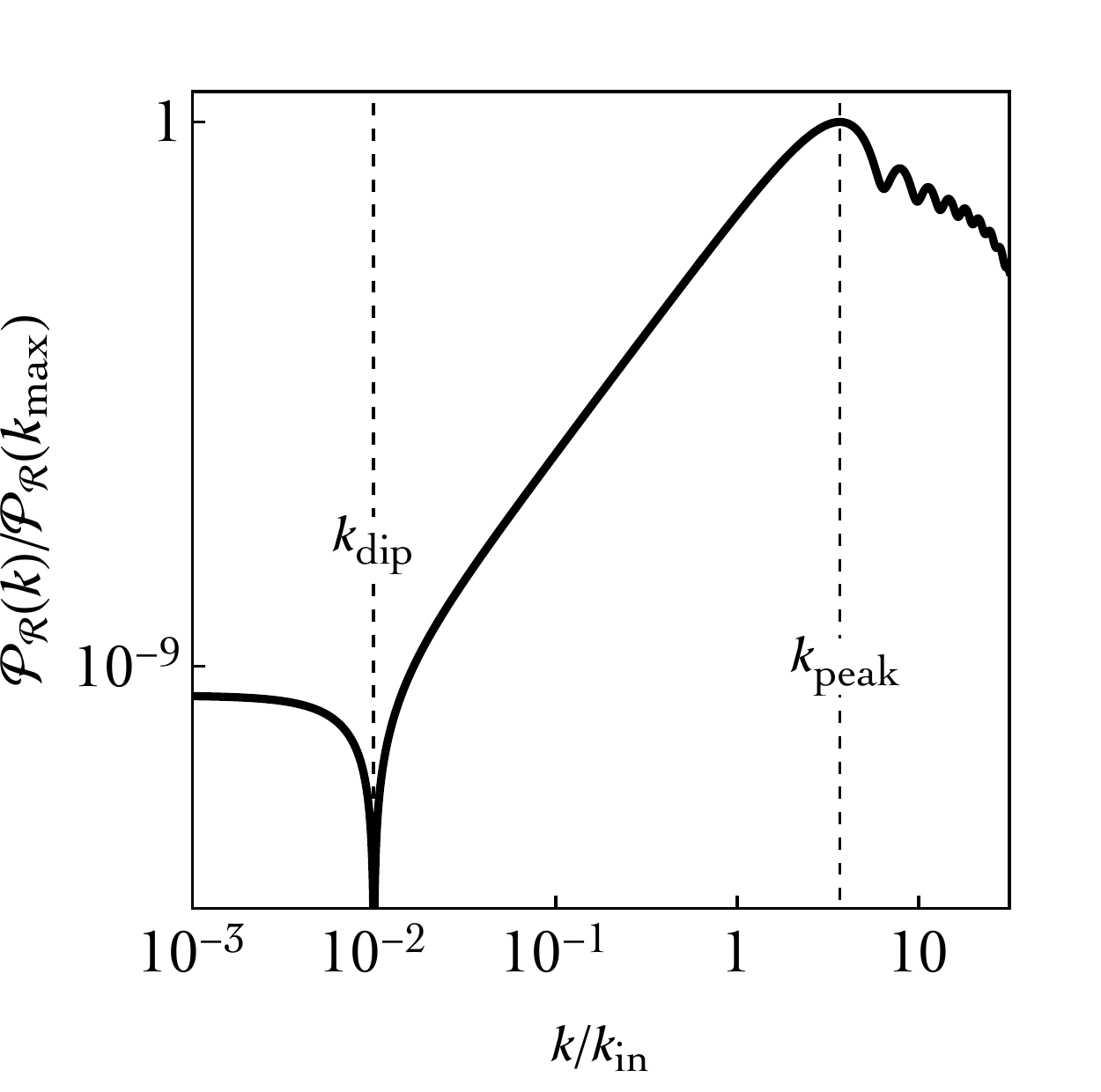}
\qquad\includegraphics[width=.48\textwidth]{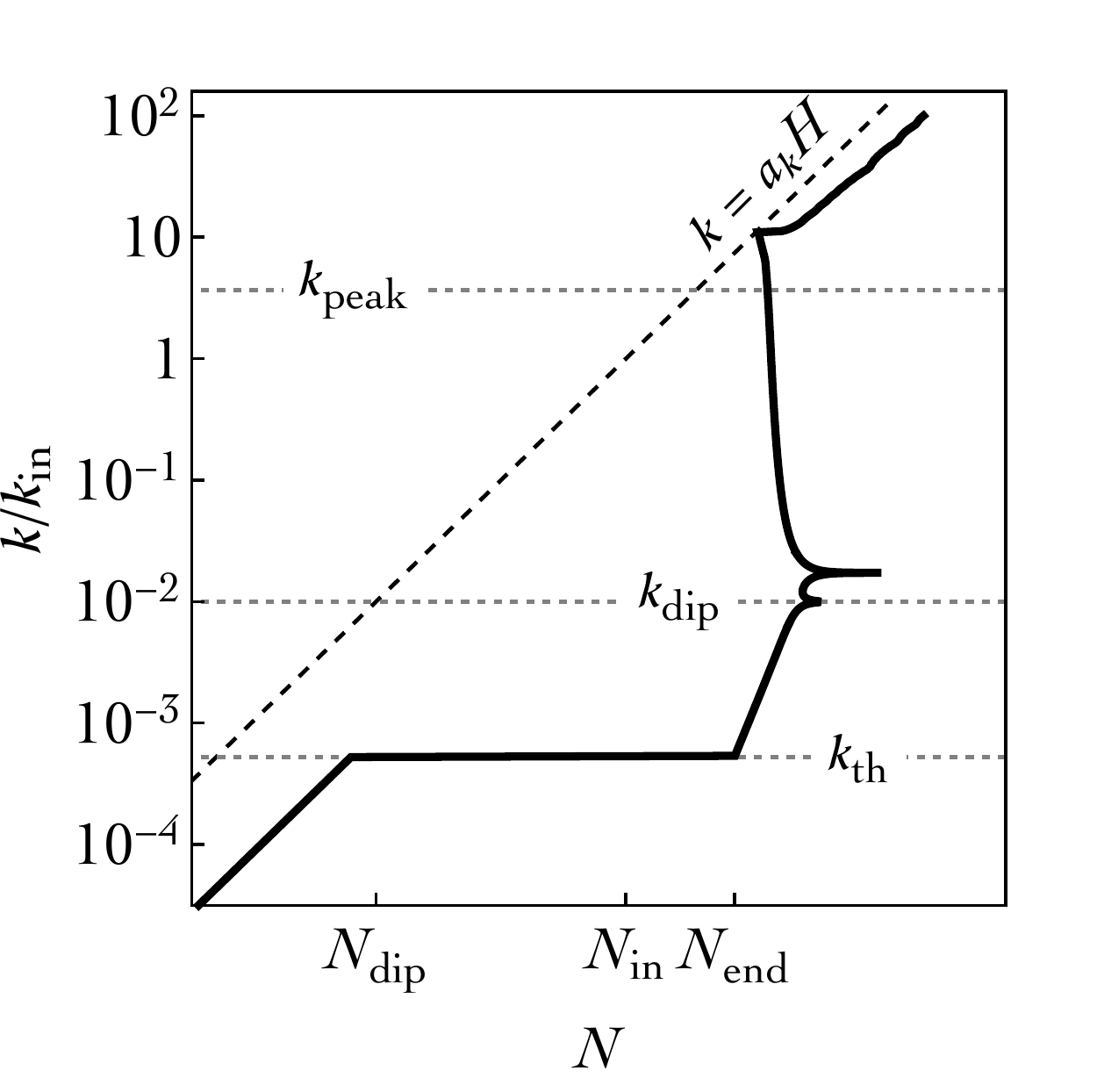}$$
\caption{\em \label{fig:Ka2} {Left panel. Power spectrum of comoving curvature perturbations as a function of the comoving scale $k$ obtained with eq\,(\ref{eq:AnalSpectrum}).
Right: $e$-fold time $N$ after which the comoving curvature perturbation in Fourier space $\mathcal{R}_k$ freezes to a constant value. 
The diagonal dashed line shows the horizon-crossing condition $k=a(\tau_k)H(\tau_k).$ Horizontal lines denote the scales at which the power spectrum of curvature perturbations
presents a dip and a maximum (see left panel). $k_{\rm th}$ denotes the minimum scale for which $\mathcal{R}_k$ becomes constant significantly after the horizon-crossing.
For both panels the parameters of the analytical model are the same as in fig.\,\ref{fig:Pot}.\vspace{0.5cm}
}
}
\end{center}
\end{figure}
We work using the spatially flat gauge. 
The perturbed line element is
\begin{align}\label{eq:PertMetric}
ds^2 = g_{\mu\nu}dx^{\mu}dx^{\nu} = (1+2\Phi)dt^2 - 2a(\partial_i B)dx^i dt - a^2\delta_{ij}dx^i dx^j\,,
\end{align}
where $\Phi = \Phi(t,\vec{x})$ and $B=B(t,\vec{x})$ are functions of the time and space coordinates, since perturbations describe departure from the homogeneous and isotropic situation.
The equation of motion of the inflaton field $\phi(t,\vec{x})$
in the perturbed metric $g_{\mu\nu}$ 
of eq.\,(\ref{eq:PertMetric}) takes the form 
\begin{align}
 \frac{d\phi}{dt} & =  \frac{(1+\Phi)}{a^3}\pi + \frac{1}{a}(\partial_i B)(\partial_i\phi)\,,\label{eq:HamSys1}    \\
\frac{d\pi}{dt}  & =    \frac{1}{a}\partial_i\left[(\partial_i B)\pi\right] + a\triangle\phi + a \partial_i\left[\Phi(\partial_i \phi)\right] - a^3(1+\Phi)\frac{dV}{d\phi}\,,\label{eq:HamSys2} 
\end{align}
where we use the Hamiltonian formalism and we use the notation $\triangle\equiv\partial_i\partial_i$. 
In the context of the familiar perturbative description\,\cite{Guth:1982ec}, the splitting
\begin{align}\label{eq:StandardDec}
\phi(t,\vec{x}) = \phi_{\rm cl}(t) + \delta\phi(t,\vec{x})\,
\end{align}
is used, where we indicate with $\phi_{\rm cl}(t)$ the classical homogeneous and isotropic 
background --that is, the solution discussed in this section, in eqs.\,(\ref{eq:FieldSol1},\,\ref{eq:FieldSol2},\,\ref{eq:FieldSol3})-- while $\delta\phi(t,\vec{x})$ is 
a small inhomogeneous perturbation. 
The metric perturbations in spatially flat gauge  satisfy the system of equations (derived, respectively, from the $^{0}_{\,\,i}$ and $^{0}_{\,\,0}$ components of the perturbed Einstein's field equations $\delta G^{\mu}_{\,\,\nu} = \delta T^{\mu}_{\,\,\nu}$ with $\delta G^{\mu}_{\,\,\nu}$ the perturbed part of the Einstein's tensor and $\delta T^{\mu}_{\,\,\nu}$ the perturbed part of the inflaton energy-momentum tensor):
\begin{align}
H\Phi & = \frac{1}{2}\left(\frac{d\phi_{\rm cl}}{dt}\right)\delta\phi\,,\label{eq:PertEFQ1} \\
-\frac{\triangle B}{a} & = 3H\Phi + \frac{1}{2H}\left.\frac{dV}{d\phi}\right|_{\phi_{\rm cl}}\delta\phi
-\frac{\Phi}{2H}\left(\frac{d\phi_{\rm cl}}{dt}\right)^2 + \frac{1}{2H}\left(\frac{d\phi_{\rm cl}}{dt}\right)
\left(\frac{d\delta\phi}{dt}\right)\,.
 \label{eq:PertEFQ2}
\end{align}
By combining eqs.\,(\ref{eq:HamSys1},\,\ref{eq:HamSys2}) with eqs.\,(\ref{eq:PertEFQ1},\,\ref{eq:PertEFQ2}), we obtain, at {linear} order in the  
perturbations, the field equation
\begin{align}
\frac{d^2\delta\phi}{dt^2} + 3H\frac{d\delta\phi}{dt} - \frac{\triangle\delta\phi}{a^2} +
\bigg\{\left.\frac{d^2V}{d\phi^2}\right|_{\phi_{\rm cl}} \underbrace{-\frac{1}{a^3}\frac{d}{dt}\bigg[
\frac{a^3}{H}\left(\frac{d\phi_{\rm cl}}{dt}\right)^2
\bigg]}_{\rm from\,metric\,perturbations}\bigg\}\delta\phi=0\,,
\end{align}
where the second term in the curly brackets is generated by the coupling of the inflaton with metric perturbations.
If we now define (in spatially flat gauge) $u\equiv a\delta\phi$ and change the variable from cosmic $t$ to conformal $\tau$ time (defined by means of $dt/d\tau = a$) we find
\begin{align}
 \frac{d^2 u}{d\tau^2} & =  \left(\triangle +\frac{1}{z}\frac{d^2z}{d\tau^2}\right)u\,,\label{eq:MS}\\
 \frac{1}{z}\frac{d^2 z}{d\tau^2} & = a^2H^2\left[(1+\epsilon-\eta)(2-\eta)+\frac{1}{aH}\left(\frac{d\epsilon}{d\tau}-\frac{d\eta}{d\tau}\right)\right]\,,\label{eq:TimeDepPot}
\end{align}
with $z\equiv (1/H)(d\phi_{\rm cl}/d\tau)$.
The field $u$ can be quantized by defining the operator
\begin{align}
\hat{u}(\tau,\vec{x}) = \int\frac{d^3 k}{(2\pi)^{3/2}}\left[
u_{\vec{k}}(\tau)a_{\vec{k}}e^{+i\vec{k}\cdot\vec{x}} + u_{\vec{k}}^{*}(\tau)a^{\dag}_{\vec{k}}e^{-i\vec{k}\cdot\vec{x}}
\right]\,,
\end{align}
with the annihilation and creation operators that satisfy the commutation relations of bosonic fields 
\begin{equation}
[a_{\vec{k}},a_{\vec{k}^{\prime}}] =[a_{\vec{k}}^{\dag},a_{\vec{k}^{\prime}}^{\dag}] =0\,,~~~~~~~[a_{\vec{k}},a_{\vec{k}^{\prime}}^{\dag}] = \delta^{(3)}(\vec{k}-\vec{k}^{\prime})\,,~~~~~~~a_{\vec{k}}|0\rangle = 0\,,
\end{equation}
and
where the vacuum condition  defines the scalar field's Fock space. The equation of motion for each mode $u_k(\tau)$ takes the form of a Schr\"odinger equation
\begin{align}\label{eq:MSmode}
\frac{d^2u_k}{d\tau^2} + \left(k^2-\frac{1}{z}\frac{d^2z}{d\tau^2}\right)u_k = 0\,,
\end{align}
with $z^{-1}(d^2z/d\tau^2)$ {acting} like a time-dependent potential.
Notice that eq.\,(\ref{eq:MSmode}) depends on $\vec{k}$ only through its modulus $k\equiv |\vec{k}|$, and, therefore,  
the same is true {of} its solution. 
For this reason,  we simply use the notation $u_k$ in place of $u_{\vec{k}}$.
The solution of eq.\,(\ref{eq:MSmode}) in each of the three regions $i=\rm I,II,III$ can be written as:
\begin{align}\label{eq:MSmodeSol}
u^{\rm (i)}_{k}(\tau) = \alpha^{\rm i}_k v_{k}(\tau) + \beta^{\,\rm i}_k v^*_{k}(\tau),
\end{align}
with
\begin{align}\label{eq:HankelGen}
v_{k}(\tau) = \frac{\sqrt{\pi}}{2}e^{i(\nu + 1/2)\pi/2}\sqrt{-\tau}H_{\nu}^{(1)}(-k\tau)\,,
\end{align}
where $H_{\nu}^{(1)}$ is the Hankel function of the first kind.
The complex coefficients $\alpha_k^{{\rm i}}$ and $\beta_k^{\,{\rm i}}$ satisfy the Wronskian condition $|\alpha_k^{{\rm i}}|^2 - |\beta_k^{\,{\rm i}}|^2 =1$.
We impose  Bunch-Davies initial conditions $\alpha^{\rm I}_k=1$ and $\beta^{\rm I}_k=0$ in the first region. The coefficients in the regions II and III are found imposing  continuity of the comoving curvature perturbation in Fourier space $\mathcal{R}_k$ and its derivative,  which in the spatially flat gauge is $\mathcal{R}_k = u_k/z.$
Finally, the two-point correlation function of the comoving curvature perturbation 
 is given by
\begin{align}
\langle 0|\mathcal{R}_k\mathcal{R}_{k^{\prime}}|0\rangle = \frac{|u_k|^2}{z^2}\delta^{(3)}(\vec{k}-\vec{k}^{\prime})
\equiv \frac{\mathcal{P}_{\mathcal{R}}}{4\pi k^3}(2\pi)^3\delta^{(3)}(\vec{k}-\vec{k}^{\prime})~~~~
\Longrightarrow~~~~\mathcal{P}_{\mathcal{R}}(\tau,k) = \frac{k^3}{2\pi^2}\frac{|u_k(\tau)|^2}{z(\tau)^2}\,,
\end{align}
where $|0\rangle$ is the vacuum quantum state of the system.
The power spectrum $\mathcal{P}_{\mathcal{R}}(k) \equiv \lim_{k\tau\to 0^-}\mathcal{P}_{\mathcal{R}}(\tau,k)$ 
is defined and computed on super-Hubble scales after the time at which the Fourier modes  $\mathcal{R}_k$ remain frozen to their final constant value, which is kept until their horizon re-entry. 
In the left panel of fig.\,\ref{fig:Ka2} we show the power spectrum $\mathcal{P}_{\mathcal{R}}(k)$ computed for the 
analytical model discussed in this section. Working at the leading order in $\epsilon,$ and therefore considering $H$ constant with $H^2 = V_*/3$, we find the following analytical expression:\footnote{In deriving this expression we have also made use of the identity
\begin{equation}
\label{eq:AsyHankel}
\lim_{x\rightarrow0}x^nH_n^{(1)}(x)=-{i\Gamma(n)}2^n/\pi.
\end{equation}}
\begin{align}\label{eq:AnalSpectrum}
\mathcal{P}_{\mathcal{R}}(k) & = \frac{H^2}{8\pi^3\epsilon_{\rm I}}
\frac{\Gamma(\nu_{\rm III})^2}{2^{1-2\nu_{\rm III}}}
e^{2\Delta N(\eta_{\rm II}-\eta_{\rm III})}
\left(\frac{k_{\rm in}}{k}\right)^{2\nu_{\rm III}-3}\left\{
|\alpha_k^{\rm III}|^2 + |\beta_k^{\rm III}|^2 - \left[\alpha_k^{\rm III}(\beta_k^{\rm III})^*e^{i\pi(\nu_{\rm III}+1/2)}+c.c.\right]
\right\}\,,
\end{align}
where $\nu = \sqrt{9/4 -\eta(3-\eta)}$, $\Gamma(\nu)$ is the Euler gamma function with argument $\nu$, and $\alpha_k^{\rm{i}}$ and $\beta_k^{\,\rm{i}}$ are complex coefficients, functions of the comoving wavenumber $k$, whose explicit expressions --in terms of Hankel functions of first and second kind-- are quite lengthy and depend on the specific values of $\Delta N$ and $\eta$ in regions II and III.
For integer values of $\eta_{\rm II} $ and $\eta_{\rm III}$, the Hankel functions admit close analytical expressions.
In the left panel of fig.\,\ref{fig:Ka2}, we use the values $\eta_{\rm II} = 4$ and $\eta_{\rm III} = -1$ (hence $\nu_{\rm III} = 5/2$). 
In this case, we find
\begin{align}
\alpha_k^{\rm III} & = \frac{1}{4x^9}\left[g_x\left(45 ie^{5\Delta N} + 15ie^{3\Delta N}x^2 +5ie^{\Delta N}x^4 + 2x^5\right)
+5ie^{2ix(1-e^{-\Delta N})+\Delta N}f_x\left(-3e^{2\Delta N}-3ie^{\Delta N}x+x^2\right)^2
\right]\,,\\
\beta_k^{\rm III} & = \frac{e^{2ix}}{4x^9}\left[
f_x\left(-45 ie^{5\Delta N} - 15ie^{3\Delta N}x^2 - 5ie^{\Delta N}x^4 + 2x^5\right) -5ie^{2ix(e^{-\Delta N}-1)+\Delta N}g_x
\left(-3e^{2\Delta N}+3ie^{\Delta N}x+x^2\right)^2
\right]\,,
\end{align}
where we defined $x\equiv k/k_{\rm in}$, $f_x\equiv 15 - 30ix -18x^2 + 4ix^3$ and $g_x\equiv -15 -12x^2-8ix^3 +2x^4$. 
The model features a large peak in the power spectrum, and the benchmark values for $\eta_{\rm II}$, $\eta_{\rm III}$ and $\Delta N$ used in the left panel of fig.\,\ref{fig:Ka2}
produce the typical enhancement (around seven orders of magnitude in between the peak amplitude of the power spectrum and the amplitude at CMB scales) that is required to generate a sizable population of PBHs {that could account for all dark matter}. 
The position of the peak of the power spectrum is related to the mass of the produced PBHs while the duration of the ultra-slow-roll phase --and, consequently, the amplitude of the peak-- is related to their abundance.

In standard slow-roll inflationary models, the comoving curvature perturbation $\mathcal{R}_k$ 
with comoving wavenumber $k$ 
freezes to a constant value
soon after the horizon-crossing time $\tau_k$ defined by $k=a(\tau_k)H(\tau_k)\equiv a_k H$,
where, as explained before, we consider  $H$ constant since we are working under the assumption $\epsilon \ll 1$.
In the presence of an ultra-slow-roll phase $\tau_k$ does not describe reliably the 
time after which perturbations stay constant. 
This is illustrated in the right panel of fig.\,\ref{fig:Ka2}, where we computed, for each $k$ on the $y$-axis, the transition time (expressed in terms of the number of $e$-folds) after which $\mathcal{R}_k$ settles to a constant value.
Despite the simplicity of the model, the result captures well the expected deviation from 
the condition $k=a_k H$ (diagonal dashed line). 
For the sake of comparison we refer to ref.\,\cite{Ballesteros:2020qam} for the analog result  in the context of a polynomial inflation model in which 100\% of dark matter consists of PBHs.

Let us close this section with one important comment.
The threshold value $k_{\rm th}$ above which the horizon crossing condition is sizably altered by the presence of the ultra-slow-roll phase (see the corresponding label in fig.\,\ref{fig:Ka2}) and 
the position of the dip $k_{\rm dip}$ admit the 
following approximate analytical expressions. As discussed in  ref.\,\cite{Ballesteros:2020qam}, the dip occurs 
for the mode with comoving wavenumber $k_{\rm dip}$ such that  {\it i)} it crosses the Hubble horizon before the beginning of the ultra-slow-roll phase and {\it ii)} the condition $|\mathcal{R}_{k_{\rm dip}}(N_{\rm end})| = 0$ is met. We find
\begin{align}\label{eq:ApproxDip}
k_{\rm dip}^2 & \approx k_{\rm in}^2\left\{\frac{e^{2\Delta N}\left[3+4\eta_{\rm II}(\eta_{\rm II}-2)\right]}{e^{2\Delta N}(2\eta_{\rm II}-1)+2e^{\Delta N(2\eta_{\rm II} - 1)}-3+2\eta_{\rm II}}\right\}
\simeq k_{\rm in}^2\,e^{-2\Delta N(\eta_{\rm II} -3/2)}\left[\frac{3+4\eta_{\rm II}(\eta_{\rm II}-2)}{2}\right]\,.
\end{align}
The comoving wavenumber $k_{\rm dip}$ crosses the Hubble horizon at time $N_{\rm dip} = N_{\rm in} - \log(k_{\rm in}/k_{\rm dip})$. 
The precise definition of $k_{\rm th}$, on the contrary, is somewhat arbitrary. As explained in ref.\,\cite{Ballesteros:2020qam}, 
for modes that cross the Hubble horizon well before the beginning of the ultra-slow-roll phase (that is for modes with comoving wavenumber $k < k_{\rm in}$)
the corresponding comoving curvature perturbation $|\mathcal{R}_k|$ freezes to a constant value that is sizably altered by the ultra-slow-roll dynamics only if the duration of the latter is long enough to compensate the 
 suppression $k/k_{\rm in}$. Consequently, for fixed $\Delta N$,  there will always be some $k \ll k_{\rm in}$ below which $|\mathcal{R}_k|$ remains constant within some accuracy $\delta\mathcal{R}$.
 In such case, we find 
 $k_{\rm th} \approx k_{\rm dip}\delta\mathcal{R}$ with $\delta\mathcal{R}\ll 1$ in general. The arbitrariness in the exact value of $k_{\rm th}$ depends on the specific choice of $\delta\mathcal{R}$.

 \section{The stochastic dynamics}\label{sec:Stoch}

\begin{figure}[t]
\begin{tikzpicture}
\node (label) at (0,0)[draw=white]{ 
       {\fd{5.75cm}{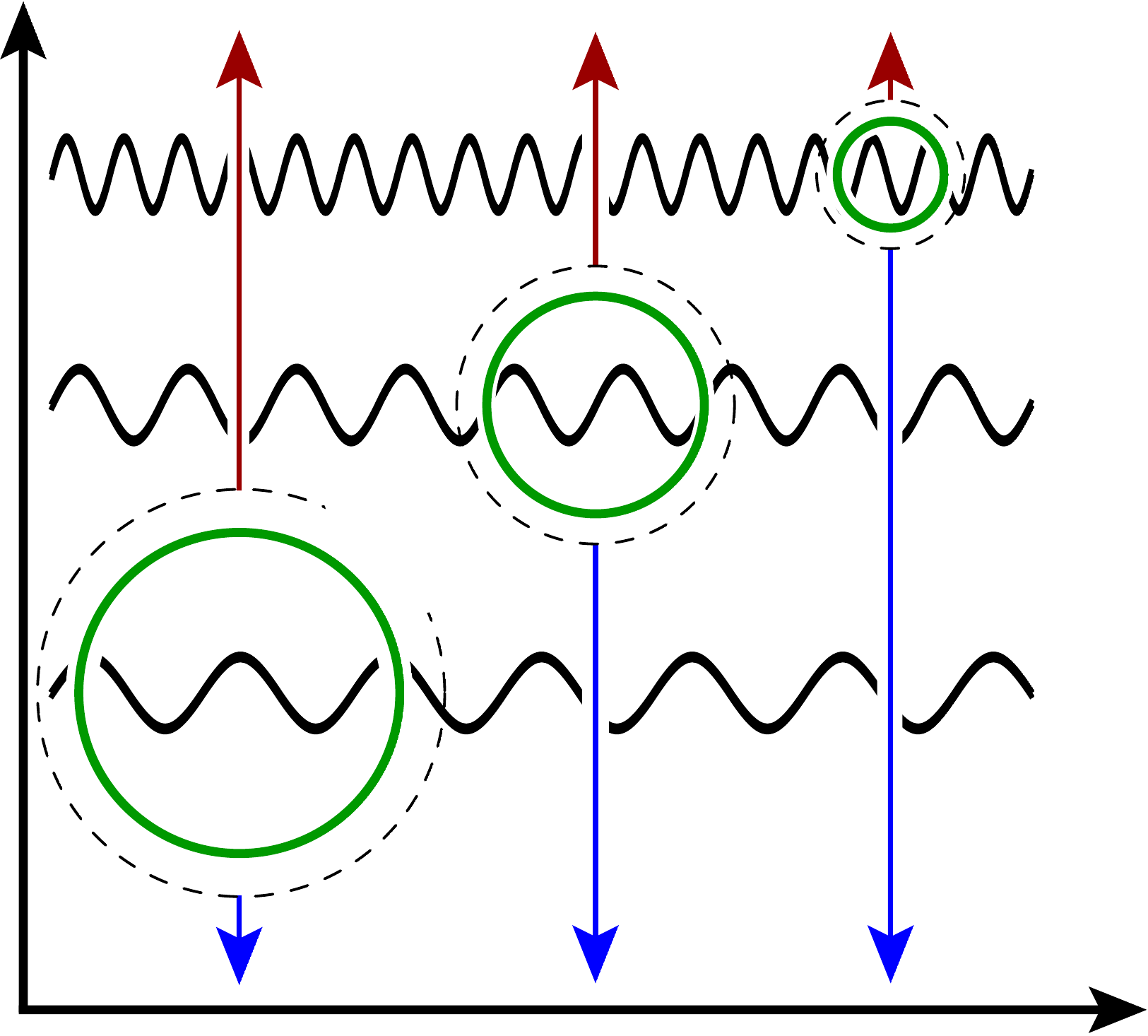}} 
      };
      \node[anchor=north] at (-3.1,2.75) {{\color{black}{\scalebox{1}{$k$}}}};
      \node[anchor=north] at (3.4,2.5) {{\color{oucrimsonred}{\scalebox{0.8}{Short-wavelength modes}}}};
      \node[anchor=north] at (3.4,-1.75) {{\color{blue}{\scalebox{0.8}{Long-wavelength modes}}}};
      \node[anchor=north] at (3.1,-2.35) {{\color{black}{\scalebox{1}{$t$}}}};
      \node[anchor=north] at (-1.4,-0.25) {{\color{forestgreen}{\scalebox{0.75}{$1/aH$}}}}; 
      \node[anchor=north] at (-.8,0.1) {{\color{black}{\scalebox{0.7}{$1/\sigma aH$}}}}; 
      \node[anchor=north] at (-1.,2.5) {{\color{oucrimsonred}{\scalebox{0.8}{$k\gtrsim k_{\sigma}$}}}};
      \node[anchor=north] at (-1.,-1.8) {{\color{blue}{\scalebox{0.8}{$k\lesssim k_{\sigma}$}}}};   
\end{tikzpicture}
\caption{\em \label{fig:SchematicWindow} 
Schematic depiction of the effect of short-wavelength fluctuations on the coarse-grained field as a function of time.}
\end{figure}
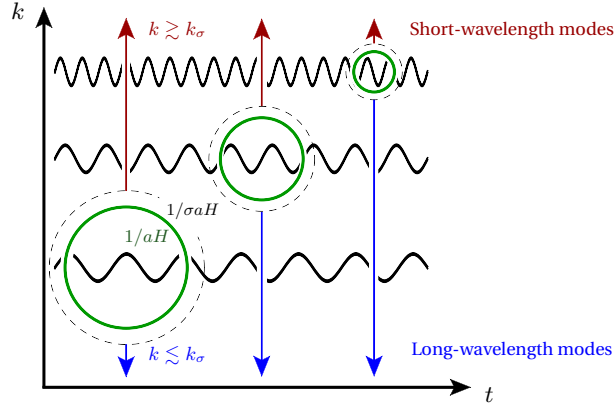

We now consider quantum fluctuations in the context of stochastic inflation.
This approach relies on an effective field theory for the long wavelength fluctuations of the inflaton field, which act as classical variables with a stochastic character due to the presence of quantum fluctuations\,\cite{Starobinsky:1986fx}.
More precisely, the inflaton field $\phi(t,\vec{x})$ in eqs.\,(\ref{eq:HamSys1},\,\ref{eq:HamSys2}) is split in two pieces: 
\begin{align}\label{eq:CoarsePhi}
\phi(t,\vec{x}) = \bar{\phi}(t)
 + 
\underbrace{\int \frac{d^3\vec{k}}{(2\pi)^{3/2}}\mathcal{W}[k-k_{\sigma}(t)]
\left[
a_{\vec{k}}\phi_{\vec{k}}(t)e^{+i\vec{x}\cdot\vec{k}}+
 {\rm h.c.}
\right]}_{\equiv \hat{\phi}_Q(t,\vec{x})}\,,
\end{align}
where $\bar{\phi}(t)$  is the coarse-grained field that
contains the long-wavelength modes
while $\hat{\phi}_Q(t,\vec{x})$ only contains the short-wavelength ones. 
The latter is treated as a perturbation. 
The short wave-length modes in the previous expression are selected  through a window  function $\mathcal{W}$ 
such that $\mathcal{W}[k-k_{\sigma}(t)] \simeq 0$  for $k\lesssim k_{\sigma}$ and 
$\mathcal{W}[k-k_{\sigma}(t)] \simeq 1$  for $k \gtrsim k_{\sigma}$, with $k_{\sigma}$ defined as $k_{\sigma} = \sigma aH$, where 
$\sigma \ll 1$ is a coarse-graining cutoff parameter.  
The crucial aspect of this description is the time-dependence of the window function $\mathcal{W}$, schematically illustrated in fig.\ \ref{fig:SchematicWindow}. As time passes by, more and more modes leave   
the short-wavelength part $\hat{\phi}_Q$ of the field to source the coarse-grained part $\bar{\phi}$ (the $k$-interval covered by the blue, downwards pointing, arrows become larger and larger).
Consequently, the dynamics of the coarse-grained field is continuously altered by the ``inflow'' of modes which cross the 
coarse-graining barrier 
as inflation proceeds. For simplicity, we choose the window function to be the Heaviside step function
\begin{equation}
\label{eq:Window}
\mathcal{W}[k-k_{\sigma}(t)] =  \theta[k-k_{\sigma}(t)]\,.
\end{equation}

If we now plug eq.\,(\ref{eq:CoarsePhi}) --together with an analogue decomposition for the conjugate momentum $\pi(t,\vec{x})$-- into eqs.\,(\ref{eq:HamSys1},\,\ref{eq:HamSys2}), we find, linearizing in $\hat{\phi}_Q$ and $\hat{\pi}_Q$, 
the system
\begin{align}
\frac{d\bar{\phi}}{dt} &= \frac{\bar{\pi}}{a^3} + \xi_{\phi}\,,\label{eq:Lang1}\\
\frac{d\bar{\pi}}{dt} &= -a^3\left.\frac{dV}{d\phi}\right|_{\bar{\phi}} + \xi_{\pi}\,,\label{eq:Lang2}
\end{align}
where we have defined the so-called noise operators
\begin{equation}\label{eq:XiDefinition}
 \xi_{\phi} \equiv - \int \frac{d^3\vec{k}}{(2\pi)^{3/2}}
\frac{d\mathcal{W}}{dt}
\left[
a_{\vec{k}}\phi_{k}e^{+i\vec{x}\cdot\vec{k}}+
a_{\vec{k}}^{\dag}\phi^*_{k}e^{-i\vec{x}\cdot\vec{k}}
\right]\,,~~~~~~
\xi_{\pi}\equiv - \int \frac{d^3\vec{k}}{(2\pi)^{3/2}}
\frac{d\mathcal{W}}{dt}
\left[
a_{\vec{k}}\pi_{k}e^{+i\vec{x}\cdot\vec{k}}+
a_{\vec{k}}^{\dag}\pi^*_{k}e^{-i\vec{x}\cdot\vec{k}}
\right]\,,
\end{equation}
and where $\phi_k(t)$ and $\pi_k(t)$ satisfy the Hamiltonian system
\begin{align}
\frac{d\phi_k}{dt} &= \frac{1}{a^3}\left(
\bar{\pi} \Phi_k  +\pi_k 
\right)\,,\label{eq:PertEq1}\\
\frac{d\pi_k}{dt} &= - \frac{\bar{\pi}}{a}k^2 B_k - a k^2\phi_k - a^3\left(
\Phi_k \left.\frac{dV}{d\phi}\right|_{\bar{\phi}} + \left.\frac{d^2 V}{d\phi^2}\right|_{\bar{\phi}}\phi_k
\right)\,.\label{eq:PertEq2}
\end{align}
In Fourier space, the metric perturbations in spatially flat gauge (see eq.\,(\ref{eq:PertMetric})) satisfy the system of equations 
\begin{align}
H\Phi_k & = \frac{1}{2}\left(\frac{d\bar{\phi}}{dt}\right)\phi_k\,,\label{eq:PertEFQ1bis} \\
\frac{k^2}{a}B_k & = 3H\Phi_k + \frac{\phi_k}{2H}\left.\frac{dV}{d\phi}\right|_{\bar{\phi}}
-\frac{\Phi_k}{2H}\left(\frac{d\bar{\phi}}{dt}\right)^2 + \frac{1}{2H}\left(\frac{d\bar{\phi}}{dt}\right)\left(\frac{d\phi_k}{dt}\right)\,.
 \label{eq:PertEFQ2bis}
\end{align}
In eq.\,(\ref{eq:XiDefinition}) the time derivative of the window function is (see eq.\,(\ref{eq:Window}))
\begin{align}
\frac{d\mathcal{W}}{dt} = \frac{d}{dt}\theta[k-k_{\sigma}(t)] = -\delta[k-k_{\sigma}(t)]\frac{dk_{\sigma}(t)}{dt}\,.
\end{align}
Notice that, as in the classical treatment of the previous section, eqs.\,(\ref{eq:PertEq1})--(\ref{eq:PertEFQ2bis}) are isotropic in the sense that $\vec{k}$ appears only through its modulus $k\equiv |\vec{k}|$. Therefore, we simply use the notation $\phi_k$, $\pi_k$, $\Phi_k$, $B_k$.
By combining these equations
we obtain
\begin{align}\label{eq:EqPerturbations}
\frac{d^2\phi_k}{dt^2} + 3H\frac{d\phi_k}{dt} + \frac{k^2}{a^2}\phi_k +
\left\{\left.\frac{d^2V}{d\phi^2}\right|_{\bar{\phi}}-\frac{1}{a^3}\frac{d}{dt}\left[
\frac{a^3}{H}\left(\frac{d\bar{\phi}}{dt}\right)^2
\right]\right\}\phi_k=0\,.
\end{align}
After defining $u_k \equiv a\phi_k$ and using the conformal time, this equation is formally analogous to eq.\,(\ref{eq:MSmode}). However, in \eq{eq:EqPerturbations} the product
$z^{-1}(d^2z/d\tau^2)$ is defined as in eq.\,(\ref{eq:TimeDepPot}) but with the Hubble parameters constructed from the coarse-grained field $\bar{\phi}$ instead of the classical field $\phi_{\rm cl}$.

Contrary to eq.\,(\ref{eq:StandardDec}), it is important to stress that both fields $\bar{\phi}$ and $\hat{\phi}_Q$ in eq.\,(\ref{eq:CoarsePhi}) have an intrinsic quantum nature. 
This is clear in the case of $\hat{\phi}_Q$ since it is a $q$-number defined in terms of creation and annihilation operators.
The same is actually true also for $\bar{\phi}$; formally, we can write
\begin{align}
\phi(t,\vec{x}) = 
\underbrace{\int \frac{d^3\vec{k}}{(2\pi)^{3/2}}\mathcal{W}[-k+k_{\sigma}(t)]
\left[
a_{\vec{k}}\phi_{\vec{k}}(t)e^{+i\vec{x}\cdot\vec{k}}+ {\rm h.c.}
\right]}_{\equiv \phi_C(t,\vec{x})}
 + 
\hat{\phi}_Q(t,\vec{x})\,,
\end{align}
where $\hat{\phi}_Q(t,\vec{x})$ is defined as in eq.\,(\ref{eq:CoarsePhi}) and where the window function in the coarse-grained part of the field selects wavenumbers with 
$k<k_{\sigma}=\sigma aH\ll aH$. On these scales, the spatial dependence in $\phi_C$ can be neglected when compared to the temporal one, and one defines $\phi_C(t,\vec{x})\rightarrow \phi_C(t)\equiv \bar{\phi}(t)$. 
In other words, the coarse-grained field $\phi_C$ can be considered homogeneous, at each time $t$, over a length 
scale $L_{\sigma}(t) \equiv (\sigma aH)^{-1}$, and
eqs.\,(\ref{eq:Lang1},\,\ref{eq:Lang2}) follow the evolution of this homogeneous patch.
Despite their intrinsic quantum nature, a classical interpretation is assigned to $\bar{\phi}$ and $\xi_{\phi,\pi}$ in  eqs.\,(\ref{eq:Lang1},\,\ref{eq:Lang2}).
In this respect, the choice of the coarse-graining cutoff wavenumber $k_{\sigma}$ plays an important role. 
 The physical picture is the following:
 Modes with comoving wavenumber $k$ that start in the Minkowski vacuum well inside the Hubble horizon ($k \gg aH$) are stretched by the expansion and become semi-classical when the condition
 $k \lesssim k_{\sigma}$ is met. In standard slow-roll inflationary models, the definition $k_{\sigma} = \sigma a H$ with $\sigma \ll 1$ is appropriate because 
 it is possible to show that soon after horizon crossing ($k \lesssim aH$) modes
  that start as Minkowski vacuum well inside the Hubble horizon
  become highly squeezed (with large occupation number, a typical signature of classical rather than quantum modes) with amplitude that is described by a Gaussian distribution\,\cite{Kiefer:1998jk}. 
 In standard slow-roll inflationary models, therefore, 
 as time passes by and inflation proceeds the quantum dynamics turns gradually into a classical one, the latter being de facto equivalent to a classical stochastic process with a Gaussian distribution. 
This is the classical interpretation that is assigned to $\bar{\phi}$.  
We remark that the  attribute  ``classical'' refers here to the fact that we can describe the system in terms of a classical random variable.

At this stage, $\xi_{\phi}$ and $\xi_{\pi}$ defined in eq.\,(\ref{eq:XiDefinition}) are still quantum objects. However, as anticipated, they also admit a classical interpretation in terms of statistical quantities. 
As a first hint, we can compute the equal-time commutator $[\xi_{\phi}(t,\vec{x}),\pi_{\phi}(t,\vec{x}^{\prime})]$. We find, in the limit $k_\sigma|\vec{x} - \vec{x}^{\prime}|\ll1$
 \begin{align}
 [\xi_{\phi}(t,\vec{x}),\xi_{\pi}(t,\vec{x}^{\prime})] \propto \phi_{k_{\sigma}}(t)\pi_{k_{\sigma}}^*(t) -  \phi^*_{k_{\sigma}}(t)\pi_{k_{\sigma}}(t) = 0\,,
 \end{align} 
 which is valid for $k_{\sigma} = \sigma a H$ with $\sigma \ll 1$. 
 Since the commutator vanishes, the variables $\xi_{\phi}$ and $\xi_{\pi}$ can be considered classical. 
 However, at this stage we cannot ascribe to them any specific numerical value since they are still defined in terms of $q$-numbers. This simply means that, from a classical point of view,
 they must be considered as stochastic variables whose statistical properties are fully determined by computing their correlation functions (that is by 
 identifying quantum expectation values with statistical moments).
 It is straightforward to check that $\langle 0|\xi_{\phi,\pi}(t,\vec{x})|0\rangle =0$.
We define the two-point correlation matrix
 \begin{align}
 \Theta(t,\vec{x};t^{\prime},\vec{x}^{\prime}) \equiv 
\left(
\begin{array}{cc}
 \langle 0|\xi_{\phi}(t,\vec{x})\xi_{\phi}(t^{\prime},\vec{x}^{\prime})|0\rangle &  \langle 0|\xi_{\phi}(t,\vec{x})\xi_{\pi}(t^{\prime},\vec{x}^{\prime})|0\rangle    \\
  \langle 0|\xi_{\pi}(t,\vec{x})\xi_{\phi}(t^{\prime},\vec{x}^{\prime})|0\rangle &    \langle 0|\xi_{\pi}(t,\vec{x})\xi_{\pi}(t^{\prime},\vec{x}^{\prime})|0\rangle
\end{array}
\right)\,,
 \end{align}
with elements 
 \begin{align}\label{eq:WhiteNoise}
  \Theta_{fg}(t,\vec{x};t^{\prime},\vec{x}^{\prime}) \equiv   \langle 0|\xi_{f}(t,\vec{x})\xi_{g}(t^{\prime},\vec{x}^{\prime})|0\rangle =
  \frac{1}{6\pi^2}\frac{dk_{\sigma}^3}{dt}
 f_{k_{\sigma}}(t)g^*_{k_{\sigma}}(t)\frac{\sin[k_{\sigma}|\vec{x} - \vec{x}^{\prime}|]}{k_{\sigma}|\vec{x} - \vec{x}^{\prime}|}\delta(t-t^{\prime})\,.
  \end{align}
The two-point correlation functions are non-zero only at equal time $t=t^{\prime}$. This property defines the so-called white noise and originates from our choice of the window function, namely a Heaviside step function in momentum space.
Other cutoffs are certainly acceptable (see, e.g., ref.\,\cite{Liguori:2004fa}) but they will lead to ``colored'' noises that unnecessarily complicate the derivation of the phase space picture which, after all, should not depend strongly on the cutoff choice. In the following, we shall restrict our analysis to the simplest case of white noise.
Furthermore, the noise is Gaussian since it is easy to check that all higher-order correlation functions can be expressed 
in terms of products of its two-point functions (and higher-order cumulants are zero). 

We are interested in the effect of the noise on length scales $\Delta x \equiv |\vec{x} - \vec{x}^{\prime}|$ over which the coarse-grained field is homogeneous, $\Delta x \ll L_{\sigma}$. 
This implies that we can evaluate eq.\,(\ref{eq:WhiteNoise}) at the same spatial point, $\vec{x} \simeq \vec{x}^{\prime}$. Eq.\,(\ref{eq:WhiteNoise}) becomes
\begin{align}\label{eq:Noise}
\Theta_{fg}(t,\vec{x};t^{\prime},\vec{x})  =  \Theta_{fg}(t)\delta(t-t^{\prime})\,,~~~~~~~
 \Theta_{fg}(t)  \equiv  \frac{d\log k_{\sigma}}{dt} \underbrace{\frac{k_{\sigma}^3}{2\pi^2}f_{k_{\sigma}}(t)g_{k_{\sigma}}^*(t)}_{\equiv \mathcal{P}_{fg}(t,k_{\sigma})}\,,
\end{align}
where $\mathcal{P}_{fg}(t,k_{\sigma})$ is the power spectrum of the fluctuations $\{\phi,\pi\}$ evaluated for each time $t$ at the corresponding coarse-graining cutoff wavenumber $k_{\sigma}$.

In summary, in the conventional stochastic interpretation we interpret eqs.\,(\ref{eq:Lang1},\,\ref{eq:Lang2}) as Langevin equations for the classical stochastic variables $\bar{\phi}$ and $\bar{\pi}$ with 
$\xi_{\phi}$ and $\xi_{\pi}$ viewed as stochastic external perturbations (classical noise) which originate from quantum fluctuations inside the horizon and with correlations specified by 
 eq.\,(\ref{eq:Noise}). 
A number of important points are worth emphasizing:
\begin{itemize}
\item [$\circ$] As a consequence of their stochastic interpretation, one can solve --in general numerically  by discretizing the time variable-- eqs.\,(\ref{eq:Lang1},\,\ref{eq:Lang2}) multiple times starting from some initial time $t_*$ up to some final time $t_0$, obtaining different solutions.
From a sufficiently large sample of stochastic realizations one can then extract a probability distribution function which effectively measures at $t_0$ the impact of field fluctuations on a scale of order $L_{\sigma}(t_0)$.
\item [$\circ$] Solving the Langevin equations (\ref{eq:Lang1},\,\ref{eq:Lang2}) without any approximation is a complicated numerical task
because the stochastic noise depends, implicitly, on the coarse-grained fields $\bar{\phi}$ and $\bar{\pi}.$
In fact the correlation functions in eq.\,(\ref{eq:Noise}) are computed from the fields $\phi_k(t)$ and $\pi_k(t),$ which in turns are obtained solving eq.\,(\ref{eq:EqPerturbations}) once the background evolution is specified. However, the latter depends on the coarse-grained fields. These are stochastic variables, and take different values for each stochastic realization of the Langevin equations. 
This implies that, in principle, at every 
time-step one should solve the Langevin equations  and re-compute the noise correlation matrix  
by integrating eq.\,(\ref{eq:EqPerturbations}) with the appropriate background functions, solutions of the Langevin equations obtained for the time-step under consideration.
With the new noise correlation matrix at hand, one can then proceed to solve the next time-step in the dynamical evolution.
It is possible to bypass this cumbersome numerical procedure if the noise does not depend on the coarse-grained fields and if $\left.dV/d\phi\right|_{\bar{\phi}}\propto \bar{\phi}^m$ with 
$m=0,1$.
In this case the Langevin system is linear and can be analytically solved by means of standard Green's function techniques\,\cite{Grain:2017dqa}. 
\item [$\circ$]  At the linear order, a solution to eqs.\,(\ref{eq:Lang1},\,\ref{eq:Lang2}) can be obtained by expanding the coarse-grained fields about their classical counterparts at first order, 
$\bar{\phi}=\phi_{\rm cl}+ \delta\phi_{\rm st}$ and $\bar{\pi}=\pi_{\rm cl}+ \delta\pi_{\rm st}$, where $\phi_{\rm cl}$ and $\pi_{\rm cl}$ are determined by 
the solution of the Langevin equations without the noise term. 
This can be thought as the first step of a recursive strategy\,\cite{Levasseur:2013ffa,Levasseur:2013tja}.
Following this approach, the noise appears, evaluated on the classical trajectory, in the equations for $\delta\phi_{\rm st}$ and $\delta\pi_{\rm st}$.  
Notice that, despite the similarity of these definitions with those of the standard perturbative description (see eq.\,(\ref{eq:StandardDec})), 
the variables $\delta\phi_{\rm st}$ and $\delta\pi_{\rm st}$ retain here their statistical meaning precisely because of the presence of the noise terms in the equations governing their evolution.
It is, therefore, only by computing the corresponding statistical moments that one can extract information about the distribution of perturbations. This is the approach that we shall take in the rest of this paper. 
Notice also that, working within this approximation, ref.\,\cite{Ezquiaga:2018gbw} 
found that the power spectrum of comoving curvature perturbations in stochastic inflation with an USR phase is significantly different from that obtained by means of the conventional perturbative approach.
As we will show later, we find instead that the power spectra computed in both approaches agree at the linear level, in contrast with the result of \cite{Ezquiaga:2018gbw}.
\item [$\circ$] Until now, we have formulated
the problem using the cosmic time $t$ as time variable. 
In the context of stochastic inflation, using another time variable is not without consequences, and actually leads to physically different stochastic processes with different probability distributions. 
This is for instance the case if one considers the number of $e$-folds $N$ instead of the cosmic time $t$.
The reason is related to the fact that changing variable from $t$ to $N$ involves $H$ which is 
a stochastic variable since it is a function of $\bar{\phi}$.
In ref.\,\cite{Vennin:2015hra}, it was argued 
 that the number of $e$-folds $N$ is  the time variable which allows one to consistently connect stochastic inflation with results from QFT on curved space-times (see also refs.\,\cite{Finelli:2010sh,Finelli:2008zg}). 
 This issue is not really relevant for us since all our results will be derived under the assumption that $H$ is constant (an approximation that is justified in the context of the model discussed in section\,\ref{sec:Mod}).
 Nevertheless, formulating the stochastic dynamics in terms of the number of $e$-folds can help elucidating the 
 physical interpretation of some of the final equations. 
 For this reason, from now on we switch  to the description in terms of the number of $e$-folds.
 For completeness, 
 the relevant equations are modified as follows.
 The Langevin equations take the form
\begin{align}
\frac{d\bar{\phi}}{dN} &= \bar{\pi} + \xi_{\phi}\,,\\
\frac{d\bar{\pi}}{dN} &= -(3-\epsilon)\bar{\pi} -\frac{1}{H^2}\left.\frac{dV}{d\phi}\right|_{\bar{\phi}} + \xi_{\pi}\,,
\end{align} 
where, compared with eqs.\,(\ref{eq:Lang1},\,\ref{eq:Lang2}), we have rescaled the conjugate momentum  
according to $\bar{\pi}/(a^3 H)\to \bar{\pi}$. 
This rescaling allows a more direct identification of $\bar{\pi}$ with the inflaton velocity.
The noise operators are
\begin{equation}
 \xi_{\phi} \equiv - \int \frac{d^3\vec{k}}{(2\pi)^{3/2}}
\frac{d\mathcal{W}}{dN}
\left[
a_{\vec{k}}\phi_{k}e^{+i\vec{x}\cdot\vec{k}}+
a_{\vec{k}}^{\dag}\phi^*_{k}e^{-i\vec{x}\cdot\vec{k}}
\right]\,,~~~~
\xi_{\pi}\equiv - \int \frac{d^3\vec{k}}{(2\pi)^{3/2}}
\frac{d\mathcal{W}}{dN}
\left[
a_{\vec{k}}\pi_{k}e^{+i\vec{x}\cdot\vec{k}}+
a_{\vec{k}}^{\dag}\pi^*_{k}e^{-i\vec{x}\cdot\vec{k}}
\right]\,,
\end{equation}
where now $\phi_k$ and $\pi_k$ are given by
\begin{align}
\frac{d\phi_k}{dN} &=
\bar{\pi} \Phi_k  +\pi_k\label{eq:PhaseSpaceN} \,,\\
\frac{d\pi_k}{dN} &= 
-(3-\epsilon)\pi_k
- \frac{\bar{\pi}k^2}{aH} B_k - \frac{k^2}{(aH)^2}\phi_k - \frac{1}{H^2}\left(
\Phi_k \left.\frac{dV}{d\phi}\right|_{\bar{\phi}} + \left.\frac{d^2 V}{d\phi^2}\right|_{\bar{\phi}}\phi_k
\right)\,.
\end{align}
If we eliminate the metric perturbations by means of the Einstein field equations, we find
\begin{align}
\frac{d^2\phi_k}{dN^2} + (3-\epsilon)\frac{d\phi_k}{dN} + \phi_k
\bigg[
\frac{k^2}{(aH)^2} + \frac{(3-\epsilon)}{V}\left.\frac{d^2 V}{d\phi^2}\right|_{\bar{\phi}} 
-\underbrace{2\epsilon(3+\epsilon-2\eta)}_{\rm metric\,perturbations}
\bigg]= 0\,,
\end{align}
where now the effect of metric perturbations enters implicitly by means of the last term in the square brackets.
For the noise correlation function in eq.\,(\ref{eq:Noise}), we have
\begin{align}\label{eq:NoiseFold}
 \Theta_{fg}(N)  \equiv  \frac{d\log k_{\sigma}}{dN}\frac{k_{\sigma}^3}{2\pi^2}f_{k_{\sigma}}(N)g_{k_{\sigma}}^*(N)\,,
\end{align} 
where $k_{\sigma} = k_{\sigma}(N)$.
\end{itemize}
From the discussion presented in this section, it is clear that a proper definition of the coarse-graining cutoff wavenumber $k_{\sigma}$ is necessary in order to have a correct interpretation of the stochastic dynamics.
In standard slow-roll inflationary models, 
the choice $k_{\sigma} = \sigma aH$ is well motivated. 
However, in the presence of an ultra-slow-roll phase the horizon-crossing condition ($\sigma\sim 1$) --as discussed in section\,\ref{sec:Mod}-- does not offer a correct description of the dynamics of the perturbations (in particular, it does not always describe the time after which perturbations stay constant). 
It is, therefore, important to re-think about the appropriate definition of $k_{\sigma}$ in models that feature an ultra-slow-roll phase. Clearly, this is crucial to correctly compute the time-evolution of the correlation matrix in 
eq.\,(\ref{eq:Noise}). 
Let us further motivate this point from a more quantitative perspective. The evolution of the scalar perturbations is controlled, in the Heisenberg picture, by the Hamiltonian operator 
(see appendix\,\ref{app:BasicDef})
\begin{align}
\hat{\mathcal{H}}(\tau) &= \frac{1}{2}\int d^3\vec{k}\bigg\{
k\underbrace{\bigg[a_{\vec{k}}(\tau)
a^{\dag}_{\vec{k}}(\tau) + a^{\dag}_{-\vec{k}}(\tau)a_{-\vec{k}}(\tau)
\bigg]}_{\rm collection\,of\,harmonic\,oscillators} + \frac{i}{z}\frac{dz}{d\tau}
\underbrace{\bigg[
a^{\dag}_{-\vec{k}}(\tau)a^{\dag}_{\vec{k}}(\tau) - a_{\vec{k}}(\tau)a_{-\vec{k}}(\tau)
\bigg]}_{\rm interacting\,term\,(pair\,creation)}
\bigg\}\,.
\end{align}
The first term in square brackets  is the standard part describing a collection of free harmonic oscillators. The second term in square brackets is an interacting term between the scalar field and 
the classical gravitational background (it vanishes in flat space-time). The interaction is described by the product of two creation operators for the mode $\vec{k}$ and $-\vec{k}$ and it represents the 
production of pairs of quanta with opposite momentum (that is, consequently and as it should be, conserved) during the cosmological expansion. 
In terms of the Hubble parameters, we find
\begin{align}\label{eq:InteractingTerm}
 \frac{1}{z}\frac{dz}{d\tau} & = aH(1+\epsilon-\eta) \simeq
\left\{
\begin{array}{ccc}
 aH & & {\rm slow\,roll\,phase\,with\,}\epsilon\simeq\eta \ll 1\,,    \\
  & &    \\
aH(1-\eta)  & &   {\rm ultra\,slow\,roll\,phase\,with\,}\epsilon \ll 1\,.
\end{array}
\right.
\end{align}
During a standard phase of slow-roll evolution, the relative importance of the interacting term is controlled by the relation between $k$ and $aH$. 
For $k \ll aH$, that is after the mode with comoving wavenumber $k$ crosses outside the Hubble horizon, the interacting term dominates and a copious pair production  enhances 
exponentially the number of quanta in the original Minkowski vacuum that, consequently, undergoes a quantum-to-classical transition. 
This justifies the standard definition $k_{\sigma} = \sigma aH$ for the coarse-graining cutoff wavenumber.
When slow-roll is violated, the description of the quantum to classical-transition is more involved.
Consider, for instance, the model discussed in section\,\ref{sec:Mod}. In region\,II, we have $\eta_{\rm II} \geqslant 3$ and the interacting term in eq.\,(\ref{eq:InteractingTerm}) flips sign compared to the standard slow-roll case. A more detailed analysis, therefore, seems necessary to better understand the quantum-to-classical transition in the presence of an ultra-slow-roll phase. In the next section we will argue that the definition $k_\sigma=\sigma aH$ can still be used in the presence of an ultra-slow-roll phase, provided that the cutoff parameter $\sigma$ is 
chosen to be small enough to allow for the classicalization of the relevant modes to occur.

 \section{Results and discussion}\label{sec:Res}

\subsection{Quantum-to-classical transition in the presence of an ultra-slow-roll phase}\label{sec:Q2C}

\begin{figure}[t]
\begin{center}
$$\includegraphics[width=.45\textwidth]{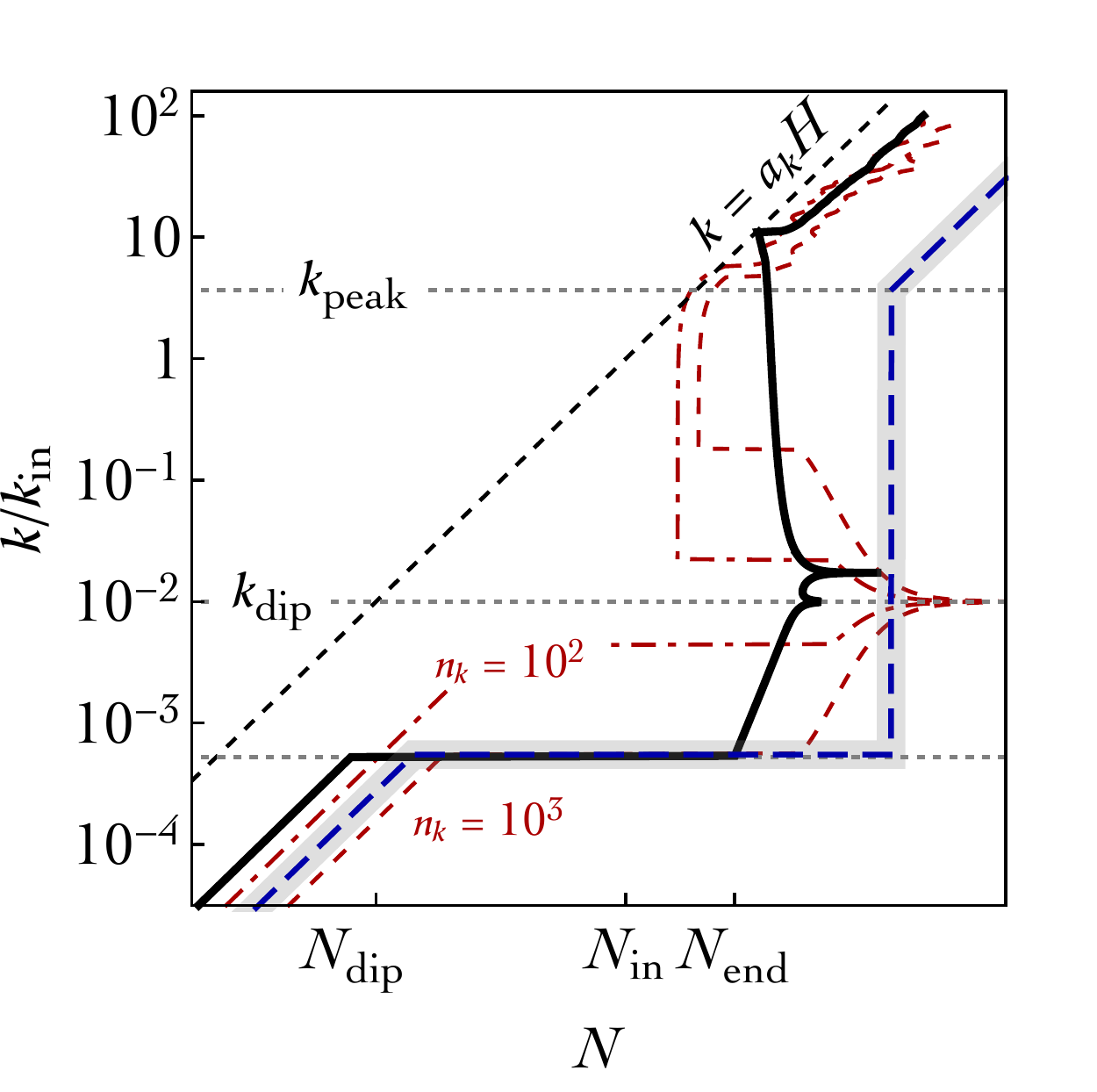}$$
\caption{\em \label{fig:OccupationCrossing} 
$e$-fold time $N$ at which the comoving curvature perturbation $\mathcal{R}_k$ becomes constant (black, solid), together with the $e$-fold times at which the occupation numbers reach certain values (in red): dashed ($n_k=10^3$) and dot-dashed ($n_k=10^2$). The occupation numbers grow exponentially only after the curvature perturbation corresponding to the same $k$ freezes out. 
Thus, in the presence of an ultra-slow-roll phase, the classicalization of the modes occurs at $k=\sigma aH$ only if $\sigma$ is small enough. The blue-dashed line represents a proxy for the quantum to classical transition, which takes into account the effect of the USR phase on the growth of the occupation numbers. It indicates that the quantum to classical transition can be completed only after the USR phase has ended for modes whose $n_k$ is affected by this phase.
}
\end{center}
\end{figure}

In the Heisenberg picture, the time-dependent occupation number $n_k(\tau)$ is defined, for each mode $k$, by the expectation value in the original vacuum state of the time-dependent particle 
number operator $a^{\dag}_{\vec{k}}(\tau)a_{\vec{k}}(\tau)$. 
We find (see appendix\, \ref{app:BasicDef} for details)
\begin{align}
n^{(\rm i)}_{k}(\tau) = \frac{\pi }{8}(-k\tau)\left\{
|\tilde{\alpha}^{\rm \,i}_k|^2 +
|\tilde{\beta}^{\rm \,i}_k|^2 + \left[
\tilde{\alpha}^{\rm \,i}_k(\tilde{\beta}^{\rm \,i}_k)^*e^{i\pi(\nu_{\rm i}+1/2)} + c.c.
\right]
\right\}\,,
\end{align}
where (omitting for simplicity the argument $-k\tau$ of the Hankel functions)
\begin{align}
|\tilde{\alpha}^{\rm \,i}_k|^2 +
|\tilde{\beta}^{\rm \,i}_k|^2 & = -\frac{4}{\pi(-k\tau)} + \left(|\alpha^{\rm \,i}_k|^2 +
|\beta^{\rm \,i}_k|^2\right)\left\{(1+\kappa_{\rm i}^2)H_{\nu_{\rm i}}^{(1)}H_{\nu_{\rm i}}^{(2)}+
H_{\nu_{\rm i}-1}^{(1)}H_{\nu_{\rm i}-1}^{(2)} + \kappa_{\rm i}\left[
H_{\nu_{\rm i}}^{(1)}H_{\nu_{\rm i}-1}^{(2)} + H_{\nu_{\rm i}}^{(2)}H_{\nu_{\rm i}-1}^{(1)}
\right]
\right\}\,,\\
\tilde{\alpha}^{\rm \,i}_k(\tilde{\beta}^{\rm \,i}_k)^* & = \alpha^{\rm \,i}_k(\beta^{\rm \,i}_k)^*\left[
(1+\kappa_{\rm i}^2)H_{\nu_{\rm i}}^{(1)}H_{\nu_{\rm i}}^{(1)} + H_{\nu_{\rm i}-1}^{(1)}H_{\nu_{\rm i}-1}^{(1)} + 2\kappa_{\rm i} H_{\nu_{\rm i}}^{(1)}H_{\nu_{\rm i}-1}^{(1)}
\right]\,,
\end{align}
with the definition $\kappa_{\rm i} \equiv \left(3/2 -\nu_{\rm i} -\eta_{\rm i}  \right)/(-k\tau)$.
The index ${\rm i}$ assumes values ${\rm i\,=I,II,III}$ depending on which one of the three regions of our model is crossed  during the time evolution. 
We use 
\begin{align}
-k\tau = xe^{N_{\rm in} - N}\,,~~~ x\equiv k/k_{\rm in}\,,
\end{align}
to convert comoving time into $e$-fold time, using $k_{\rm in} = a_{\rm in}H$ as a reference wavenumber. 
For a given $k$ we can compute, in terms of $N_{\rm in}$, the $e$-fold time $N_k$ at which we have horizon crossing: $N_k = N_{\rm in} + \log x$. 
In region\,I, we find the exact result
\begin{align}\label{eq:OccI}
n_k^{({\rm I})}(N) = \frac{1}{4x^2}e^{2(N-N_{\rm in})}\,,~~~~N\leqslant N_{\rm in}\,,
\end{align}
and the occupation number grows exponentially, as expected in the case of standard slow-roll inflation with small Hubble parameters. Let us now consider region\,II. 
The analytical expression for the occupation number is much more complicated but we can still learn something if we take the limit $-k\tau \ll 1$. In region\,II, 
this implies that we are considering $e$-fold time $N$ such that $x \ll e^{N-N_{\rm in}}$ (with $N_{\rm in} \leqslant N \leqslant N_{\rm end}$) for a given $x$. We find 
\begin{align}\label{eq:OccII}
n_{x \ll e^{N-N_{\rm in}}}^{({\rm II})} \propto e^{(2\nu_{\rm II} + 1)N}\,,~~~~N_{\rm in} \leqslant  N \leqslant N_{\rm end}\,.
\end{align}
In this limit, we still have an exponential growth. If we take our benchmark value $\eta_{\rm II} = 4$ (hence $\nu_{\rm II} = 5/2$) we find 
that the occupation number grows as $\propto e^{6N}$, which is much faster than the growth shown in eq.\,(\ref{eq:OccI}). 
We can take a similar limit also in region\,III where we find
\begin{align}\label{eq:OccIII}
n_{x \ll e^{N-N_{\rm in}}}^{({\rm III})} \propto e^{(2\nu_{\rm III} -1)N}\,,~~~~N_{\rm end} \leqslant  N\,.
\end{align}
If we take our benchmark value $\eta_{\rm III} = -1$ (hence $\nu_{\rm III} = 5/2$) we find 
that the occupation number grows $\propto e^{4N}$, which is faster than the one shown in eq.\,(\ref{eq:OccI}) (but slower than the growth in region\,II).
\begin{figure}[t]
\begin{center}
$$\includegraphics[width=.48\textwidth]{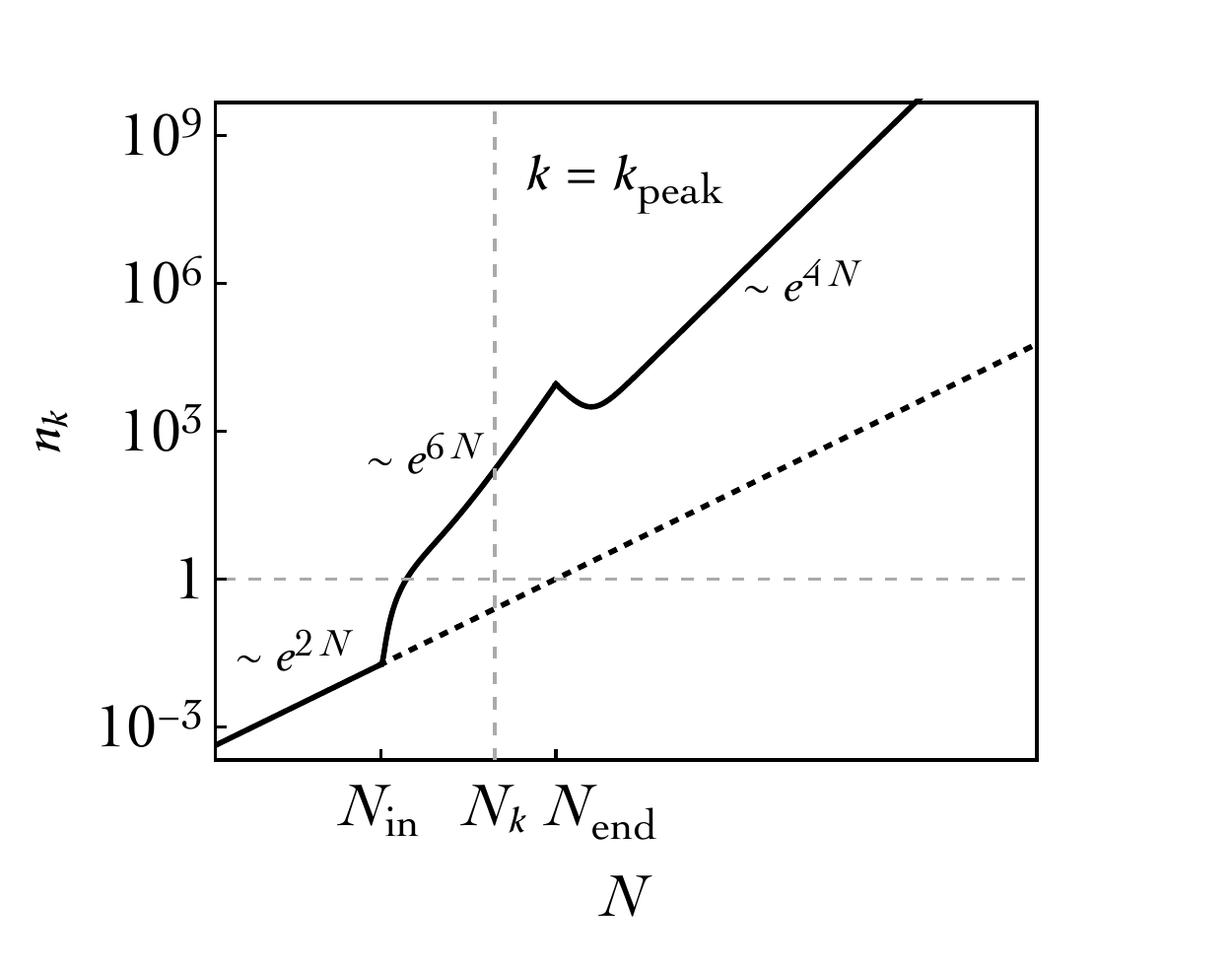}
\qquad\includegraphics[width=.48\textwidth]{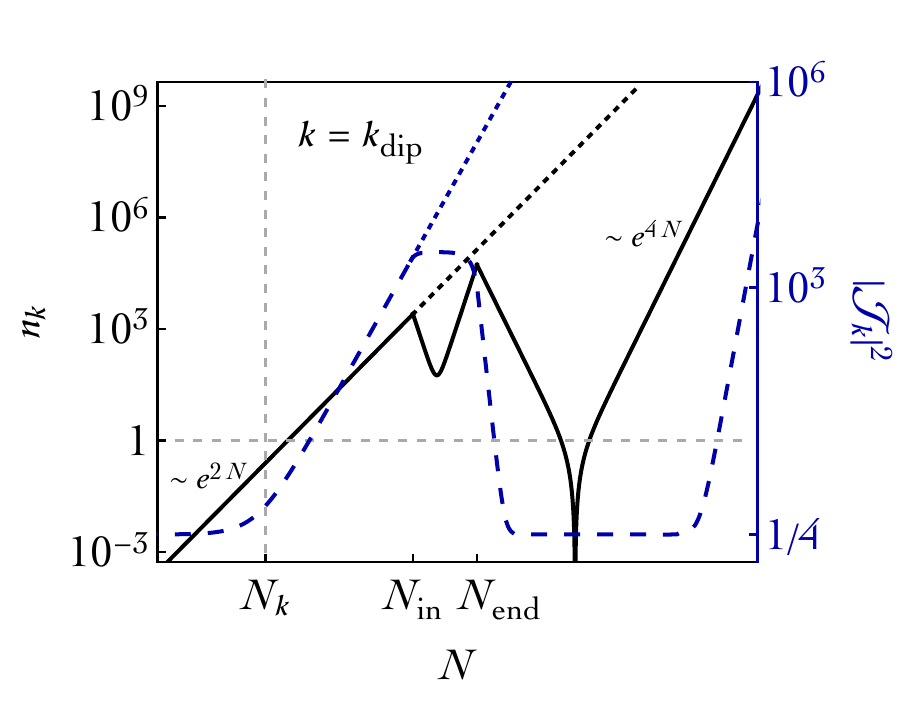}$$
\caption{\em \label{fig:OccupationNumber} 
Time evolution of the occupation number density as function of the number of $e$-folds for two representative modes $k_{\rm peak}$ (left panel) and $k_{\rm dip}$ (right panel).
$N_k$ denotes the horizon crossing $e$-fold time. 
The blue line in the right panel corresponds to the evolution of $|\mathcal{J}_k(\tau)|^2,$ see appendix\, \ref{app:BasicDef} for details.
 }
\end{center}
\end{figure}
In the left panel of fig.\,\ref{fig:OccupationNumber} we show the time evolution of the occupation number for the mode corresponding to the peak
of the comoving power spectrum, $k = k_{\rm peak}$.
The three exponential behaviors computed in eqs.\,(\ref{eq:OccI},\,\ref{eq:OccII},\,\ref{eq:OccIII}) are evident. 
For this mode, it is clear that soon after horizon crossing (vertical dashed line at $e$-fold time $N_k$) we have $n_k \gg 1$ and the mode undergoes a quantum-to-classical transition. 
More care is needed in the case of modes with comoving wavenumbers close to $k_{\rm dip}$.
As discussed in section\, \ref{sec:Mod} and more in depth in ref.\,\cite{Ballesteros:2020qam}, these modes cross the Hubble horizon before the beginning of the ultra-slow-roll phase. In this stage of the evolution the comoving curvature perturbation $\mathcal{R}_k$ is a superposition of a constant mode and a negligible decaying mode. The  latter, once the ultra-slow-roll phase starts,  becomes an exponentially growing mode rather than a decaying one.
Crucially, the sign is opposite with respect to the one of the constant mode.\footnote{Here we refer to the real or imaginary part of the complex $\mathcal{R}_k$.} This implies that the growing mode is able to cancel the constant contribution if the duration of the ultra-slow-roll phase is long enough. Finally, when the ultra-slow-roll phase ends, $\mathcal{R}_k$ settles to its final and constant value.
These dynamics explain the presence of the dip in the power spectrum of $\mathcal{R}.$
Qualitatively, a similar picture can be drawn also for the occupation number. It can be significantly reduced during the ultra-slow-roll phase and after that it starts growing exponentially again. In this case, however, the dynamics are more complicated, and the occupation number density can present a second dip during the last regime (region\,III).
An example of the evolution is shown in the right panel of fig.\,\ref{fig:OccupationNumber}.
To corroborate our analysis on the quantum-to-classical transition we investigate the evolution of another quantity, $\mathcal{J}_k(\tau),$ which can be related to the 
anti-commutator of the operator $\hat{u}$ and its conjugate momentum $\hat{p}$:
\begin{align}
\big|\big\langle\big\{\hat{u}(\tau,\vec{k}),\hat{p}^{\dag}(\tau,\vec{k})\big\}\big\rangle\big|^2  & = 4|\mathcal{J}_k(\tau)|^2 - 1\,,\label{eq:CommMainText}
\end{align}
A detailed analysis is performed in appendix\, \ref{app:BasicDef}, and here we only summarize our main findings.
One can show that $|\mathcal{J}_k(\tau)|^2 \geqslant 1/4$, and the minimum value for this quantity is attained in the far past, $\tau\rightarrow-\infty$, when the mode is in the Bunch-Davies vacuum state.
This means that initially the anti-commutator vanishes and the state admits a pure quantum description. During standard slow-roll inflation, and after horizon crossing, $\mathcal{J}_k(\tau)$ grows exponentially to large values, a signal of the quantum-to-classical transition.
Instead, during the ultra-slow-roll phase and the subsequent region\,III this growth can be significantly altered. The precise evolution depends on the specific comoving wavenumber. For modes close to $k_{\rm dip},$ $\mathcal{J}_k(\tau)$ can drop and even present a second dip in the region\,III. The time evolution for the wavenumber $k_{\rm dip}$  is shown by the dashed blue line in the right panel of fig.\,\ref{fig:OccupationNumber}.
Sufficiently long after the end of the ultra-slow-roll phase, $|\mathcal{J}_k(\tau)|^2$ grows exponentially to large values. This suggests that only at these times one can firmly consider the state as semi-classical.
These results are illustrated in fig.\,\ref{fig:OccupationCrossing}. The black solid line shows the $e$-fold time at which the comoving curvature perturbation $\mathcal{R}_k$ becomes constant while red lines are isocontours of $n_k$ for two representative values, $n_k=10^2$ and $n_k=10^3.$

Summarizing, this discussion suggests the following interpretation:

\begin{itemize}

\item [$\circ$] Modes with $k \lesssim k_{\rm th}$ are not affected by the presence of the ultra-slow-roll phase. Their quantum-to-classical transition, therefore, follows the standard picture and occurs soon after horizon crossing.

\item [$\circ$] Modes with $k_{\rm th} \lesssim k \lesssim k_{\rm peak}$ are sizably affected 
by the ultra-slow-roll phase. 
The exact $e$-fold time at which a given mode with comoving wavenumber $k$ undergoes the quantum-to-classical transition is difficult to compute (and even define) precisely. 
The simplest option is to assume that for all these modes the quantum-to-classical transition is completed only 
after the end of the ultra-slow-roll phase. This case is better motivated in appendix\,\ref{app:BasicDef}.
For illustrative purposes, this choice corresponds to the dashed blue line in fig.\,\ref{fig:OccupationCrossing}.

\item [$\circ$] For modes with $k \gtrsim k_{\rm peak}$ the quantum-to-classical transition follows the standard picture and occurs soon after horizon crossing after the end of the ultra-slow-roll phase.

\end{itemize}

\subsection{Stochastic noise}\label{sec:QN}

We now move on to the computation of the noise correlation matrix in eq.\,(\ref{eq:NoiseFold}). 
 When $k_{\sigma}$ is constant, $\Theta_{fg}(N) = 0$. 
 A non-trivial result arises from the time-dependent value $k_{\sigma} = \sigma aH$  for the coarse-graining cutoff wavenumber. We find the following general expressions
\begin{align}
\Theta^{(\rm i)}_{\phi\phi}(N) &= \frac{H^2}{8\pi}
\sigma^{3}\left\{
\left(|\alpha^{\rm \,i}_{k_{\sigma}}|^2+|\beta^{\rm \,i}_{k_{\sigma}}|^2\right)
H^{(1)}_{\nu_{\rm i}}
H^{(2)}_{\nu_{\rm i}}+
\left[
\alpha^{\rm \,i}_{k_{\sigma}}(\beta^{\rm \,i}_{k_{\sigma}})^* 
H^{(1)}_{\nu_{\rm i}}
H^{(1)}_{\nu_{\rm i}}
e^{i\pi(\nu_{\rm i}+1/2)} + c.c.
\right]
\right\}\,,\label{eq:RumoreI} \\
\Theta^{(\rm i)}_{\phi\pi}(N) &= -
 \frac{H^2}{8\pi}
\sigma^{3}\bigg\{
|\alpha^{\rm \,i}_{k_{\sigma}}|^2H^{(1)}_{\nu_{\rm i}}\left[-\left(\nu_{\rm i}-\frac{3}{2}\right)
H^{(2)}_{\nu_{\rm i}}+\sigma H^{(2)}_{\nu_{\rm i}-1}
\right]+
|\beta^{\rm \,i}_{k_{\sigma}}|^2H^{(2)}_{\nu_{\rm i}}\left[-\left(\nu_{\rm i}-\frac{3}{2}\right)
H^{(1)}_{\nu_{\rm i}}+\sigma H^{(1)}_{\nu_{\rm i}-1}
\right] \nonumber \\ &
\hspace{1.5cm}+
\alpha^{\rm \,i}_{k_{\sigma}}(\beta^{\rm \,i}_{k_{\sigma}})^* 
H^{(1)}_{\nu_{\rm i}}\left[-\left(\nu_{\rm i}-\frac{3}{2}\right)
H^{(1)}_{\nu_{\rm i}}+\sigma H^{(1)}_{\nu_{\rm i}-1}
\right] 
e^{i\pi(\nu_{\rm i}+1/2)} + c.c.
\bigg\}\,, \label{eq:RumoreIII}\\
\Theta^{(\rm i)}_{\pi\pi}(N) &= 
\frac{H^2}{8\pi}
\sigma^3\bigg\{
\left(|\alpha^{\rm \,i}_{k_{\sigma}}|^2+|\beta^{\rm \,i}_{k_{\sigma}}|^2\right)\left[-\left(\nu_{\rm i}-\frac{3}{2}\right)H^{(1)}_{\nu_{\rm i}}+\sigma H^{(1)}_{\nu_{\rm i}-1}
\right]\left[-\left(\nu_{\rm i}-\frac{3}{2}\right)H^{(2)}_{\nu_{\rm i}}+\sigma H^{(2)}_{\nu_{\rm i}-1}
\right]\nonumber\\
&\hspace{1.5cm}+\;
\alpha^{\rm \,i}_{k_{\sigma}}(\beta^{\rm \,i}_{k_{\sigma}})^* 
e^{i\pi(\nu_{\rm i}+1/2)}\left[-\left(\nu_{\rm i}-\frac{3}{2}\right)H^{(1)}_{\nu_{\rm i}}+\sigma H^{(1)}_{\nu_{\rm i}-1}
\right]^2 + c.c.
\bigg\}\,, \label{eq:RumoreII}
\end{align}
with ${\rm i\,=I,II,III}$ depending on which one of the three regions of our model is crossed at $e$-fold time $N$ (see the $x$-axis labels in fig.\,\ref{fig:OccupationCrossing} for a graphical representation). 
In eqs.\,(\ref{eq:RumoreI},\,\ref{eq:RumoreII},\,\ref{eq:RumoreIII}) the argument of the Hankel functions, that we omit for simplicity, is $\sigma$.
At lowest order in $\sigma$ the noise correlation functions admit simple analytical expressions, see table\, \ref{eq:NoiseTable}.

\begin{table}[htp]
\begin{center}
\begin{tabular}{||c||c|c|c||}
\hline\hline
$\sigma\rightarrow0$ & \textbf{region\,I} & \textbf{region\,II} & \textbf{region\,III} \\
\hline\hline
\multirow{2}{*}{$\Theta_{\phi\phi}(N)$}  & \multirow{2}{*}{$\frac{H^2}{4\pi^2}$} & \multirow{2}{*}{$\frac{H^2}{4\pi^2}e^{-2\eta_{\rm II}(N-N_{\rm in})}$} & \multirow{2}{*}{$\frac{H^2}{4\pi^2}e^{-2\eta_{\rm III}(N-N_{\rm end})-2\eta_{\rm II}\Delta N}$} \\
 & &  &  \\
 \hline
\multirow{2}{*}{$\Theta_{\pi\phi}(N)$}  & \multirow{2}{*}{0} & 
\multirow{2}{*}{$-\frac{H^2}{4\pi^2}\,\eta_{\rm II}\,e^{-2\eta_{\rm II}(N-N_{\rm in})}$}  & \multirow{2}{*}{$-\frac{H^2}{4\pi^2}\,\eta_{\rm III}\,
e^{-2\eta_{\rm III}(N-N_{\rm end})-2\eta_{\rm II}\Delta N}$} \\
  & & & \\ \hline
  \multirow{2}{*}{$\Theta_{\pi\pi}(N)$}  & \multirow{2}{*}{0} & 
\multirow{2}{*}{$\frac{H^2}{4\pi^2}\,\eta_{\rm II}^2\,e^{-2\eta_{\rm II}(N-N_{\rm in})}$}  & \multirow{2}{*}{$\frac{H^2}{4\pi^2}\,
\eta_{\rm III}^2\,
e^{-2\eta_{\rm III}(N-N_{\rm end})-2\eta_{\rm II}\Delta N}$} \\ 
  & & & \\  \hline\hline
\end{tabular}
\end{center}\vspace{-0.35cm}
\caption{{\it Noise correlation functions in the limit $\sigma\to 0$.}}\label{eq:NoiseTable}
\end{table}

To solve the stochastic dynamics, we consider, as anticipated in section\,\ref{sec:Stoch}, the expansion 
of the coarse-grained field about its classical counterpart at first order, namely
$\bar{\phi}=\phi_{\rm cl}+ \delta\phi_{\rm st}$ and $\bar{\pi}=\pi_{\rm cl}+ \delta\pi_{\rm st}$, where $\phi_{\rm cl}$ and $\pi_{\rm cl}$ define the classical trajectory. The latter is the classical solution computed in section\,\ref{sec:Mod} while
$\delta\phi_{\rm st}$ and $\delta\pi_{\rm st}$ are, on the contrary, statistical variables.  
This simply follows from the fact that if we substitute $\bar{\phi}=\phi_{\rm cl}+ \delta\phi_{\rm st}$ and $\bar{\pi}=\pi_{\rm cl}+ \delta\pi_{\rm st}$ in the Langevin equations, then $\phi_{\rm cl}$ and $\pi_{\rm cl}$ solve, by definition, the system without the noise terms while $\xi_{\phi}$ and $\xi_{\pi}$ enter in the equations for $\delta\phi_{\rm st}$ and $\delta\pi_{\rm st}$.

To interpret $\delta\phi_{\rm st}$ and $\delta\pi_{\rm st}$, therefore, we have to compute their statistical moments. 
The latter, in full generality, are given by 
\begin{align}
\langle \delta\phi_{\rm st}^n\delta\pi_{\rm st}^m\rangle(N) & = 
\int d\bar{\phi}d\bar{\pi}\,[\bar{\phi}-\phi_{\rm cl}(N)]^n[\bar{\pi}-\pi_{\rm cl}(N)]^m P(\bar{\phi},\bar{\pi},N)\,,
\end{align}
where $P(\Phi,N)$ is the phase-space probability density for the coarse-grained variables $\bar{\phi}$ and 
 $\bar{\pi}$ that solves the Fokker-Planck equation (see e.g.\,\cite{Grain:2017dqa,Starobinsky:1994bd,Tolley:2008na})
   \begin{align}\label{eq:FP}
  \frac{\partial P(\Phi,N)}{\partial N} & = -\sum_{A=1}^{2}\frac{\partial}{\partial \Phi_A}
  \left[\mathcal{D}_A P(\Phi,N)\right]
  +\frac{1}{2}\sum_{A,B=1}^{2}D_{AB}(N)\frac{\partial^2 P(\Phi,N)}{\partial\Phi_A\partial\Phi_B}
   \,,
   \end{align}
   where $\Phi\equiv (\bar{\phi},\bar{\pi})^{\rm T}$. In eq.\,(\ref{eq:FP}), $\mathcal{D}$ is the drift vector 
   with components (the notation $_{\{1,2\}=\{\phi,\pi\}}$ is understood for matrix indices)
    \begin{align}\label{eq:Dirft}
  \mathcal{D}_{\phi} = \bar{\pi}\,,~~~~~~~~~~~~~~\mathcal{D}_{\pi} =-(3-\epsilon_{\bar{\phi}})\left(
  \bar{\pi} + \left.\frac{d\log V}{d\phi}\right|_{\bar{\phi}}
  \right)\,,
    \end{align}  
    and $D$ is the diffusion matrix
   \begin{align}\label{eq:DiffusionMatrix}
D(N) \equiv \frac{\mathbb{1}}{2}\left[\Theta^{(\rm i)}_{\phi\phi}(N)+\Theta^{(\rm i)}_{\pi\pi}(N)\right]+
\frac{\sigma_1}{2}\left[\Theta^{(\rm i)}_{\phi\pi}(N)+\Theta^{(\rm i)}_{\pi\phi}(N)\right] +
\frac{\sigma_3}{2}\left[\Theta^{(\rm i)}_{\phi\phi}(N)-\Theta^{(\rm i)}_{\pi\pi}(N)\right]\,,
\end{align}
with $\sigma_{i=1,2,3}$ the Pauli matrices. 
The drift term in the Fokker-Planck equation describes the deterministic  part  of  the  dynamics while the diffusion term gives
 the stochastic one.  
 It is important for what follows to remark that in eq.\,(\ref{eq:Dirft}) $\epsilon_{\bar{\phi}}$ indicates 
 the Hubble parameter evaluated on the coarse-grained field $\bar{\phi}$, namely $2\epsilon_{\bar{\phi}} = 
 (d\bar{\phi}/dN)^2 = \bar{\pi}^2$.
 
  By using the Fokker-Planck equation, it is possible to write 
 the equations describing the evolution of the statistical moments. 
 Let us focus on the two-point statistical correlators which are relevant for the computation of the power spectrum. We find
  \begin{align}
 \frac{d}{dN}\langle \delta\phi_{\rm st}^2\rangle & = 2\langle \delta\phi_{\rm st}\delta\pi_{\rm st}\rangle +
  D_{\phi\phi}\,,\label{eq:StatMom1}\\
\frac{d}{dN}\langle \delta\phi_{\rm st}\delta\pi_{\rm st}\rangle & = 
-(3-\epsilon_{\rm cl})\left.\frac{d^2\log V}{d\phi^2}\right|_{\phi_{\rm cl}}\langle \delta\phi_{\rm st}^2\rangle 
+\left[
-3(1-\epsilon_{\rm cl}) + \pi_{\rm cl}\left.\frac{d\log V}{d\phi}\right|_{\phi_{\rm cl}}
\right]\langle \delta\phi_{\rm st}\delta\pi_{\rm st}\rangle + \langle \delta\pi_{\rm st}^2\rangle + D_{\phi\pi}
 \,,\label{eq:StatMom2}\\
\frac{d}{dN}\langle \delta\pi_{\rm st}^2\rangle & = 2\left[
-3(1-\epsilon_{\rm cl}) + \pi_{\rm cl}\left.\frac{d\log V}{d\phi}\right|_{\phi_{\rm cl}}
\right]\langle \delta\pi_{\rm st}^2\rangle - 2(3-\epsilon_{\rm cl})\left.\frac{d^2\log V}{d\phi^2}\right|_{\phi_{\rm cl}}
\langle \delta\phi_{\rm st}\delta\pi_{\rm st}\rangle + D_{\pi\pi}\,.\label{eq:StatMom3}
 \end{align}
where, to be crystal-clear with our notation, we indicate with $\epsilon_{\rm cl}$ the Hubble parameter $\epsilon$ evaluated on the classical trajectory, namely $2\epsilon_{\rm cl} = (d\phi_{\rm cl}/dN)^2$. 
Eqs.\,(\ref{eq:StatMom1},\,\ref{eq:StatMom2},\,\ref{eq:StatMom3}) are of general validity. 
They are obtained by expanding the components of the drift vector around the classical trajectory
 \begin{align}
  \mathcal{D}_{\phi} & = \pi_{\rm cl} + \delta\pi_{\rm st}\,,\\
  \mathcal{D}_{\pi} & = \left[-3(1-\epsilon_{\rm cl}) + 
  \pi_{\rm cl}\left.\frac{d\log V}{d\phi}\right|_{\phi_{\rm cl}}\right](1+\delta\pi_{\rm st}) 
  - (3-\epsilon_{\rm cl})\left.\frac{d^2\log V}{d\phi^2}\right|_{\phi_{\rm cl}}\delta\phi_{\rm st}\,,\label{eq:DeltaPi1}
  \end{align}
and using integration by parts in the Fokker-Planck equation (under the assumption that, by definition, the phase-space probability density $P$ decays fast enough at infinity so that the boundary terms in the integration by parts vanish).

To rewrite more explicitly eqs.\,(\ref{eq:StatMom1},\,\ref{eq:StatMom2},\,\ref{eq:StatMom3}) one can use the following relation
\begin{align}
\left.\frac{d^2\log V}{d\phi^2}\right|_{\phi_{\rm cl}} = \frac{1}{3}(3 - \eta)\eta\,.
\end{align}
We find 
   \begin{align}
 \frac{d}{dN}\langle \delta\phi_{\rm st}^2\rangle_{\rm (i)} & = 2\langle\delta\phi_{\rm st}\delta\pi_{\rm st}\rangle_{\rm (i)} + D_{\phi\phi}^{({\rm i})}\,,
 \\
\frac{d}{dN}\langle \delta\phi_{\rm st}\delta\pi_{\rm st}\rangle_{\rm (i)} & =
-(3-\eta_{\rm i})\eta_{\rm i}\,\langle \delta\phi_{\rm st}^2\rangle_{\rm (i)} -3\langle \delta\phi_{\rm st}\delta\pi_{\rm st}\rangle 
+ \langle \delta\pi_{\rm st}^2\rangle_{\rm (i)} + D_{\phi\pi}^{({\rm i})}
 \,,
\\
\frac{d}{dN}\langle \delta\pi_{\rm st}^2\rangle_{\rm (i)} & =  -6\langle \delta\pi_{\rm st}^2\rangle_{\rm (i)} 
-2(3-\eta_{\rm i})\eta_{\rm i}\,\langle \delta\phi_{\rm st}\delta\pi_{\rm st}\rangle_{\rm (i)}
+ D_{\pi\pi}^{({\rm i})}
\,.
 \end{align}

This system can then be solved in each one of the three regions (denoted with subscripts ${\rm i}$) by starting from some classical phase-space configuration at some reference initial time.
We focus here on the behavior of the solutions in region\,III. 
By starting from a classical phase-space configuration at $N_{\rm end}$ (since in region\,II the noise decreases exponentially):
\begin{equation}
 \langle \delta\phi_{\rm st}^2\rangle_{\rm III}\big|_{N_{\rm end}} = 0\,,~~~~~
 \langle \delta\phi_{\rm st}\delta\pi_{\rm st}\rangle_{\rm III}\big|_{N_{\rm end}} = 0\,,~~~~~
 \langle \delta\pi_{\rm st}^2\rangle_{\rm III}\big|_{N_{\rm end}} = 0\,,
\end{equation}
and
 using the expressions for the noise at lowest order in $\sigma$ in table\, \ref{eq:NoiseTable}, we find
  \begin{align}
  \langle \delta\phi_{\rm st}^2\rangle_{\rm III} & = \frac{H^2}{4\pi^2}j_{\eta_{\rm II},\eta_{\rm III}}(N-N_{\sigma})e^{-2\eta_{\rm III}N}\,,\label{eq:MajorI}\\
  \langle \delta\phi_{\rm st}\delta\pi_{\rm st}\rangle_{\rm III} & = 
  -\frac{H^2}{4\pi^2}j_{\eta_{\rm II},\eta_{\rm III}}\eta_{\rm III}(N-N_{\sigma})e^{-2\eta_{\rm III}N}\,,
  \label{eq:MajorII}\\  
\langle \delta\pi_{\rm st}^2\rangle_{\rm III} & =  
\frac{H^2}{4\pi^2}j_{\eta_{\rm II},\eta_{\rm III}}\eta_{\rm III}^2(N-N_{\sigma})e^{-2\eta_{\rm III}N}\label{eq:MajorIII}\,.
  \end{align}
with $j_{\eta_{\rm II},\eta_{\rm III}} = e^{2(\eta_{\rm III}N_{\rm end}-\eta_{\rm II}\Delta N)}.$ 
One can check that from a generic initial condition the solution will evolve exponentially fast towards eqs.\,(\ref{eq:MajorI},\,\ref{eq:MajorII},\,\ref{eq:MajorIII}).
The field diffuses in all three directions in phase-space but, crucially, with very precise relations among the two-point statistical correlators.

Before moving on, it is also worth noting that the solution in region I is
    \begin{align}
 \langle \delta\phi_{\rm st}^2\rangle_{\rm I} = \frac{H^2}{4\pi^2}(N-N_*)\,,~~~~~
 \langle \delta\phi_{\rm st}\delta\pi_{\rm st}\rangle_{\rm I} = 0\,,~~~~~
 \langle \delta\pi_{\rm st}^2\rangle_{\rm I} = 0\,.
\end{align} 
This is of course nothing but the standard result in slow-roll inflation according to which the inflaton field only diffuses along the $_{\phi\phi}$ direction.

\subsection{The power spectrum of comoving curvature perturbations}\label{sec:PowerSpectrumStoch}

The power spectrum of comoving curvature perturbations is defined by the Fourier transform of the two-point correlation function of $\mathcal{R}$
\begin{align}
\langle \mathcal{R}(\vec{x}_1)\mathcal{R}(\vec{x}_2)\rangle = 
\int\frac{d^3\vec{k}_1}{(2\pi)^{{3/2}}}\frac{d^3\vec{k}_2}{(2\pi)^{{3/2}}}
e^{i\vec{k}_1\vec{x}_1 + i\vec{k}_2\vec{x}_2}\Delta_{\mathcal{R}}(\vec{k}_1,\vec{k}_2)\,,
\end{align} 
where the left-hand side indicates an ensemble average in the sense discussed in the previous section.
In stochastic inflation (see discussion in section\,\ref{sec:Stoch}), 
we are interested in computing the effect of quantum fluctuations (interpreted as classical noise) on length scales over which the coarse grained field is homogeneous. In the previous equation, this implies that 
$\Delta_{\mathcal{R}}(\vec{k}_1,\vec{k}_2) = \delta^{(3)}(\vec{k}_1 + \vec{k}_2)\Delta_{\mathcal{R}}(k)$ with $\Delta_{\mathcal{R}}(k)$ that can be considered as function of the modulus $k_1 = k_2\equiv k$ only. Furthermore, we can restrict the computation to correlators evaluated at the same spatial point. The integration over $k$ will be limited only to the long wavelength interval 
$k\in [0,k_{\sigma}]$. We find
\begin{align}
\langle \mathcal{R}^2\rangle \equiv \langle \mathcal{R}(\vec{x})\mathcal{R}(\vec{x})\rangle = \int_0^{k_{\sigma}}
\frac{dk}{k}\mathcal{P}_{\mathcal{R}}(k)\,,
\end{align}
where we introduced the conventional definition of dimensionless power spectrum 
$\mathcal{P}_{\mathcal{R}}(k) \equiv k^3\Delta_{\mathcal{R}}(k)/2\pi^2$. 
If we now change variable from $k$ to the number of $e$-folds by means of $k=a_k H$ with $dk/k = dN_k$ and take derivatives on both sides, we find that we can write
\begin{align}\label{eq:PowerSpectrumStoch}
\mathcal{P}_{\mathcal{R}}(k) = \frac{d}{dN}\langle\mathcal{R}^2\rangle 
= 
\frac{1}{2\epsilon_{\rm cl}}\left[
\frac{d}{dN}\langle \delta\phi_{\rm st}^2\rangle - 2(\epsilon_{\rm cl} -\eta_{\rm cl})\langle \delta\phi_{\rm st}^2\rangle
\right]
\,,~~{\rm at\,time}\hspace{0.cm}
\resizebox{40mm}{!}{
\parbox{30mm}{
\begin{tikzpicture}[]
\node (label) at (0,0)[draw=white]{ 
       {\fd{3cm}{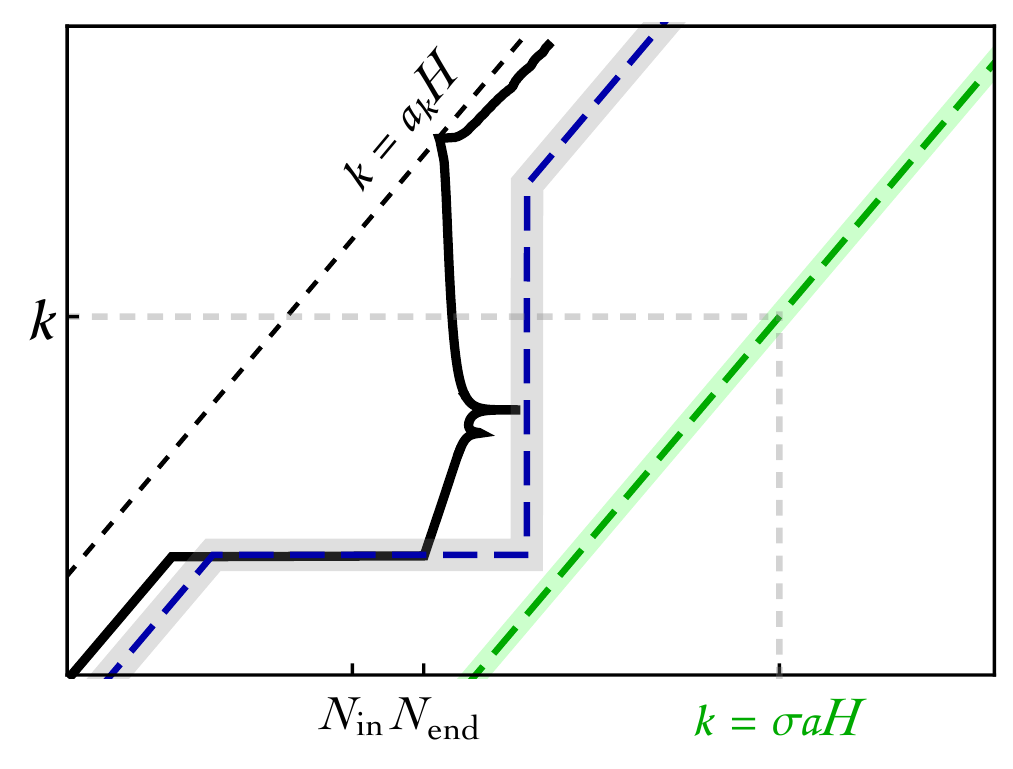}} 
      };
\end{tikzpicture}
}}
\end{align} 
The right-hand side has to be evaluated at time $N_{\sigma}$, with $\sigma$  
small enough to allow for classicalization of the mode $k$ for which the power spectrum is computed. 
Even though we cannot compute precisely when this transition happens in the presence of the ultra-slow-roll phase, 
we know--following from our discussion in section\,\ref{sec:Q2C}--that, at least for $k\geqslant k_{\rm th}$, 
the safest choice is to compute  the power spectrum in region\,III after the end of the ultra-slow-roll phase, as 
shown in the inset plot in eq.\,(\ref{eq:PowerSpectrumStoch}). 
In the conventional background+perturbation splitting approach, this corresponds to the usual 
prescription according to which the power spectrum has to be evaluated after the perturbation $\mathcal{R}_k$ 
associated with the 
wavenumber $k$ freezes to the final constant value that it maintains until its horizon re-entry after the end of inflation.
We can, therefore, write
\begin{align}\label{eq:MasterP}
\mathcal{P}_{\mathcal{R}}(k) =
\frac{1}{2\epsilon_{\rm III}}\left\{
D^{\rm (III)}_{\phi\phi}
+2\left(\langle \delta\phi_{\rm st}\delta\pi_{\rm st}\rangle_{\rm III}
+\eta_{\rm III}\,\langle \delta\phi_{\rm st}^2\rangle_{\rm III}
\right)
\right\}\Big|_{k=k_\sigma}\,,
\end{align}
where we have used eq.\,(\ref{eq:StatMom1}) to rewrite $d\langle\delta\phi_{\rm st}^2\rangle/dN$.
We can compute the power spectrum in eq.\,(\ref{eq:MasterP}) analytically by working at the lowest order in $\sigma$.

In region\,III, as a consequence of eqs.\,(\ref{eq:MajorI},\,\ref{eq:MajorII}) 
we find that $\langle \delta\phi_{\rm st}\delta\pi_{\rm st}\rangle_{\rm III}$ and $\eta_{\rm III}\,\langle \delta\phi_{\rm st}^2\rangle_{\rm III}$ cancel out. 
Using eq.\,(\ref{eq:EpsilonRegionIII}) and eq.\,(\ref{eq:RumoreI}), we find 
(recall that the argument of the Hankel function in eq.\,(\ref{eq:RumoreI}) is $\sigma$)
\begin{align}
\mathcal{P}_{\mathcal{R}}(k) = \frac{H^2}{8\pi\epsilon_{\rm I}} & 
e^{2\eta_{\rm II}\Delta N}e^{2\eta_{\rm III}(N-N_{\rm end})}
\sigma^{3}\times\nonumber \\& \left\{
\left(|\alpha^{\rm \rm III}_{k_{\sigma}}|^2+|\beta^{\rm \rm III}_{k_{\sigma}}|^2\right) 
H^{(1)}_{\nu_{\rm III}}
H^{(2)}_{\nu_{\rm III}}
+\left[
\alpha^{\rm \rm III}_{k_{\sigma}}(\beta^{\rm \rm III}_{k_{\sigma}})^*
H^{(1)}_{\nu_{\rm III}}H^{(1)}_{\nu_{\rm III}}
 e^{i\pi(\nu_{\rm \rm III}+1/2)} + c.c.
\right]
\right\}\,,
\end{align}
with $k=k_{\sigma}$ (so that we can consider $\alpha^{\rm \rm III}_{k_{\sigma}} = \alpha^{\rm \rm III}_{k}$ and 
$\beta^{\rm \rm III}_{k_{\sigma}}=\beta^{\rm \rm III}_{k}$). 
Using, at the leading order in $\sigma$, the asymptotic expression in eq.\,(\ref{eq:AsyHankel}) and the relation $(k/k_{\rm in}) = \sigma e^{(N-N_{\rm in})}$ to eliminate the remaining overall inverse power of $\sigma$, we find (remember that in region\,III with $\eta_{\rm III}$ negative we have $2\eta_{\rm III} = (3-2\nu_{\rm III})$)
\begin{align}\label{eq:AnalSpectrum2}
\mathcal{P}_{\mathcal{R}}(k) = \frac{H^2}{8\pi^3\epsilon_{\rm I}}\frac{\Gamma(\nu_{\rm III})^2}
{2^{1-2\nu_{\rm III}}}
e^{2(\eta_{\rm II}-\eta_{\rm III})\Delta N}\left(\frac{k}{k_{\rm in}}\right)^{2\eta_{\rm III}}\left\{
|\alpha^{\rm III}_{k}|^2+|\beta^{\rm III}_{k}|^2 
-\left[
\alpha^{\rm III}_{k}(\beta^{\rm III}_{k})^* e^{i\pi(\nu_{\rm III}+1/2)} + c.c.
\right]
\right\}\,,
\end{align}
which coincides with eq.\,(\ref{eq:AnalSpectrum}).
We conclude that the computation of the power spectrum obtained in the context of stochastic inflation matches precisely, even in the presence of an ultra-slow-roll phase,
the result obtained by means of the conventional perturbative approach. 
This conclusion disagrees
with the claim presented in ref.\,\cite{Ezquiaga:2018gbw}. 
Translated into our notation, ref.\,\cite{Ezquiaga:2018gbw} claims that {\it i)} the first term proportional to $D^{\rm (III)}_{\phi\phi}$ in eq.\,(\ref{eq:MasterP}) reproduces the power spectrum obtained by solving the 
 Mukhanov-Sasaki equation and {\it ii)} the term $\langle \delta\phi_{\rm st}\delta\pi_{\rm st}\rangle_{\rm III}
+\eta_{\rm III}\,\langle \delta\phi_{\rm st}^2\rangle_{\rm III}$ also contributes to the power spectrum, leading to an additional enhancement at its peak.
Our analytical approach shows that {\it i)} is correct. As far as {\it ii)} is concerned, 
on the contrary, we find that a precise cancellation between $\langle \delta\phi_{\rm st}\delta\pi_{\rm st}\rangle_{\rm III}$
and $\eta_{\rm III}\,\langle \delta\phi_{\rm st}^2\rangle_{\rm III}$ makes their sum disappear from the final result.

Two aspects of our result deserve further analysis. 
The first one is that we have worked at the leading order in $\sigma$, and one may wonder what happens if higher-order terms are included. 
This is a non-trivial question which is intimately related to the exact meaning of $\sigma$ in the stochastic approach. 
Second, we have assumed the analytical toy model introduced in section\,\ref{sec:Mod} whereas
ref.\,\cite{Ezquiaga:2018gbw} discussed an explicit numerical analysis in the context of a more realistic model (more specifically, the one introduced in ref.\,\cite{Garcia-Bellido:2017mdw}).
Even though our analytical description captures well all relevant features of the power spectrum in the presence of an approximate stationary inflection point, it remains an approximation. For instance, 
in more realistic models $\epsilon_{\rm I}$ is not constant (equivalently, $\eta_{\rm I} < 0$) and it may also reach $O(1)$ values during the 
inflationary dynamics before the beginning of the USR phase. 

\subsection{Numerical analysis}

In order to put our analytical results on firmer ground and answer the previous questions, we now consider a numerical analysis in the context of the class of models discussed in ref.\,\cite{Ballesteros:2020qam}:
the inflaton potential in the Einstein frame takes the form
\begin{align}\label{eq:TempHDO1}
V(\phi) \equiv \tilde{V}[\varphi(\phi)]\,,~~~~ {\rm with}~~
\tilde{V}(\varphi) = 
\frac{1}{(1+\xi \varphi^2)^2}\left(a_2\varphi^2 + a_3\varphi^3 + a_4\varphi^4 + \sum_{n=5}^{\mathcal{N}}a_n\varphi^n\right)\,,~~~
\frac{d\phi}{d\varphi} = \frac{\sqrt{1+\xi\varphi^2(1 + 6\xi)}}{1+\xi\varphi^2}\,,
\end{align}  
where $\phi$ is the canonically normalized inflaton field and $\xi$ the non-minimal coupling to gravity (defined for the scalar field $\varphi$ in the Jordan frame). 
The potential $V(\phi)$ in eq.\,(\ref{eq:TempHDO1}) features the presence of an approximate stationary inflection point 
as the consequence of a tuning between the coefficients of the quadratic and cubic terms, which are taken to be of the same order and opposite in sign. The presence of higher-dimensional operators 
(which in general are not forbidden and should be included) is not crucial for the production of PBHs but helps improving the fit of the model against CMB observables\,\cite{Ballesteros:2020qam}.

In the following, we consider in our numerical analysis a specific choice of coefficients for the potential in eq.\,(\ref{eq:TempHDO1}) (to be precise, we adopt the same inflationary solution indicated in ref.\,\cite{Ballesteros:2020qam} with a cyan star; this solution is in agreement with CMB observables and generates the correct abundance of dark matter in the present-day Universe in the form of PBHs). 
\begin{figure}[!htb!]
\begin{center}
$$\includegraphics[width=.45\textwidth]{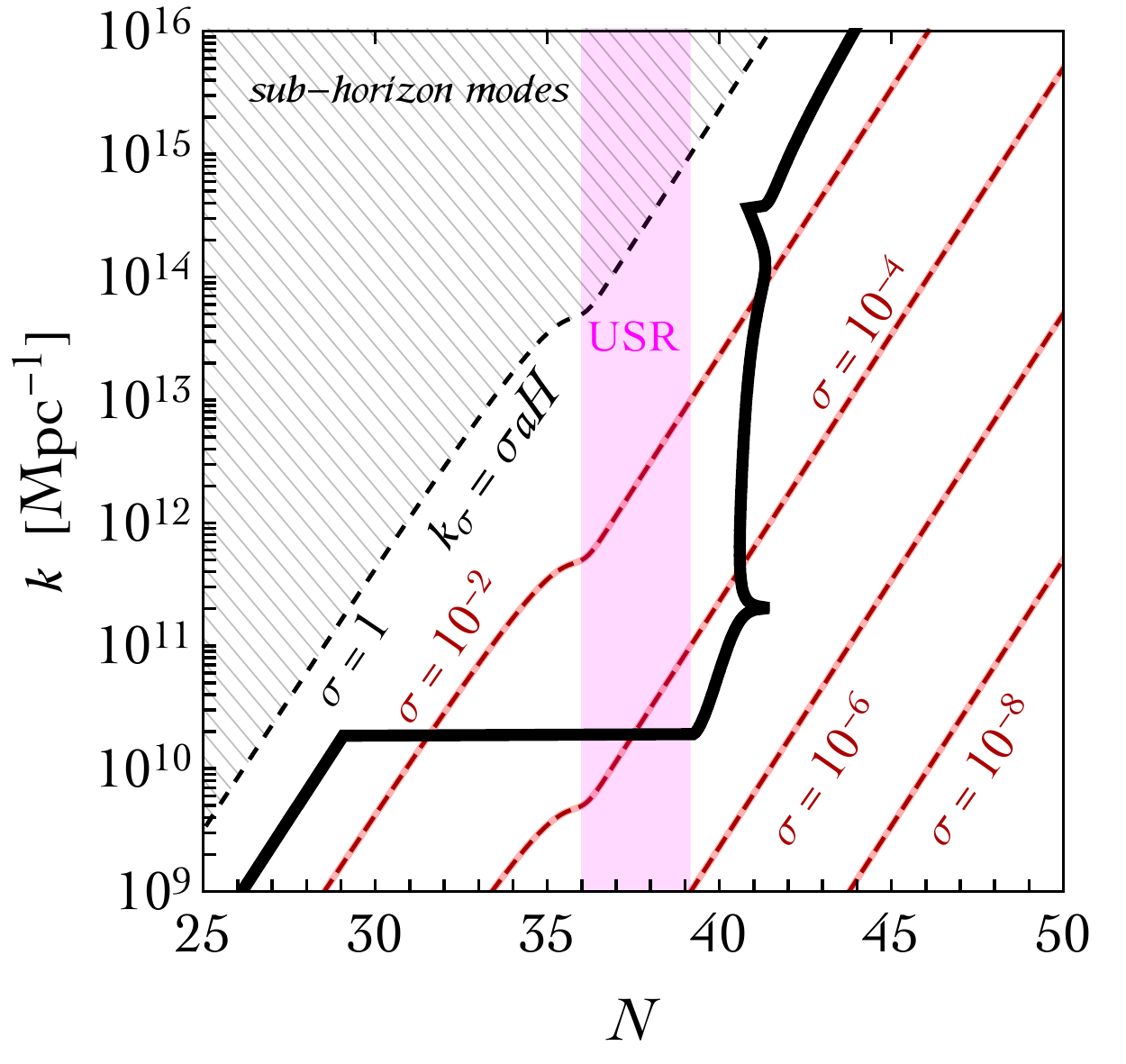}
\qquad\includegraphics[width=.45\textwidth]{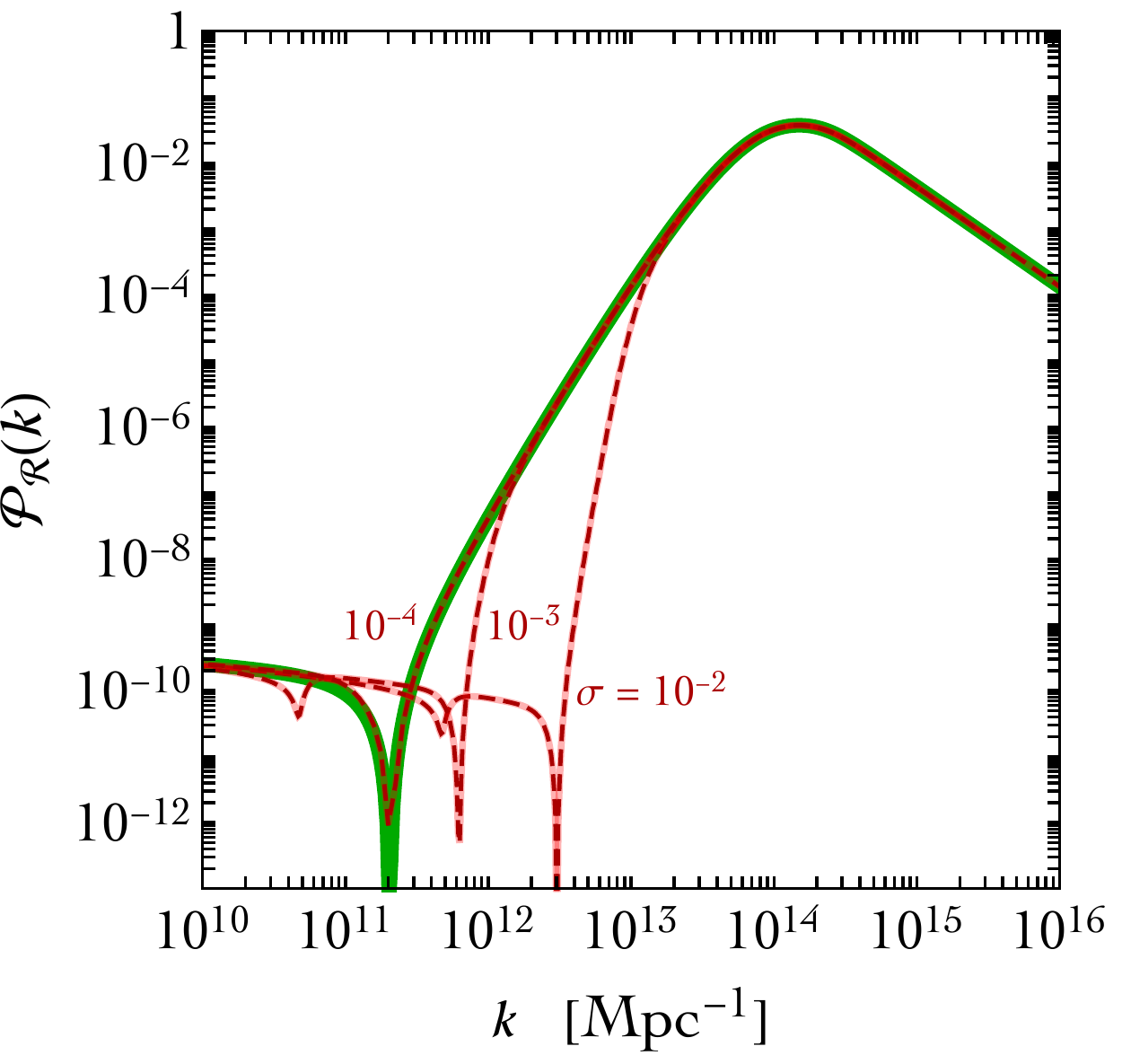}$$
\caption{\em \label{fig:NumericalCheck} 
Left panel. Transition time (solid black line) for the comoving wavenumbers $k$ on the $y$-axis 
as function of the number of $e$-folds. This plot refers to the numerical model discussed in ref.\,\cite{Ballesteros:2020qam} (see text for details). 
Diagonal dashed lines correspond to the condition $k=k_{\sigma} = \sigma aH$ for different $\sigma$.
The vertical band shaded in magenta covers the $e$-fold interval during which the USR phase takes place.
Right panel. Power spectrum obtained, for the model discussed in ref.\,\cite{Ballesteros:2020qam}, by solving numerically the Mukhanov-Sasaki equation (solid green line). 
We also show the value of $D_{\phi\phi}/(2\epsilon)$ for different $\sigma$ (red dashed lines). 
As discussed in the text, this quantity reproduces, in the limit of small $\sigma$, 
the power spectrum computed by means of the Mukhanov-Sasaki equation. 
Notice that $D_{\phi\phi}/(2\epsilon)$ is a function of the number of $e$-folds, but for fixed $\sigma$ 
the dependence on N can be translated into a $k$-dependence as discussed in eq.\,(\ref{eq:PowerSpectrumStoch}).
}
\end{center}
\end{figure} 

In the left panel of fig.\,\ref{fig:NumericalCheck} we show for each comoving wavenumber $k$ 
the transition time (solid black line) after which the corresponding mode freezes to its final constant value. 
This plot has to be compared with the right panel of fig.\,\ref{fig:Ka2}, and the agreement is evident. 
The region shaded in magenta marks the interval of $e$-folds during which the USR phase takes place (that is the interval 
between $N_{\rm in}$ and $N_{\rm end}$ in our analytical toy model). 
We also show contours of $k_{\sigma} = \sigma aH$ for different values of $\sigma$ with $\sigma=1$ corresponding to the horizon crossing condition. In the right panel of fig.\,\ref{fig:NumericalCheck} we show the power spectrum obtained by solving numerically the 
Mukhanov-Sasaki equation (solid green line). This plot has to be compared with the left panel of fig.\,\ref{fig:Ka2}, and shows that, as anticipated, the analytical approximation
captures all relevant features of the numerical solution.

\begin{figure}[!htb!]
\begin{center}
$$\includegraphics[width=.45\textwidth]{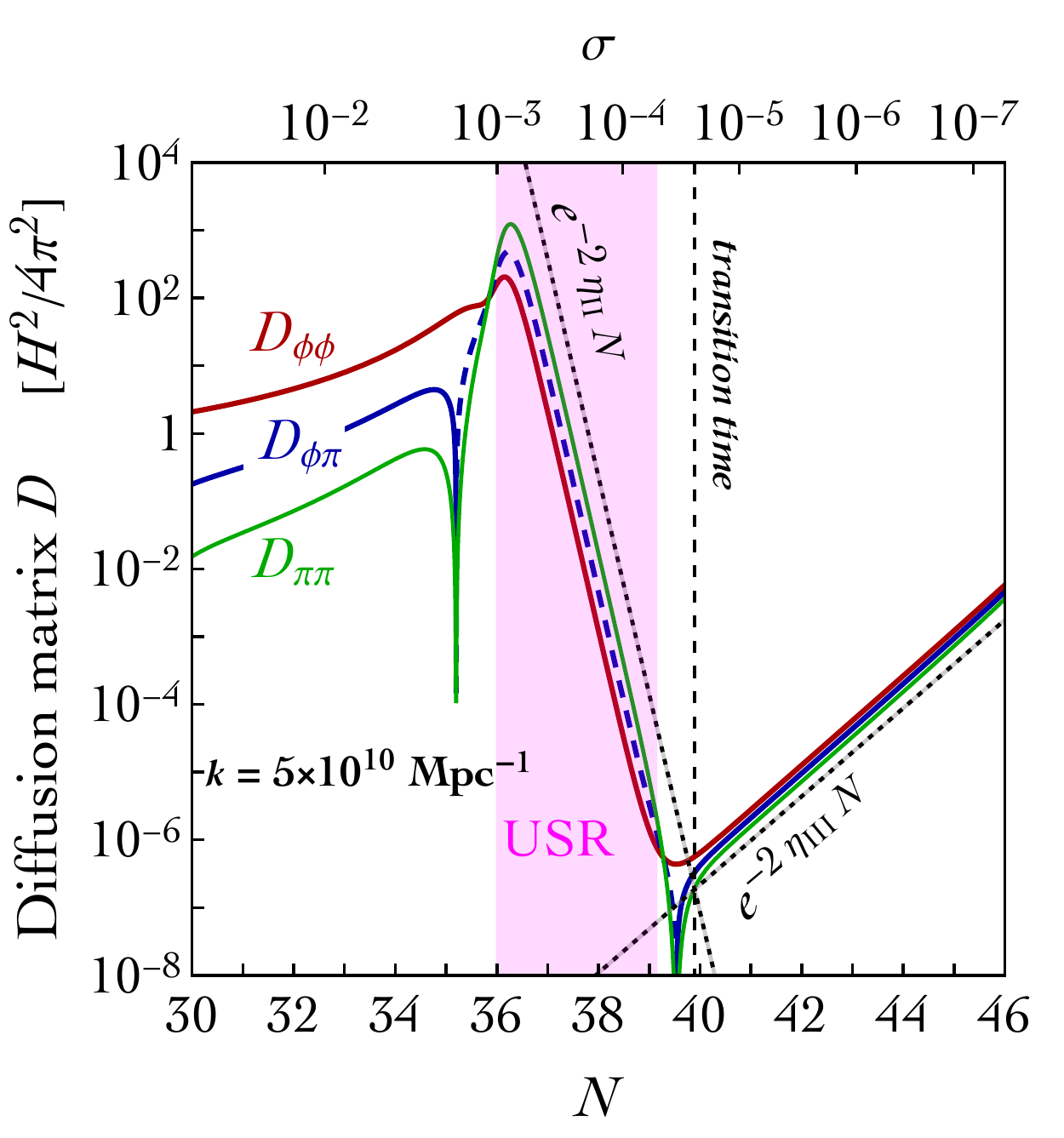}
\qquad\includegraphics[width=.45\textwidth]{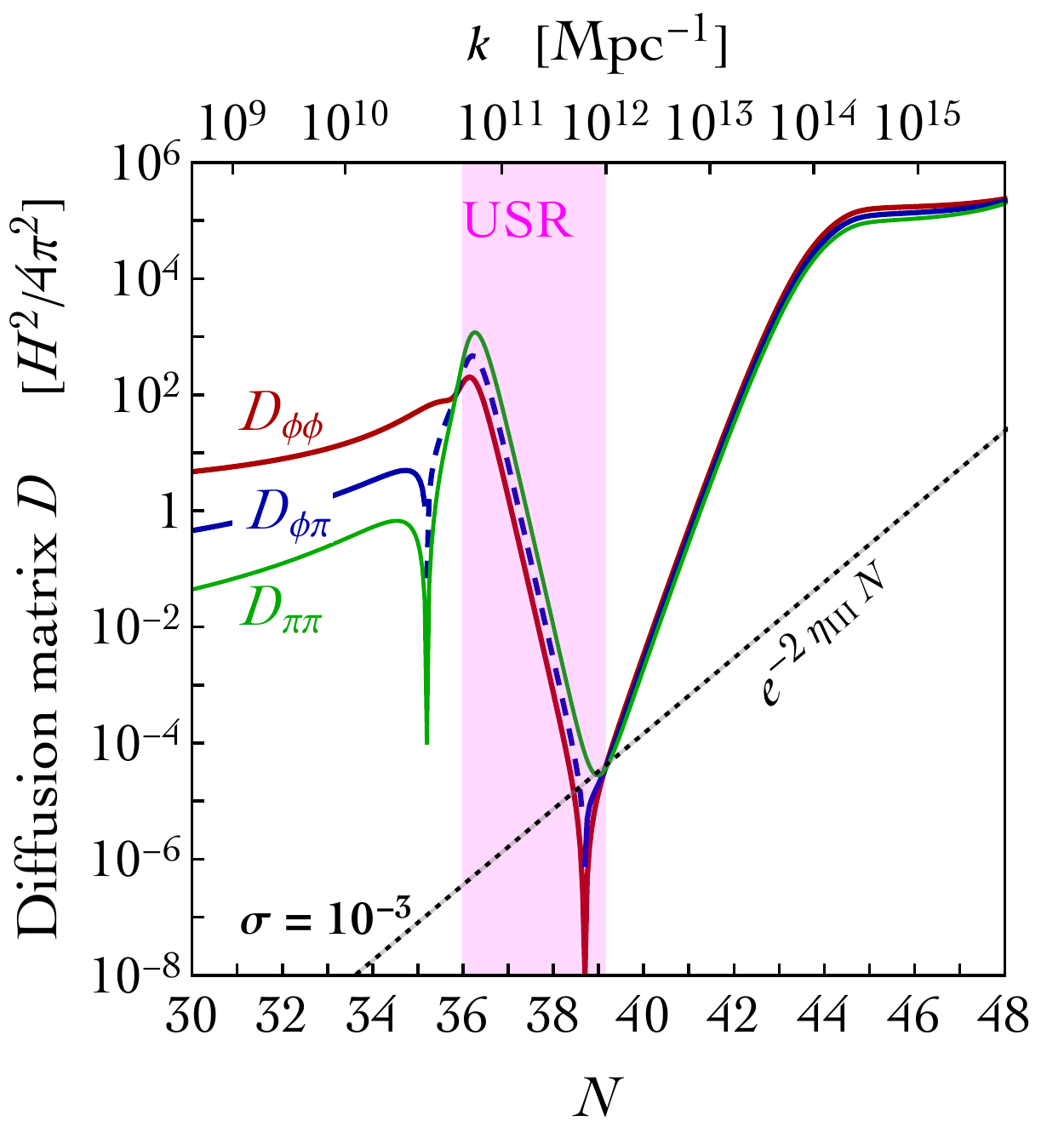}$$
\caption{\em \label{fig:DiffusionMatrix} 
Left panel. Elements of the diffusion matrix $D(N)$ in eq.\,(\ref{eq:DiffusionMatrix}) computed for 
the numerical model discussed in ref.\,\cite{Ballesteros:2020qam}. 
We fix $k=5\times 10^{10}$ Mpc$^{-1}$ and show our results as function of $N$ (equivalently, as function of $\sigma$). 
The dashed part of $D_{\phi\pi}$ indicates negative values.
Right panel. Same as in the left panel but for fixed $\sigma = 10^{-3}$. See text for discussion. 
The dotted black lines correspond to the analytical results in table\,\ref{eq:NoiseTable} derived in the $\sigma \to 0$ limit.
 }
\end{center}
\end{figure}

The next step consists of the computation of the elements of the diffusion matrix $D(N)$ in eq.\,(\ref{eq:DiffusionMatrix}).  
The latter are defined in terms of the noise correlators in eq.\,(\ref{eq:NoiseFold}) which take the form
\begin{align}\label{eq:NoiseCorrelatorFull}
 \Theta_{fg}(N)  = \frac{(1-\epsilon)}{2\pi^2}k_{\sigma}^3f_{k_{\sigma}}(N)g_{k_{\sigma}}^*(N)\,,
\end{align}
with $f_k,g_k = \phi_k,\pi_k$ and $k=k_{\sigma}=\sigma aH$ (with both $a$ and $H$ that depend on $N$).
Notice that the relation between $\phi_k$ and $\pi_k$ is given by eq.\,(\ref{eq:PhaseSpaceN}) and one should also include the gravitational potential term (which is proportional to $\epsilon$ and gives a sub-leading contribution in the analytical model). 
Compared to our analytical analysis, things are a little bit more complicated now because we want to keep $\sigma$ non-zero and finite.
In fact, unless one takes the limit $\sigma\to 0$, the diffusion matrix will depend, in addition to $N$, also on the specific choice of $\sigma$. To be more precise, we have three parameters ($k$, $N$ and $\sigma$) related by $k = k_{\sigma} = \sigma aH$.
We may interpret the time evolution in two different ways:
\begin{itemize}
\item [$\circ$] We keep the comoving wavenumber $k = k_{\sigma} = \sigma aH$ fixed. 
In this case the value of $\sigma$ is not fixed, but to each $N$ will correspond a different $\sigma$ such that the condition $k = k_{\sigma} = \sigma aH$ is verified. The outcome of this procedure is illustrated in the left panel of fig.\,\ref{fig:DiffusionMatrix} where we plot the components 
of $D(N)$ as a function of $N$ for the representative value $k=5\times 10^{10}$ Mpc$^{-1}$.  
This choice 
corresponds to a mode which crosses all three relevant regions in its dynamical evolution after horizon crossing (which happens at $N_k \simeq 28$, before the beginning of the USR phase). For each $N$ on the bottom-side $x$-axis the corresponding $\sigma$ is shown on the top-side $x$-axis. 
To guide the eye, one can just draw {a horizontal} line in the left panel of fig.\,\ref{fig:NumericalCheck} at the specific value of $k$ analyzed, and see at which $N$ each value of $\sigma aH$ is crossed. 
This way of reasoning can be useful for the computation of the power spectrum since the latter is evaluated as function of the comoving wavenumber $k$.

It is worth pausing for a number of comments. 
In region\,I (that is for $N\lesssim 36$, before the beginning of the USR phase) we have 
a sizable value of $\sigma$. 
Consequently, the approximation given in table\,\ref{eq:NoiseTable} in which $D_{\phi\phi}=H^2/4\pi^2$ 
and $D_{\phi\pi}=D_{\pi\pi} = 0$ is not valid since it is obtained in the $\sigma\to 0$ limit (and for constant $H$ in the toy model, while in the numerical model the evolution of $H$ is not neglected). 
Nevertheless, notice that the numerical result satisfies the expected analytical hierarchy since 
at the lowest non-vanishing order in $\sigma$ one finds $D_{\phi\pi} = O(\sigma^2)$ and $D_{\pi\pi} = O(\sigma^4)$.
As an additional interesting observation, notice that for $k=5\times 10^{10}$ Mpc$^{-1}$ 
the modes travel through the USR phase and enter  region\,III (that is, for $N\gtrsim 40$) 
for relatively small values of $\sigma$, namely $\sigma \lesssim 10^{-5}$. 
These values of $\sigma$ are so small that our analytical  approximations, strictly valid in the
 $\sigma \to 0$ limit, are now perfectly recovered (see table\,\ref{eq:NoiseTable}). 
 The same is true for the analytical results obtained in region\,II during which the USR takes place. This is shown by the dotted lines in the right panel of fig.\ \ref{fig:NumericalCheck}.

\item [$\circ$] We keep $\sigma$ fixed. In this case the value of $k$ is not fixed, but to each $N$ will correspond a different $k$ according to $k = k_{\sigma} = \sigma aH$. The outcome of this procedure is illustrated in the right panel of fig.\,\ref{fig:DiffusionMatrix} for the specific value
 $\sigma = 10^{-3}$. For each $N$, the corresponding $k$ is indicated on the top-side $x$-axis.
 In other words, instead of following {a horizontal} line for fixed $k$, we are now moving in diagonal 
for fixed $\sigma$ in the left panel of fig.\,\ref{fig:NumericalCheck}. 
Notice that in this case we are considering a sizable value of $\sigma$. Indeed, in region\,III we find 
that the 
analytical approximation in table\,\ref{eq:NoiseTable}, obtained in the $\sigma \to 0$ limit, 
does not match the numerical result.

\end{itemize}
These numerical results for the diffusion matrix $D(N)$ fully confirm our numerical estimate. 
The main result of this 
analysis is the following:
In the presence of an USR phase, taking the diffusion matrix to be as if 
slow-roll applied
(that is $D_{\phi\phi} = H^2/4\pi^2$ and $D_{\phi\pi} = D_{\pi\pi} = 0$)
is incorrect.\footnote{This is, for instance, what was done in ref.\,\cite{Biagetti:2018pjj}}
On the contrary, we find that the noise correlation functions drop exponentially fast during USR. Our physical interpretation of this result is given in section\,\ref{sec:Q2C}.
The USR dynamics tend to stop the ``inflow'' of modes that transit from the short- to the long-wavelength part 
of the inflaton field fluctuations, and delay it until after the end of the USR phase.
 This is because the classicalization of the quantum modes does not follow the usual horizon-crossing condition in 
the presence of USR but gets completed only after the latter ends.

Our results for the diffusion matrix disagree with those
found in ref.\,\cite{Ezquiaga:2018gbw}
in which $D_{\phi\phi}$ features a peak (instead of an exponential suppression) when the USR phase is crossed. In addition, we have been able to recover the alleged enhancement of the power spectrum from quantum diffusion reported in \cite{Ezquiaga:2018gbw}  by computing the spectrum at the horizon crossing condition $k=aH$ (that is, for $\sigma = 1$). As we have discussed, this prescription is incorrect and, instead, a sufficiently small value of $\sigma$ must be taken for each mode.
Indeed, we can evaluate the power spectrum in the stochastic approach (at small $\sigma$) for the numerical model by means of eq.\,(\ref{eq:PowerSpectrumStoch}). In particular, we can compute the contribution given by  $D_{\phi\phi}/(2\epsilon)$.
To this end, for a given comoving wavenumber $k$ 
we just need to evaluate, as explained in the discussion below eq.\,(\ref{eq:PowerSpectrumStoch}), 
the value of $D_{\phi\phi}$ corresponding to the number of $e$-folds fixed by the choice of $\sigma$.  
The plot in the left panel of fig.\,\ref{fig:DiffusionMatrix} is particularly illustrative.
For $k=5\times 10^{10}$ Mpc$^{-1}$, we can read immediately the value of  $D_{\phi\phi}$ for a given $N$ 
(equivalently, $\sigma$). 
From this plot, we also see that for
$\sigma \lesssim 10^{-5}$ (or smaller) the value of 
$D_{\phi\phi}$ 
follows the expected functional form $\propto e^{-2\eta N}$;
 consequently, the ratio $D_{\phi\phi}/(2\epsilon)$ settles to a constant value independent of $\sigma$ since 
 $\epsilon$ evolves in the same way as $D_{\phi\phi}$ after the USR phase (and the exponential dependence on $N$ cancels out in the ratio). 
 
 In the right panel of fig.\,\ref{fig:NumericalCheck} we show the outcome of this procedure for different values of 
 $\sigma$. 
 For $\sigma \lesssim 10^{-4}$ we find sizable deviations with respect to 
 the power spectrum obtained by solving the Mukhanov-Sasaki equation. 
 These deviations are concentrated between the dip and the peak of the power spectrum. 
Inspecting the left panel of fig.\,\ref{fig:NumericalCheck}, this is not surprising since 
 for $\sigma \gtrsim 10^{-4}$ we are including in the computation modes which are not yet classical. 
 On the other hand, if we take $\sigma \lesssim 10^{-5}$ we find that 
 $D_{\phi\phi}/2\epsilon$ quickly recovers the expected power spectrum. 
 These numerical results confirms what we found analytically in eqs.\,(\ref{eq:MasterP},\,\ref{eq:AnalSpectrum2}).
 
 \begin{figure}[!htb!]
\begin{center}
$$\includegraphics[width=.45\textwidth]{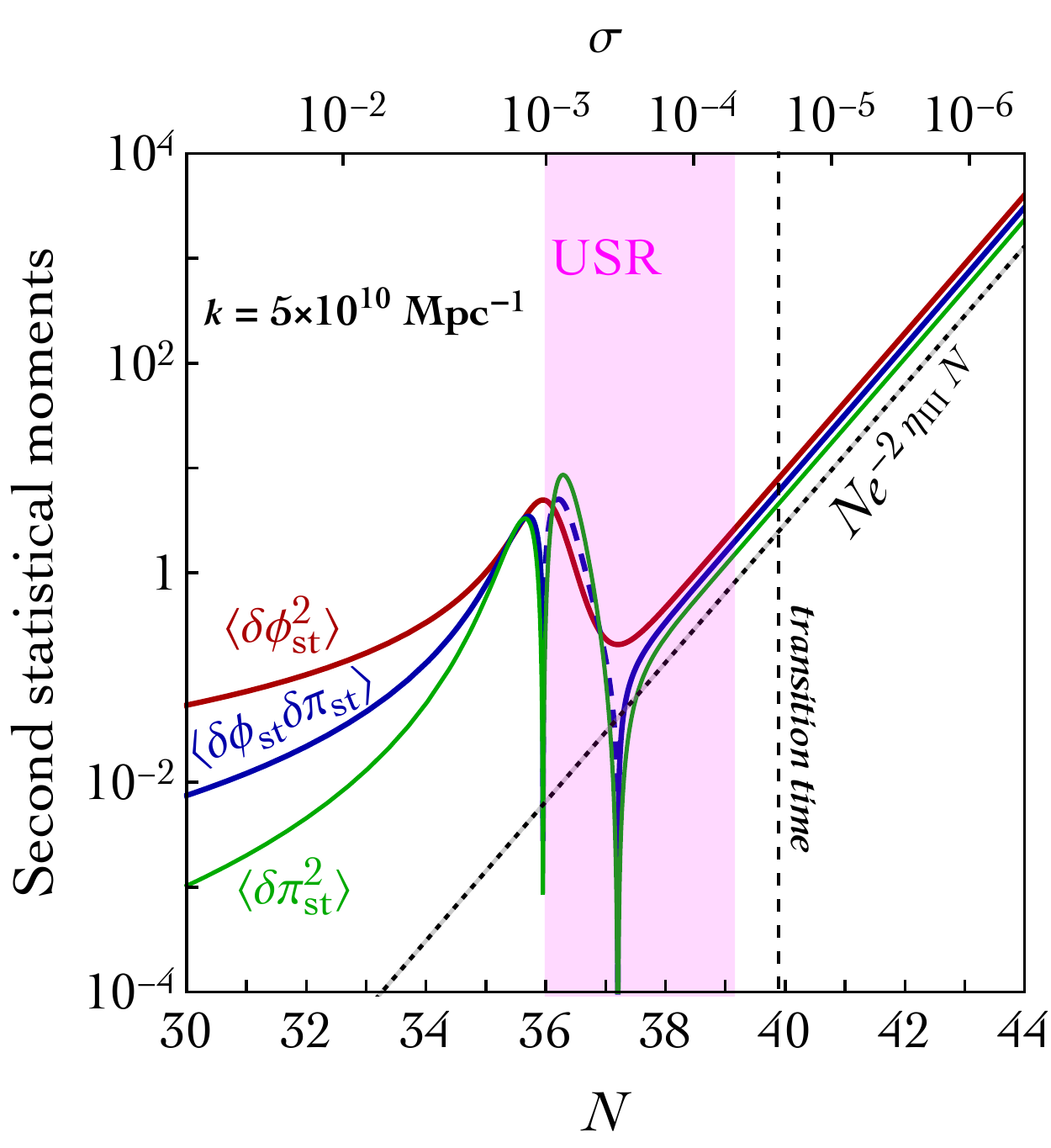}
\qquad\includegraphics[width=.45\textwidth]{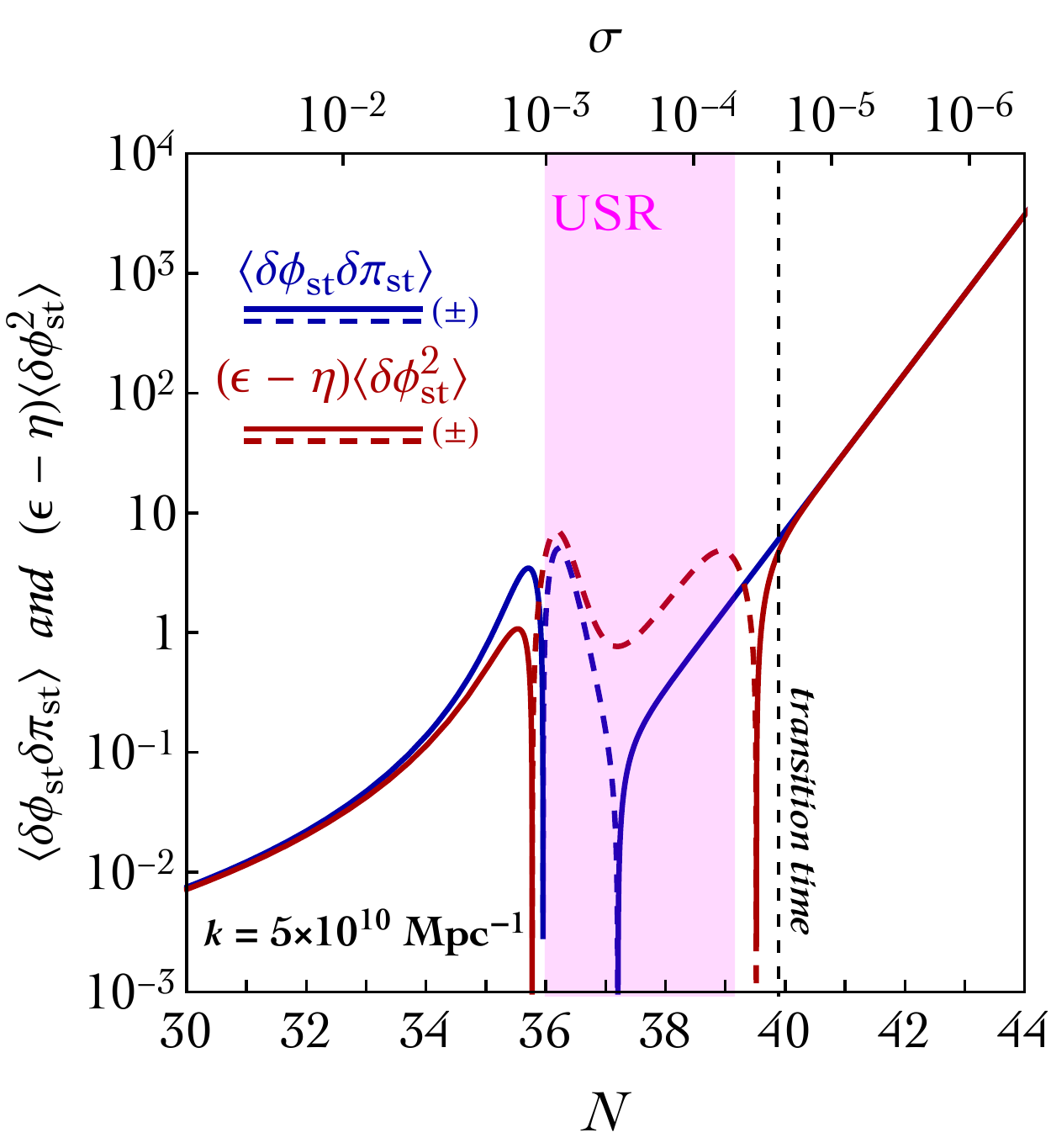}$$
\caption{\em \label{fig:SecondStatisticalMoments} 
Left panel. Numerical solutions of  eqs.\,(\ref{eq:StatMom1},\,\ref{eq:StatMom2},\,\ref{eq:StatMom3}) for 
the numerical model discussed in ref.\,\cite{Ballesteros:2020qam}. 
We fix $k=5\times 10^{10}$ Mpc$^{-1}$ and show our results as function of $N$ (equivalently, as function of $\sigma$). 
The dashed part of $\langle \delta\phi_{\rm st}\delta\pi_{\rm st}\rangle$ indicates negative values.
Right panel. Time evolution of the two terms $\langle \delta\phi_{\rm st}\delta\pi_{\rm st}\rangle$ and 
$(\epsilon-\eta)\langle \delta\phi_{\rm st}^2\rangle$ for $k=5\times 10^{10}$ Mpc$^{-1}$.
 }
\end{center}
\end{figure} 
 
What remains to be checked is the contribution coming from the combination 
$\langle \delta\phi_{\rm st}\delta\pi_{\rm st}\rangle - 
(\epsilon-\eta)\langle \delta\phi_{\rm st}^2\rangle$ which vanishes in the analytical approximation. In order to do so, we solve numerically the system formed by eqs.\,(\ref{eq:StatMom1},\,\ref{eq:StatMom2},\,\ref{eq:StatMom3}). 
We use vanishing initial conditions for the fields but we checked that our conclusions are not affected by this choice. 
Compared to eqs.\,(\ref{eq:MajorI},\,\ref{eq:MajorII},\,\ref{eq:MajorIII}) (obtained 
in the limit $\sigma \to 0$), the situation is now more complicated because we work at finite $\sigma$, and 
the solutions $\langle \delta\phi_{\rm st}^2\rangle$, 
$\langle \delta\phi_{\rm st}\delta\pi_{\rm st}\rangle$ and $\langle\delta\pi_{\rm st}^2\rangle$ 
will in general depend on $N$ and $\sigma$. As before, it is instructive to fix the value of $k$ and consider the solutions as functions of $N$ (and thus $\sigma$). 
In the left panel of fig.\,\ref{fig:SecondStatisticalMoments} we show 
$\langle \delta\phi_{\rm st}^2\rangle$, 
$\langle \delta\phi_{\rm st}\delta\pi_{\rm st}\rangle$ and $\langle\delta\pi_{\rm st}^2\rangle$ as functions of $N$ for
$k=5\times 10^{10}$ Mpc$^{-1}$.  
The logic of the plot follows what we already discussed for the left panel of fig.\,\ref{fig:DiffusionMatrix}. 
In region\,III, after the USR phase, the solutions settle down to their analytical estimates given in 
 eqs.\,(\ref{eq:MajorI},\,\ref{eq:MajorII},\,\ref{eq:MajorIII}).
 In the right panel of fig.\,\ref{fig:SecondStatisticalMoments} we investigate the behavior of 
 the combination $\langle \delta\phi_{\rm st}\delta\pi_{\rm st}\rangle - 
(\epsilon-\eta)\langle \delta\phi_{\rm st}^2\rangle$. 
For definiteness, we fix again $k=5\times 10^{10}$ Mpc$^{-1}$ and plot separately 
$\langle \delta\phi_{\rm st}\delta\pi_{\rm st}\rangle$ and $(\epsilon-\eta)\langle \delta\phi_{\rm st}^2\rangle$ (see labels in the figure). 
The take-home message of this plot is clear. After crossing the transition time into the classical regime after the end of the USR phase, 
$\langle \delta\phi_{\rm st}\delta\pi_{\rm st}\rangle$ and 
$\langle \delta\phi_{\rm st}^2\rangle$ settle to their asymptotic functional forms $\propto Ne^{-2\eta N}$ 
and the difference 
$\langle \delta\phi_{\rm st}\delta\pi_{\rm st}\rangle - 
(\epsilon-\eta)\langle \delta\phi_{\rm st}^2\rangle$ recovers the cancellation discussed in the analytical model. 
For the mode under consideration ($k=5\times 10^{10}$ Mpc$^{-1}$) this requires $\sigma \lesssim 10^{-5}$. 
If one takes a value of $\sigma$ that is too large (say, $\sigma \simeq 10^{-4}$ in this specific case) the difference 
$\langle \delta\phi_{\rm st}\delta\pi_{\rm st}\rangle - 
(\epsilon-\eta)\langle \delta\phi_{\rm st}^2\rangle$ gives a non-vanishing contribution to the power spectrum.
 This is a spurious contribution to the power spectrum since, as it is clear from the left panel of fig.\,\ref{fig:NumericalCheck}, 
this amounts to including modes that are still sizably affected by the USR phase. 
Such a choice of not-small-enough $\sigma$ is at odds with the very logic
of stochastic inflation in which the effect of quantum corrections perturb the coarse-grained field configuration only after their classicalization, when they can be interpreted as classical noise as we have argued in previous sections.
 
This analysis can be repeated without surprises for all modes that contribute to the enhancement of the power spectrum. 
After the end of the USR phase, when modes affected by the USR phase freeze to their final constant value and the quantum-to-classical transition takes place, the difference $\langle \delta\phi_{\rm st}\delta\pi_{\rm st}\rangle - 
(\epsilon-\eta)\langle \delta\phi_{\rm st}^2\rangle$ gives a vanishing contribution as we already found
 in the analytical discussion. 
The precise value of $\sigma$ at which this happens depends on the specific $k$ analyzed, as it is clear 
from the left panel in fig.\,\ref{fig:NumericalCheck}. For instance, modes close to the peak only require values as small as $\sigma = 10^{-4}$, but for modes close to the dip, even smaller values are needed. In conclusion, the numerical analysis of the model introduced in ref.\,\cite{Ballesteros:2020qam} 
confirms our analytical results derived for the toy model.

The bottom line is that the power spectrum of comoving curvature perturbations computed in stochastic inflation at the linear order  matches precisely the result obtained by solving the Mukhanov-Sasaki equation within the more conventional 
analysis based on the perturbative approach. 
 This result is not surprising per se, since it was already well-established in the context of slow-roll inflation, see e.g.\,\cite{Vennin:2015hra}.
 The non-trivial point of our analysis is that we have extended its validity to the case in which an USR phase is present, which is relevant for the formation of PBHs in many models of inflation.
 As a byproduct of our analysis, we have also clarified the role of the
stochastic noise and the issue of the quantum-to-classical transition in this scenario.

\subsection{Non-Gaussianities}\label{sec:Nongaussianities}

So far the discussion has been focused on the two-point statistical correlators. 
This is enough to compute the power spectrum of comoving curvature perturbations but it is legitimate to ask what changes when higher-order correlators are included in the analysis.
Unfortunately, the approach developed in section\,\ref{sec:QN} and applied in section\,\ref{sec:PowerSpectrumStoch} to the computation of the power spectrum is not well-suited here. 
The reason is that section\,\ref{sec:QN} is based on the linear expansion 
$\bar{\phi}=\phi_{\rm cl}+ \delta\phi_{\rm st}$ and $\bar{\pi}=\pi_{\rm cl}+ \delta\pi_{\rm st}$ while 
non-Gaussianities, 
measure deviations from linearity.
One can check that a generalization of
eqs.\,(\ref{eq:StatMom1},\,\ref{eq:StatMom2},\,\ref{eq:StatMom3}) to the third and fourth order in the 
statistical moments leads to the solutions $\langle \delta\phi_{\rm st}^3\rangle = 0$ and 
$\langle \delta\phi_{\rm st}^4\rangle = 3\langle \delta\phi_{\rm st}^2\rangle^2$ 
which is precisely what we expect for a Gaussian statistic.\footnote{This is in contrast with the findings of ref.\,\cite{Ezquiaga:2018gbw}, which also uses the linear expansion.}
This implies that the comoving curvature perturbation, defined at this order  by 
$\mathcal{R} = \delta\phi_{\rm st}/\sqrt{2\epsilon_{\rm cl}}$, is Gaussian.

The formalism adopted so far, therefore, is totally fine as long as we are only interested in the computation of the 
power spectrum but, to gain some insight about non-Gaussianities, we need to stray from the beaten path.
One obvious way to overcome this issue would be to extend our analysis 
to the second order in the stochastic perturbative expansion of the coarse-grained fields.  
Let us sketch what would be the result if such approach were pursued. 
We consider the second-order expansion 
$\bar{\phi}=\phi_{\rm cl}+ \delta\phi_{{\rm st},1} + \delta\phi_{{\rm st},2},$ 
where $\phi_{\rm cl}$, together with $\pi_{\rm cl}$, solves the deterministic Langevin equations 
without the noise term while the quantities $\delta\phi_{{\rm st},1}$ and $\delta\phi_{{\rm st},2}$ are 
stochastic variables of, respectively, first and second order in the noise. We write 
\begin{align} \label{ng1}
\mathcal{R} = (\delta\phi_{{\rm st},1}+ \delta\phi_{{\rm st},2}-\langle \delta\phi_{{\rm st},2}\rangle)
/\sqrt{2\epsilon_{\rm cl}}
\end{align} 
so that
$\langle \mathcal{R} \rangle = 0$. 
Furthermore, based on the previous consideration, the first-order term 
$\mathcal{R}_{\mathcal{G}} \equiv \delta\phi_{{\rm st},1}/\sqrt{2\epsilon_{\rm cl}}$ is Gaussian.
In the expression \eq{ng1}
we have a Gaussian term plus a correction which is formally, and by construction, 
of order $\delta\phi_{{\rm st},2}-\langle \delta\phi_{{\rm st},2}\rangle \sim O(\delta\phi_{{\rm st},1}^2)$.
We can recast \eq{ng1}
 in the conventional form
\begin{align}\label{eq:LocalBispectrum}
\mathcal{R} = \mathcal{R}_{\mathcal{G}} + \frac{3f_{\rm NL}}{5}
\left(\mathcal{R}_{\mathcal{G}}^2 - \langle \mathcal{R}_{\mathcal{G}}^2\rangle\right)\,,~~~~
{\rm with}~~\mathcal{R}_{\mathcal{G}} = \frac{\delta\phi_{{\rm st},1}}{\sqrt{2\epsilon_{\rm cl}}}\,,~~~~~
~~~f_{\rm NL}^2 = \frac{25}{9}(2\epsilon_{\rm cl})\left[
\frac{\langle \delta\phi_{{\rm st},2}^2\rangle - 
\langle \delta\phi_{{\rm st},2}\rangle^2}{
\langle \delta\phi_{{\rm st},1}^4\rangle - 
\langle \delta\phi_{{\rm st},1}^2\rangle^2
}
\right]\,.
\end{align}

From this simple analysis we learn a number of things. 
First, it confirms that non-Gaussianity of the comoving curvature perturbation emerges as a non-linear effect 
that can be captured only going beyond the first order in the stochastic perturbative
 expansion.
Second, implementing the expansion 
$\bar{\phi}=\phi_{\rm cl}+ \delta\phi_{{\rm st},1} + \delta\phi_{{\rm st},2}$ seems to capture 
only a specific kind of non-Gaussianities, that are of the so-called {\rm local} type (defined by the first  
expression in eq.\,(\ref{eq:LocalBispectrum})\,\cite{Komatsu:2001rj}).
Let us consider the three-point correlation function in momentum space:
\begin{align}\label{eq:Bispectrum}
\langle \mathcal{R}_{k_1}\mathcal{R}_{k_2}\mathcal{R}_{k_3} \rangle = (2\pi)^3\delta^{(3)}(\vec{k}_1+\vec{k}_2+\vec{k}_3)
B_{\mathcal{R}}(k_1,k_2,k_3)\,,
\end{align}
where $B_{\mathcal{R}}(k_1,k_2,k_3)$ defines the bispectrum. 
As it is well-known, local-type non-Gaussianities leave a specific imprint in the bispectrum and a direct computation in Fourier space gives
\begin{align}\label{eq:LocalBispectrumFourier}
B^{\rm \,loc}_{\mathcal{R}}(k_1,k_2,k_3) = (6f_{\rm NL}/5)[\Delta_{\mathcal{R}}(k_1)\Delta_{\mathcal{R}}(k_2) + 
\Delta_{\mathcal{R}}(k_1)\Delta_{\mathcal{R}}(k_3) +
\Delta_{\mathcal{R}}(k_2)\Delta_{\mathcal{R}}(k_3)]\,.
\end{align}
In the local limit, therefore, the bispectrum acquires a simple factorizable form. 
Furthermore, in the squeezed limit ($k_1=k_2\equiv k\gg k_3$) two of the three momenta are indistinguishable, and 
we have $B^{\rm loc,sq}_{\mathcal{R}}(k,k,k_3) = (12\tilde{f}_{\rm NL}/5)\Delta_{\mathcal{R}}(k)\Delta_{\mathcal{R}}(k_3)$.

Notice that, to the best of our knowledge,  describing non-Gaussianities in the local limit
seems to be a limitation of the stochastic inflation approach. 
For instance, ref.\,\cite{Vennin:2015hra} computed non-Gaussianities in the context of stochastic inflation by means of the 
so-called $\delta N$ formalism (see ref.\,\cite{Abolhasani:2019cqw} for a review) 
but also in this case the computation is limited to the 
local bispectrum and, furthermore, is not valid for arbitrary configurations of the 
momenta since the squeezed limit is assumed.\footnote{See, however, ref.\,\cite{Tada:2016pmk} in which the author discuss, in the context of the $\delta N$ formalism, 
a general way that allows to obtain more generic bispectra (that is, without assuming the local limit but still working in the squeezed configuration).} 
However, whether the presence of a USR phase gives rise to non-Gaussianities which are exclusively of local type (and, therefore, computable within the formalism of the second-order expansion 
$\bar{\phi}=\phi_{\rm cl}+ \delta\phi_{{\rm st},1} + \delta\phi_{{\rm st},2}$ by means of eq.\,(\ref{eq:LocalBispectrum}) which would give the value of $f_{\rm NL}$ in eq.\,(\ref{eq:LocalBispectrumFourier})) is not something that can be a priori established.

A different approach to the problem has been pursued in refs.\,\cite{Ezquiaga:2019ftu,Pattison:2017mbe}.
The idea is to compute directly the probability distribution function (PDF) of the comoving curvature perturbation $\mathcal{R},$ abandoning the perturbative approach described before.
In the stochastic $\delta N$ formalism the comoving curvature perturbations can be related to the variations in the number of $e$-folds of expansion induced by the quantum fluctuations of the inflation.
In practice, the PDF of $\mathcal{R}$ is obtained from the PDF of the number of $e$-folds, and the
latter is computed solving a Fokker-Planck equation.
Following this method and focusing on slow-roll dynamics, ref.\,\cite{Ezquiaga:2019ftu} finds that the PDF of $\mathcal{R}$ has highly non-Gaussian tails, which can not be described with standard non-Gaussian expansions.
This implies that the abundance of PBHs largely deviates from the estimates obtained with the Gaussian statistics.
It would be interesting to extend the analysis of ref.\,\cite{Ezquiaga:2019ftu} to the class of models discussed here, featuring an epoch of slow-roll violation.
The role of non-Gaussianities in the context of inflationary models with an USR phase deserves a dedicated analysis which will be presented elsewhere.

\section{Summary and conclusions}\label{sec:Conc}
In this work we have analyzed the role of quantum diffusion during inflation. More specifically we have focused on an inflationary scenario featuring an USR phase, which is relevant for the production of PBHs.
Below we summarize the main findings and novelties of our study.
\begin{itemize}

\item [$\circ$] The back-reaction of quantum fluctuations on the inflationary dynamics is described in the context of stochastic inflation. In this approach the long wavelength perturbations of the inflaton are sourced by the modes which cross a coarse-graining scale, $k\lesssim k_{\sigma}=\sigma\,aH$ with $\sigma<1,$ and become semi-classical, behaving as an external noise source, see eqs.\,(\ref{eq:Lang1},\,\ref{eq:Lang2},\,\ref{eq:XiDefinition}). In this way the evolution of the inflaton is treated as a diffusion process.
Clearly, to properly model the stochastic dynamics it is crucial to understand the quantum-to-classical transition.
We have considered two diagnostic tools to study such a regime, namely the time-dependent occupation number and a quantity related to the anti-commutator of the field perturbation operator and its conjugate momentum (see section\,\ref{sec:Q2C} and appendix\, \ref{app:BasicDef} for details). For each comoving wavelength $k$, these functions vanish in the original Minkowski vacuum and, in standard slow-roll, take large values once the mode crosses the Hubble horizon, a signal of classicalization.
We have shown that this picture can be significantly altered in presence of an USR phase. For peaked spectra and for comoving scales from around the dip and up to the peak of the power spectrum of comoving curvature perturbations, the quantum-to-classical transition is delayed, and completes only when the USR phase ends. This goes in parallel with the evolution of the comoving curvature perturbation $\mathcal{R}_{k},$ which, for these modes, freezes to a constant value only when the USR phase is over.

\item [$\circ$] We have solved the classical dynamics in eqs.\,(\ref{eq:Lang1},\,\ref{eq:Lang2}) expanding at first order the coarse-grained inflaton field $\bar{\phi}$ and its conjugate momentum $\bar{\pi}$ around their classical trajectories, $\bar{\phi}=\phi_{\rm cl}+ \delta\phi_{\rm st}$ and $\bar{\pi}=\pi_{\rm cl}+ \delta\pi_{\rm st}.$
The evolution of the perturbations $\delta\pi_{\rm st}$ and $\delta\phi_{\rm st}$ is governed by the noise correlation matrix. The latter, at leading order in $\sigma$ and in standard slow-roll, has a simple structure: it is constant in time
and non-vanishing only in the $_{\phi\phi}$ direction. Once again, this result dramatically changes during the USR phase. We have found that during this stage of the inflationary evolution, the inflaton diffuses in all three directions in phase-space, and the noise correlation functions fall exponentially fast. The reason is that, as explained above, classicalization is delayed during the USR phase, and this implies that the inflow of modes which source the noise correlation functions is stopped. 

\item [$\circ$] Finally, we have computed the power spectrum of comoving curvature perturbations at linear order in stochastic inflation. We have demonstrated that it matches the standard result obtained with the more conventional approach based on {the splitting into background and perturbations}.
This outcome was already known in the case of slow-roll inflation. We have shown that it holds also in presence of an USR phase, which is the key ingredient for the formation of PBHs in a broad class of models of single-field inflation.
A proper treatment of the quantum-to-classical transition is crucial to understand this result. 
We reiterate that in presence of an USR phase, the usual horizon crossing condition is not a good proxy for the classicalization of the modes.
Indeed, we have shown that following this incorrect prescription (i.e. $\sigma=1$), spurious features appear in the power spectrum of comoving curvature perturbations, as it occurs in ref.\,\cite{Ezquiaga:2018gbw}.  
Notice also that the power spectrum does not depend on the coarse-graining parameter $\sigma$ (as it should be) as long as this quantity is taken small enough to guarantee the classicalization of the modes.

Our results have been obtained in the context of an analytical model, which, despite its simplicity and approximations, captures remarkably well the inflationary dynamics with a transient USR phase, and allow for a transparent 
treatment.  We have then confirmed our results by means of a full numerical analysis, performed in the context of a concrete inflationary model which generates all the DM in the form of PBHs.   

An analysis of the power spectrum of curvature perturbations in stochastic inflation at linear order in fluctuations for $\eta = 3$ was performed in ref.\,\cite{Cruces:2018cvq}. We agree with the conclusion expressed in ref.\,\cite{Cruces:2018cvq} that the power spectrum of comoving curvature perturbations is not altered by the stochastic dynamics with respect to standard perturbation theory. Notice, however, that the results for the noise correlation matrix of ref.\,\cite{Cruces:2018cvq} match the ones we have obtained for Region I, shown in Table \ref{eq:NoiseTable}. We have extended their analysis by including the Regions II and III, relevant for PBH formation.

\item [$\circ$] The focus of our work {has been} the calculation of the two-point function in stochastic inflation. It would be interesting to extend the analysis to higher order correlators, in order to determine the non-Gaussian corrections. For this purpose, one should go beyond the linear expansion that we have adopted in this paper. Its is worth mentioning a couple of possible directions.
One possible strategy could be the one discussed in section\,\ref{sec:Stoch}. One could solve multiple times the Langevin eqs.\,(\ref{eq:Lang1},\,\ref{eq:Lang2}) and from these stochastic realizations infer the PDF of the field fluctuations (similarly to ref.\,\cite{Biagetti:2018pjj} but implementing our results for the two-point correlators of the noise function). Another approach, adopted in ref.\,\cite{Ezquiaga:2019ftu} in the context of slow-roll inflation, is to compute the PDF of the comoving curvature perturbations using the stochastic $\delta N$ formalism. It would be {worthwhile to extend} the analysis of ref.\,\cite{Ezquiaga:2019ftu} to scenarios where slow-roll is violated, as {would be the case} in presence of an USR phase. We leave these studies for future work.
\end{itemize}

\begin{acknowledgments}
The work of G.B.\ and J.R.\ is funded by a {\it Contrato de Atracci\'on de Talento (Modalidad 1) de la Comunidad de Madrid} (Spain), with number 2017-T1/TIC-5520, by {\it MINECO} (Spain) under contract FPA2016-78022-P, {\it MCIU} (Spain) through contract PGC2018-096646-A-I00 and by the IFT UAM-CSIC Centro de Excelencia Severo Ochoa SEV-2016-0597 grant.
The research of A.U. is supported in part by the MIUR under contract 2017\,FMJFMW (``{New Avenues in Strong Dynamics},'' PRIN\,2017) 
and by the INFN grant 
``SESAMO -- 
SinergiE di SApore e Materia Oscura.''
M.T. acknowledges support from the INFN grant ``LINDARK,'' the research grant ``The Dark Universe: A Synergic Multimessenger Approach No. 2017X7X85'' funded by MIUR, and the project ``Theoretical Astroparticle Physics (TAsP)'' funded by the INFN.
\end{acknowledgments}

\appendix

\section{Quantization of scalar perturbations in the Heisenberg picture 
and quantum-to-classical transition}\label{app:BasicDef}

Consider the field perturbation $u(\tau,\vec{x})$ that satisfies, in position space, the equation of motion
\begin{align}
\left(\frac{d^2}{d\tau^2} - \triangle  - \frac{1}{z}\frac{d^2z}{d\tau^2}\right)u(\tau,\vec{x}) = 0\,.
\end{align}
To quantize the system, one first notices that  this equation of motion can be derived--modulo a total derivative--from the action\footnote{We use the identity 
\begin{align}
(u^{\prime})^2 + \frac{z^{\prime\prime}}{z}u^2 = \left(u^{\prime} - \frac{z^{\prime}}{z}u\right)^2 + \left(\frac{z^{\prime}}{z} u^2\right)^{\prime}\,.
\end{align}
}
 (interchangeably, in the following we shall also use the short-hand notation $^{\prime} \equiv d/d\tau$ and 
 $(\partial u)^2=(\partial_i u)(\partial_i u)$)
\begin{align}\label{eq:QuadraticAction}
\mathcal{S}_2 = \frac{1}{2}\int d\tau d^3\vec{x}\left[\left(
u^{\prime} - \frac{z^{\prime}}{z}u
\right)^2 - (\partial u)^2
\right]\,.
\end{align}
Defining the conjugate momentum $p = \delta\mathcal{S}_2/\delta u^{\prime}$, we obtain $p = u^{\prime} - (z^{\prime}/z)u$. 
Consequently, the Hamiltonian is
\begin{align}
\mathcal{H}(\tau) = \frac{1}{2}\int d^3\vec{x}\left[
p^2 + (\partial u)^2 + \frac{2z^{\prime}}{z}pu
\right]\,.
\end{align}
We promote $u(\tau,\vec{x})$ and $p(\tau,\vec{x})$ to quantum operators $\hat{u}(\tau,\vec{x})$ and $\hat{p}(\tau,\vec{x})$ with equal-time commutation relations $[\hat{u}(\tau,\vec{x}),
\hat{p}(\tau,\vec{x}^{\prime})] = i\delta^{(3)}(\vec{x}-\vec{x}^{\prime})$. 
In Fourier space, we find the following Hamiltonian operator\footnote{We use the following $2\pi$-convention for the Fourier transform
\begin{align}\label{eq:RealityCond}
\hat{f}(\tau,\vec{x}) = \int\frac{d^3\vec{k}}{(2\pi)^{3/2}} \hat{f}(\tau,\vec{k})e^{i\vec{k}\cdot\vec{x}}\,.
\end{align}
For operators describing  real fields (i.e. $\hat{f}(\tau,\vec{x}) = \hat{f}^{\dag}(\tau,\vec{x})$) we have $\hat{f}(\tau,\vec{k}) = \hat{f}^{\dag}(\tau,-\vec{k})$.}
\begin{align}\label{eq:MainH}
\hat{\mathcal{H}}(\tau) = \frac{1}{2}\int d^3\vec{k}\left\{
\hat{p}(\tau,\vec{k})\hat{p}^{\dag}(\tau,\vec{k}) + k^2\hat{u}(\tau,\vec{k})\hat{u}^{\dag}(\tau,\vec{k}) + 
\frac{z^{\prime}}{z}\left[
\hat{p}(\tau,\vec{k})\hat{u}^{\dag}(\tau,\vec{k}) + \hat{u}(\tau,\vec{k})\hat{p}^{\dag}(\tau,\vec{k})
\right]
\right\}\,,
\end{align}
with commutation relations $[\hat{u}(\tau,\vec{k}),\hat{p}(\tau,\vec{k}^{\prime})] = i\delta^{(3)}(\vec{k}+\vec{k}^{\prime})$ and 
$[\hat{u}(\tau,\vec{k}),\hat{p}^{\dag}(\tau,\vec{k}^{\prime})] = i\delta^{(3)}(\vec{k}-\vec{k}^{\prime})$. 
We study the evolution of the system in the Heisenberg picture. To this end, we introduce the time-dependent ladder operators\footnote{Eqs.\,(\ref{eq:FieldOp},\,\ref{eq:MomentumOp}) follow from 
the usual form of the lowering operator
\begin{align}
a_{\vec{k}}(\tau) = \sqrt{\frac{k}{2}}\hat{u}(\tau,\vec{k}) + \frac{i}{\sqrt{2k}}\hat{p}(\tau,\vec{k})\,,
\end{align}
after noticing that 
\begin{align}
a^{\dag}_{\vec{k}}(\tau) = \sqrt{\frac{k}{2}}\hat{u}^{\dag}(\tau,\vec{k}) - \frac{i}{\sqrt{2k}}\hat{p}^{\dag}(\tau,\vec{k}) 
=\sqrt{\frac{k}{2}}\hat{u}(\tau,-\vec{k}) - \frac{i}{\sqrt{2k}}\hat{p}(\tau,-\vec{k}) 
 \,,
\end{align}
where the last step follows from the comment below eq.\,(\ref{eq:RealityCond}).
}
\begin{align}
\hat{u}(\tau,\vec{k}) & = \frac{1}{\sqrt{2k}}\left[a_{\vec{k}}(\tau) + a^{\dag}_{-\vec{k}}(\tau)\right]\,,	\label{eq:FieldOp}	\\
\hat{p}(\tau,\vec{k}) & = -i\sqrt{\frac{k}{2}}\left[a_{\vec{k}}(\tau) - a^{\dag}_{-\vec{k}}(\tau)\right]\,,	\label{eq:MomentumOp}
\end{align}
which have equal-time commutation relations $[a_{\vec{k}}(\tau),a^{\dag}_{\vec{k}^{\prime}}(\tau)] = \delta^{(3)}(\vec{k}-\vec{k}^{\prime})$. 
The Hamiltonian takes the form
\begin{align}
\hat{\mathcal{H}}(\tau) &= \frac{1}{2}\int d^3\vec{k}\bigg\{
k\bigg[a_{\vec{k}}(\tau)
a^{\dag}_{\vec{k}}(\tau) + a^{\dag}_{-\vec{k}}(\tau)a_{-\vec{k}}(\tau)
\bigg] + \frac{i}{z}\frac{dz}{d\tau}
\bigg[
a^{\dag}_{-\vec{k}}(\tau)a^{\dag}_{\vec{k}}(\tau) - a_{\vec{k}}(\tau)a_{-\vec{k}}(\tau)
\bigg]
\bigg\}\,.
\end{align}
The Heisenberg equations can be written, in close matrix form
\begin{align}
\frac{d}{d\tau}
\left(
\begin{array}{c}
 a_{\vec{k}}(\tau)   \\
   \\
  a^{\dag}_{-\vec{k}}(\tau)
\end{array}
\right) & =  
\left(
\begin{array}{cc}
 -ik &   z^{\prime}/z \\
 &  \\
 z^{\prime}/z  & ik
\end{array}
\right)\left(
\begin{array}{c}
 a_{\vec{k}}(\tau)   \\
   \\
  a^{\dag}_{-\vec{k}}(\tau)
\end{array}
\right)\,.
\end{align}
The off-diagonal terms are responsible for particle creation in curved space-time.
To solve this system, we use a Bogoliubov transformation. Starting from some given initial condition at conformal time $\tau_{\star}$, the ladder operators at conformal time $\tau$
can be written as:
\begin{align}
a_{\vec{k}}(\tau) & = y_{\vec{k}}(\tau)a_{\vec{k}}(\tau_{\star}) +  w_{\vec{k}}(\tau)a^{\dag}_{-\vec{k}}(\tau_{\star})\,,\label{eq:Ladder1}\\
a^{\dag}_{-\vec{k}}(\tau) & = y^*_{\vec{k}}(\tau)a^{\dag}_{-\vec{k}}(\tau_{\star}) +  w^*_{\vec{k}}(\tau)a_{\vec{k}}(\tau_{\star})\,,\label{eq:Ladder2}
\end{align}
with the condition $|y_{\vec{k}}(\tau)|^2 - |w_{\vec{k}}(\tau)|^2 = 1$ which follows from the fact that the commutation relations among ladder operators must be preserved during the unitary evolution.
From eqs.\,(\ref{eq:FieldOp},\,\ref{eq:MomentumOp}), we find
\begin{align}
\hat{u}(\tau,\vec{k}) & =  u_{\vec{k}}(\tau)a_{\vec{k}}(\tau_{\star}) +  u^*_{\vec{k}}(\tau)a^{\dag}_{-\vec{k}}(\tau_{\star}) \,,	\label{eq:FieldOp2}	\\
\hat{p}(\tau,\vec{k}) & =  -i\left[p_{\vec{k}}(\tau)a_{\vec{k}}(\tau_{\star}) -  p^*_{\vec{k}}(\tau)a^{\dag}_{-\vec{k}}(\tau_{\star})\right]\,,                                  \label{eq:MomentumOp2}
\end{align}
where we defined
\begin{align}
u_{\vec{k}}(\tau) & \equiv \frac{1}{\sqrt{2k}}\left[y_{\vec{k}}(\tau) + w^*_{\vec{k}}(\tau)\right]\,,\label{eq:Inv1} \\
p_{\vec{k}}(\tau) & \equiv \sqrt{\frac{k}{2}}\left[y_{\vec{k}}(\tau) - w^*_{\vec{k}}(\tau)\right]\,.\label{eq:Inv2}
\end{align}
It is easy to see that $u_{\vec{k}}$ and $p_{\vec{k}}$ satisfy the following equations
\begin{align}
\frac{d^2u_k}{d\tau^2} + \left(k^2 - \frac{1}{z}\frac{d^2z}{d\tau^2}\right)u_k = 0\,,~~~~~&{\rm with}~~~~u_k(\tau_{\star}) = \frac{1}{\sqrt{2k}}\,,\label{eq:Mode1}\\
p_k(\tau) = i\left[\frac{du_k}{d\tau} - \frac{1}{z}\frac{dz}{d\tau}u_k\right]\,,~~~~~&{\rm with}~~~~p_k(\tau_{\star}) = \sqrt{\frac{k}{2}}\,.\label{eq:Mode2}
\end{align}
Notice that from the condition $|y_{\vec{k}}(\tau)|^2 - |w_{\vec{k}}(\tau)|^2 = 1$ we have, by means of eqs.\,(\ref{eq:Inv1},\,\ref{eq:Inv2}), the Wronskian condition
\begin{align}\label{eq:Wronsk}
p_k(\tau)u_k^*(\tau) + p_k^*(\tau)u_k(\tau) = 1~~~~~\Longrightarrow~~~~~i\left[\frac{du_k(\tau)}{d\tau}u_k^*(\tau)- \frac{du_k^*(\tau)}{d\tau}u_k(\tau)\right] = 1\,.
\end{align}
The strategy is to solve eqs.\,(\ref{eq:Mode1},\,\ref{eq:Mode2}) and reconstruct--by inverting eqs.\,(\ref{eq:Inv1},\,\ref{eq:Inv2})--the evolution of the ladder operators in 
eqs.\,(\ref{eq:Ladder1},\,\ref{eq:Ladder2}). In eqs.\,(\ref{eq:Mode1},\,\ref{eq:Mode2}) we switched to the notation $u_{\vec{k}}\to u_{k}$ and $p_{\vec{k}}\to p_{k}$ since the mode functions only depend on the modulus of $\vec{k}$ (the same is true for $y_{\vec{k}}$ and $w_{\vec{k}}$). The initial conditions for $u_k$ and $p_k$ follow trivially from eqs.\,(\ref{eq:Ladder1},\,\ref{eq:Ladder2}). 
As far as the initial value $\tau_{\star}$ is concerned, we assume, as customary in the context of inflation, that the system starts in the vacuum state $|0\rangle$ defined by the condition $a_{\vec{k}}(\tau_{\star})|0\rangle = 0$. 
We are now in position to compute the occupation number used in section\,\ref{sec:Q2C}.  
The time-dependent occupation number $n_k(\tau)$ is defined, for each mode $k$, by the expectation value in the original vacuum state of the time-dependent particle 
number operator $a^{\dag}_{\vec{k}}(\tau)a_{\vec{k}}(\tau)$. 
We obtain
\begin{align}\label{eq:Modeuk}
\bar{n}_k(\tau)& = \langle 0|a^{\dag}_{\vec{k}}(\tau)a_{\vec{k}}(\tau)|0\rangle = 
\langle 0|\left[
y^*_{k}(\tau)a^{\dag}_{\vec{k}}(\tau_{\star}) +  w^*_{k}(\tau)a_{-\vec{k}}(\tau_{\star})
\right]\left[
y_{k}(\tau)a_{\vec{k}}(\tau_{\star}) +  w_{k}(\tau)a^{\dag}_{-\vec{k}}(\tau_{\star})
\right]|0\rangle =  \\
&= |w_k(\tau)|^2\delta^{(3)}(0) \equiv n_k(\tau)\delta^{(3)}(0)\,,
\end{align}
where the $\delta^{(3)}$ in momentum space at zero argument has the usual interpretation of spatial volume and arises because we are computing the total number of particles rather than the number density, the latter thus
being $n_k(\tau) = |w_k(\tau)|^2$. Summarizing, if we solve the system in eqs.\,(\ref{eq:Inv1},\,\ref{eq:Inv2}) for $w_k^*$ and use the Wronskian condition, the occupation number density can be written as:
\begin{align}\label{eq:Explicitnk}
n_k(\tau) = \frac{k}{2}|u_k(\tau)|^2 + \frac{1}{2k}|p_k(\tau)|^2 -\frac{1}{2}\,.
\end{align}

In the following we shall consider another tool to investigate the quantum-to-classical transition. Let us start from the following definition
\begin{align}
\Delta\hat{u}(\tau,\vec{k}) \equiv \hat{u}(\tau,\vec{k}) - \langle 0|\hat{u}(\tau,\vec{k})|0\rangle\,,
\end{align}
together with the analogue expression involving $\hat{p}(\tau,\vec{k})$. We can compute the following expectation values
\begin{align}
\langle 0|\Delta\hat{u}(\tau,\vec{k})\Delta\hat{u}^{\dag}(\tau,\vec{k}^{\prime}) |0\rangle & = |u_k(\tau)|^2\delta^{(3)}(\vec{k} - \vec{k}^{\prime})\equiv \Delta u^2_k(\tau)\delta^{(3)}(\vec{k} - \vec{k}^{\prime})\,,\label{eq:Heis1}\\
\langle 0|\Delta\hat{p}(\tau,\vec{k})\Delta\hat{p}^{\dag}(\tau,\vec{k}^{\prime}) |0\rangle & = |p_k(\tau)|^2\delta^{(3)}(\vec{k} - \vec{k}^{\prime})\equiv \Delta p^2_k(\tau)\delta^{(3)}(\vec{k} - \vec{k}^{\prime})\,.\label{eq:Heis2}
\end{align}
Therefore we find $\Delta u^2_k(\tau)\Delta p^2_k(\tau) = |u_k(\tau)|^2|p_k(\tau)|^2 \equiv |\mathcal{J}_k(\tau)|^2$. 
By using the square of the Wronskian condition in eq.\,(\ref{eq:Wronsk}), it is easy to see that 
\begin{align}
0 \leqslant |u_k(\tau)p_k^*(\tau) - u_k^*(\tau)p_k(\tau)|^2 = 4 |\mathcal{J}_k(\tau)|^2 - 1~~~~\Longrightarrow~~~~|\mathcal{J}_k(\tau)|^2 \geqslant 1/4\,.
\end{align}
By combining this result together with eqs.\,(\ref{eq:Heis1},\,\ref{eq:Heis2}) we find the analogous of the Heisenberg uncertainty relation 
\begin{align}\label{eq:Unc}
\Delta u^2_k(\tau)\Delta p^2_k(\tau) = |\mathcal{J}_k(\tau)|^2 \geqslant 1/4\,.
\end{align}
As a consequence of eqs.\,(\ref{eq:Mode1},\,\ref{eq:Mode2}), we trivially see that  the uncertainty is minimized, as it should be, in the original vacuum state.
Furthermore, we can also compute the expectation value of the commutator and anti-commutator
\begin{align}
\langle 0|\hat{u}(\tau,\vec{k})\hat{p}^{\dag}(\tau,\vec{k}^{\prime})+\hat{p}^{\dag}(\tau,\vec{k})\hat{u}(\tau,\vec{k}^{\prime}) |0\rangle & = 
i\underbrace{\left[u_k(\tau)p_k^{*}(\tau)-u_k^*(\tau)p_k(\tau)\right]}_{\equiv \langle\{\hat{u}(\tau,\vec{k}),\hat{p}^{\dag}(\tau,\vec{k})\}\rangle}\delta^{(3)}(\vec{k} - \vec{k}^{\prime})\,,\\
\langle 0|\hat{u}(\tau,\vec{k})\hat{p}^{\dag}(\tau,\vec{k}^{\prime})-\hat{p}^{\dag}(\tau,\vec{k})\hat{u}(\tau,\vec{k}^{\prime}) |0\rangle & = 
i\underbrace{\left[u_k(\tau)p_k^{*}(\tau)+u_k^*(\tau)p_k(\tau)\right]}_{\equiv \langle[\hat{u}(\tau,\vec{k}),\hat{p}^{\dag}(\tau,\vec{k})]\rangle}\delta^{(3)}(\vec{k} - \vec{k}^{\prime})\,.
\end{align}
Using the Wronskian condition one gets:
\begin{align}
\big|\big\langle\big\{\hat{u}(\tau,\vec{k}),\hat{p}^{\dag}(\tau,\vec{k})\big\}\big\rangle\big|^2  & = 4|\mathcal{J}_k(\tau)|^2 - 1\,,\\
\big|\big\langle\big[\hat{u}(\tau,\vec{k}),\hat{p}^{\dag}(\tau,\vec{k})\big]\big\rangle\big|^2  & = 1\,,\label{eq:Anti}
\end{align}
where the last condition is of course compatible with the commutation relations below eq.\,(\ref{eq:MainH}).
The quantity $|\mathcal{J}_k(\tau)|^2$, therefore, can be considered as measuring the ``classicality'' of the vacuum state during its evolution. 
At $\tau = \tau_{\star}$, $|\mathcal{J}_k(\tau_{\star})|^2 = 1/4$ and the anti-commutator vanishes. 
In this case the state has a pure quantum interpretation.
During standard slow-roll inflationary dynamics, $|\mathcal{J}_k(\tau)|^2$ grows exponentially after horizon crossing, and the state quickly becomes semi-classical. 
The uncertainty relation in eq.\,(\ref{eq:Unc}) also grows $\propto |\mathcal{J}_k(\tau)|^2$. This is because the uncertainty in the field position $|u_k(\tau)|^2$ (defined in eq.\,(\ref{eq:Heis1})) grows 
exponentially while the uncertainty in the conjugate momentum $|p_k(\tau)|^2$ remains fixed\footnote{This follows from the exact solution of eq.\,(\ref{eq:Mode1}) during inflation with vanishingly small Hubble parameters, 
\begin{align}
u_k =  \frac{1}{\sqrt{2k}}e^{-ik(\tau-\tau_{\star})}\left(1-\frac{i}{\tau k}\right)\,,
\end{align}
with the initial time $\tau_{\star}$ defined by the flat-space limit $k\tau_{\star}\to -\infty$ where we can unambiguously identify the minimum energy state (Bunch-Davies vacuum). 
} with the phase space density that, consequently, exhibits high squeezing.
Notice that, in light of eq.\,(\ref{eq:Anti}), eq.\,(\ref{eq:Unc}) can also be formally written in the form
\begin{align}\label{eq:HeisUnc}
\Delta u^2_k(\tau)\Delta p^2_k(\tau) \geqslant \frac{1}{4}\big|\big\langle\big[\hat{u}(\tau,\vec{k}),\hat{p}^{\dag}(\tau,\vec{k})\big]\big\rangle\big|^2\,,
\end{align}
which is the analogous of the Heisenberg uncertainty principle. 

To be more concrete, we consider now, in the spirit of the model discussed in section\,\ref{sec:Mod}, an explicit computation which is relevant for the analysis presented in section\,\ref{sec:Q2C}. 
We consider modes with comoving wavenumber such that they cross the Hubble horizon before the beginning of the ultra-slow-roll phase, $N_k \ll N_{\rm in}$.
As we shall see, this working assumption allows us
to study, by means of simple analytical expressions, the issue of classicalization for modes with $k \ll k_{\rm in}$ (thus, for instance, for modes close to $k_{\rm dip}$). 
We start from the evolution equation for the mode $u_k,$ eq.\,(\ref{eq:Mode1}). We shall consider the super-Hubble limit and neglect the $k^2$ term, which leads to:
\begin{align}\label{eq:MSsuXHub}
\frac{d}{d\tau}\left[z^2\frac{d}{d\tau}\left(\frac{u_k}{z}\right)\right]=0~~~~\Longrightarrow~~~~\frac{u_k}{z} = C_1(k) + C_2(k)\int_{\tau_{\rm ref}}^{\tau}\frac{d\tau^{\prime}}{z(\tau^{\prime})^2}\,,
\end{align}
where in the $\tau$-dependent part of the solution for $u_k/z$ we integrate starting from some reference time $\tau_{\rm ref}$.
The computation of the conjugate momentum in eq.\,(\ref{eq:Mode2}) is a bit more subtle because it requires to include in eq.\,(\ref{eq:MSsuXHub}) the first $k^2$ correction.
The reason is the following. Consider for simplicity the standard scenario of slow-roll inflation with $\epsilon \simeq \eta \ll 1$.  
In this case the first term of the solution in eq.\,(\ref{eq:MSsuXHub}) is the constant mode while the second is the decaying one.
If one just takes these two solutions in eq.\,(\ref{eq:MSsuXHub}), it follows, by means of eq.\,\ref{eq:Mode2}, that $zp_k = iC_2(k)$. Therefore, disregarding at late time the decaying solution, 
one would find $|\mathcal{J}_k|^2 = |C_2(k)|^2|C_1(k)|^2$ which is just a constant value. This incorrect procedure does not describe
the exponential growth of  $|\mathcal{J}_k|^2$. To capture this behavior, we need to include the $O(k^2)$ correction to the constant mode $u_k/z = C_1(k)$. 
To this end, we write for the constant mode the ansatz $u_k/z = C_1(k)[1+F(\tau)k^2]$ using $k^2$ as expansion parameter. It follows that 
the function $F(\tau)$ solves at order $k^2$ the differential equation $F^{\prime\prime}(\tau) + (2/z)z^{\prime}F^{\prime}(\tau) +1 = 0$. 
We do not need to solve this equation since $F(\tau)$ only enters in the computation of $p_k$ through its first derivative, namely $zp_k = iC_2(k) + iC_1(k)z^2k^2F^{\prime}(\tau)$.
Solving for $F^{\prime}(\tau)$, we find
\begin{align}\label{eq:ConjMom}
zp_k = iC_2(k) - iC_1(k)k^2\int_{\tau_{\rm ref}}^{\tau}d\tau^{\prime}z(\tau^{\prime})^2\,.
\end{align}
Since in the standard scenario of slow-roll inflation under consideration $z(\tau)$ grows with time, now we see that the second term in eq.\,(\ref{eq:ConjMom}) contributes to the
exponential growth of $|\mathcal{J}_k|^2$.

In the following, let us move to the case in which the inflationary dynamics features an ultra-slow-roll phase.
Compared to the previous scenario, 
in the presence of an ultra-slow-roll phase the second term in the solution in eq.\,(\ref{eq:MSsuXHub}) represents a growing--rather than decaying--mode.
On the contrary, the second term in eq.\,(\ref{eq:ConjMom}) decays instead of growing. 
Based on these differences, we expect --at the qualitative level-- the following time evolution  for $|\mathcal{J}_k|^2$
 \begin{align}\label{eq:ArrowOfTime}
 |\mathcal{J}_k(N)|^2 ~=
\resizebox{45mm}{!}{
\parbox{45mm}{
\begin{tikzpicture}[]
\node (label) at (0,0)[draw=white]{ 
       {\fd{8cm}{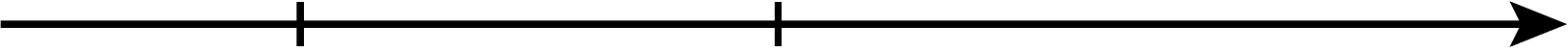}} 
      };
      \node[anchor=north] at (4.23,0) {{\color{black}{\scalebox{1}{$N$}}}};
      \node[anchor=north] at (-2.45,-0.08) {{\color{black}{\scalebox{1}{$N_{\rm in}$}}}};
      \node[anchor=north] at (-3.4,-0.1) {{\color{black}{\scalebox{1}{region\,I}}}};
      \node[anchor=north] at (-1.3,-0.1) {{\color{black}{\scalebox{1}{region\,II}}}};
      \node[anchor=north] at (2.,-0.1) {{\color{black}{\scalebox{1}{region\,III}}}};
      \node[anchor=north] at (-0.,-0.08) {{\color{black}{\scalebox{1}{$N_{\rm end}$}}}};
      \node[anchor=north] at (-3.25,0.5) {{\color{blue}{\scalebox{0.8}{exp growth}}}};
      \node[anchor=north] at (-1.25,0.5) {{\color{oucrimsonred}{\scalebox{0.8}{ultra-slow-roll}}}};
      \node[anchor=north] at (1.6+0.5,0.705) {{\color{blue}{\scalebox{0.8}{eventually a new}}}};
      \node[anchor=north] at (1.885+0.5,0.45) {{\color{blue}{\scalebox{0.8}{exp growth takes over}}}};
\end{tikzpicture}
}}~~~~~~~~~~~~~~~~~~~~~~~~~~~~~~~~~~~~~~~~~~~~~~~
\end{align}
For $N<N_{\rm in}$, $|\mathcal{J}_k|^2$ grows exponentially as just discussed in the standard case (remember that $N_k< N_{\rm in}$).
What happens during the ultra-slow-roll phase for $N_{\rm in}\leqslant N \leqslant N_{\rm end}$ depends on the details of the model and, consequently, on the explicit computation of the coefficients $C_{1,2}(k)$. 
The final exponential growth sets in only after the end of the ultra-slow-roll phase when the second term in eq.\,(\ref{eq:ConjMom}), computed for $N>N_{\rm end}$, starts dominating. 
Let us try to explore more quantitatively region\,II.
Consider our reference model in section\,\ref{sec:Mod}. 
Using the number of $e$-folds as time variable and for $N_{\rm in}\leqslant N \leqslant N_{\rm end}$, in region\,II we find
\begin{align}
\left.\frac{u_k}{z}\right|_{{\rm II}} & = \frac{iH}{\sqrt{4\epsilon_{\rm I}k^3}}\bigg\{
1-\bigg(\frac{k}{k_{\rm in}}\bigg)^2\frac{1}{(2\eta_{\rm II} - 3)}\bigg[
\underbrace{e^{(2\eta_{\rm II} - 3)(N-N_{\rm in})}}_{\rm exponential\,growth}-1
\bigg]
\bigg\}\,,\label{eq:Simpleuk}\\
\left.zp_k\right|_{{\rm II}}  & = \frac{\sqrt{\epsilon_{\rm I}k^3}}{H}\bigg(\frac{k_{\rm in}}{k}\bigg)\bigg\{
1 + \frac{1}{(1-2\eta_{\rm II})}\bigg[\underbrace{
e^{(1-2\eta_{\rm II})(N-N_{\rm in})}}_{\rm exponential\,decay}-1
\bigg]
\bigg\}\simeq \frac{\sqrt{\epsilon_{\rm I}k^3}}{H}\bigg(\frac{k_{\rm in}}{k}\bigg) \,.
\end{align}
The exponential growth (for $u_k/z$) and decay (for $zp_k$) anticipated before are now evident (remember that $\eta_{\rm II}> 3/2$). 
The growing solution is suppressed by $(k/k_{\rm in})^2$.
This means that if the duration of the ultra-slow-roll phase is not enough to compensate for this suppression, in region\,II $|\mathcal{J}_k|^2$ will remain basically constant (and equal to the value reached at $N_{\rm in}$ as a consequence of the exponential growth in region\,I, see the schematic in eq.\,(\ref{eq:ArrowOfTime})).
 After $N_{\rm end}$,  $|\mathcal{J}_k|^2$ grows exponentially
again as a consequence of the second term in eq.\,(\ref{eq:ConjMom}).
This is the typical behavior that we expect for modes with comoving wavenumber $k\lesssim k_{\rm th}$ which are not affected by the ultra-slow-roll phase (see fig.\,\ref{fig:Ka2}). 
In practice
for these modes classicalization is reached, as in the standard picture, soon after their horizon crossing before $N_{\rm in}$.

Consider now modes with $k_{\rm th}< k \ll k_{\rm in}$ whose comoving curvature perturbation --by definition of $k_{\rm th}$, see section\,\ref{sec:Mod}-- is affected by the ultra-slow-roll phase.
In this case the exponential growth in eq.\,(\ref{eq:Simpleuk}) becomes sizable and compensate the $(k/k_{\rm in})^2$ suppression as $N$ increases towards $N_{\rm end}$. 
From eq.\,(\ref{eq:Simpleuk}) we see that $u_k/z$ (and thus $|\mathcal{J}_k|^2$) actually vanishes for 
\begin{align}\label{eq:approx0}
\mathcal{N}_0 = N_{\rm in} + \frac{1}{(2\eta_{\rm II} - 3)}\log\left[1+\left(\frac{k_{\rm in}}{k}\right)^2(2\eta_{\rm II} - 3)\right]\,,
\end{align}
which is within the ultra-slow-roll phase if $(k/k_{\rm in})^2 > (2\eta_{\rm II} - 3)[e^{(2\eta_{\rm II} - 3)\Delta N} - 1]^{-1}$.
Of course, 
the fact that $|\mathcal{J}_k|^2$ vanishes is an artifact of the approximation used since, as we discussed, $|\mathcal{J}_k|^2\geqslant 1/4$. 
Nevertheless, the previous observation remains qualitatively true also in more complete computations where one finds that during the ultra-slow-roll phase 
$|\mathcal{J}_k|^2$ may drop down to its minimum value $|\mathcal{J}_k|^2 = 1/4$ for $N$ close to $\mathcal{N}_0$ in eq.\,(\ref{eq:approx0}).  We shall see an explicit example in a moment.
Consider now region\,III. 
We find
\begin{align}
\left.\frac{u_k}{z}\right|_{{\rm III}} & = \frac{iH}{\sqrt{4\epsilon_{\rm I}k^3}}\bigg\{
1-\frac{x^2}{(2\eta_{\rm II}-3)}\bigg[
e^{(2\eta_{\rm II}-3)\Delta N} -1
\bigg] - \frac{x^2 e^{(2\eta_{\rm II}-3)\Delta N}}{(2\eta_{\rm III} - 3)}\bigg[
\underbrace{e^{(2\eta_{\rm III}-3)(N - N_{\rm end})}}_{\rm exponential\,decay} 
-1
\bigg]
\bigg\}\label{eq:Nice1}\,,\\
\left.zp_k\right|_{{\rm III}}  & = \frac{\sqrt{\epsilon_{\rm I}k^3}}{H}\frac{1}{x}\bigg\{
1 + \frac{e^{(1-2\eta_{\rm II})\Delta N}}{(1-2\eta_{\rm III})}\bigg[
\underbrace{e^{(1-2\eta_{\rm III})(N-N_{\rm end})}}_{\rm exponential\,growth}
-1
\bigg]\mathcal{F}_k
\bigg\}\,,\label{eq:Nice2}
\end{align}
with 
\begin{align}
\mathcal{F}_k & \equiv 1- \frac{x^2}{(2\eta_{\rm II} - 3)}\bigg[e^{(2\eta_{\rm II} - 3)\Delta N} - 1\bigg]\,.
\end{align}
\begin{figure}[t]
\begin{center}
$$\includegraphics[width=.42\textwidth]{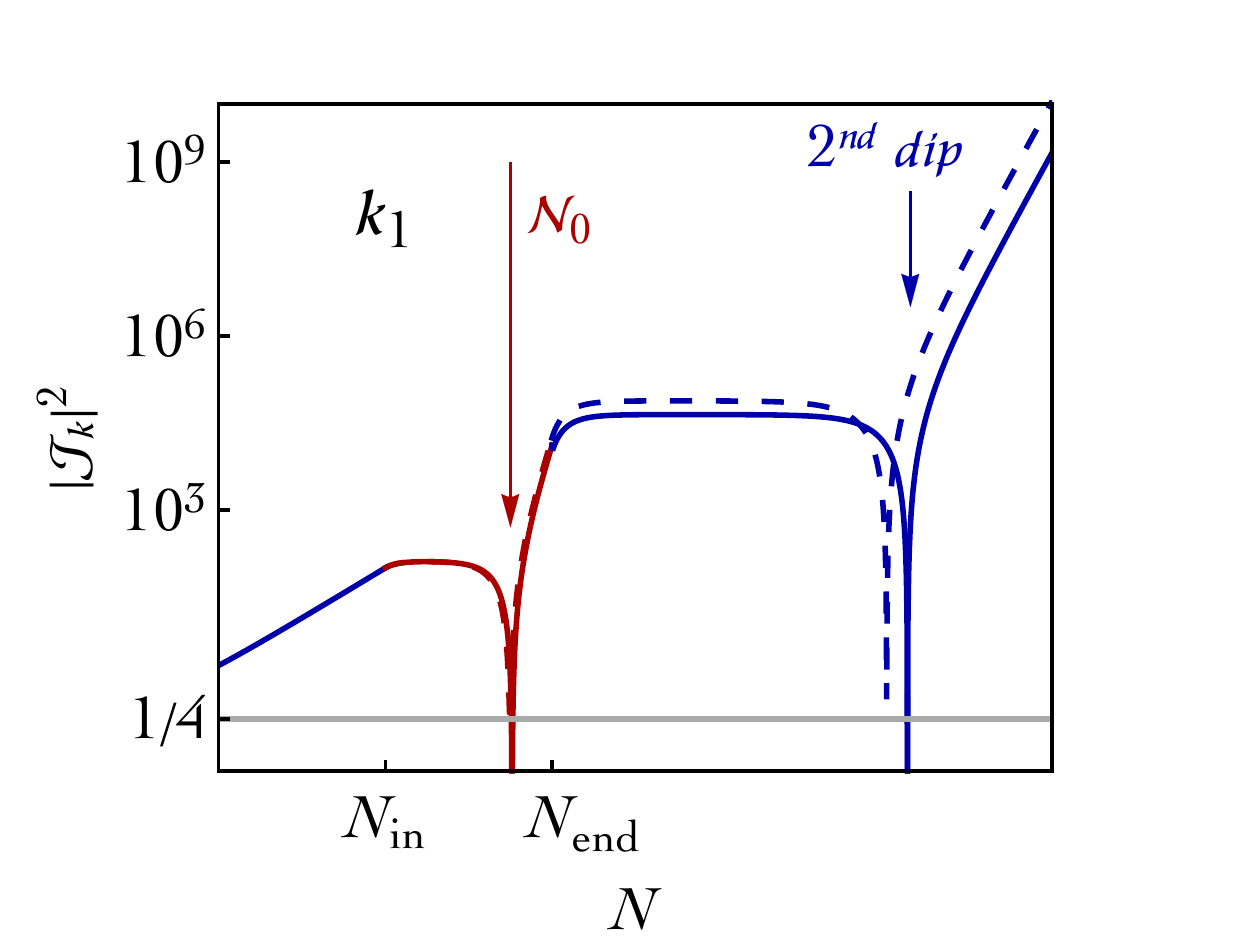}
\qquad\includegraphics[width=.42\textwidth]{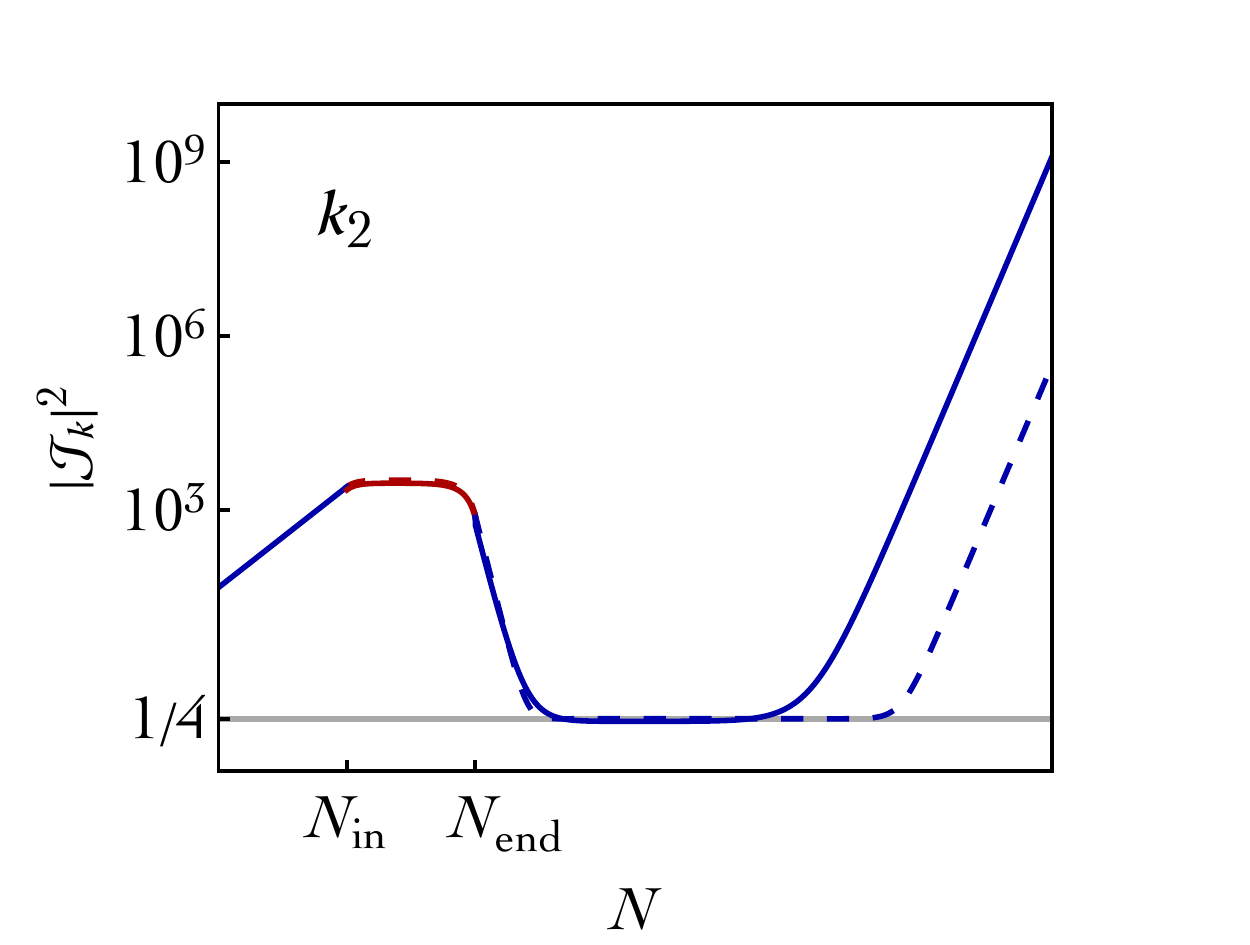}$$
\caption{\em \label{fig:ShiftTest} 
Explicit representation of the schematic illustrated in eq.\,(\ref{eq:ArrowOfTime}) for two different modes with comoving wavenumbers $k_{\rm th}< k \ll k_{\rm in}$. 
The solid lines follow the approximation discussed in section\,\ref{app:BasicDef} (see eqs.\,(\ref{eq:MSsuXHub},\,\ref{eq:ConjMom})) while, for comparison, the dashed lines are obtained with the analytical solutions 
derived in section\,\ref{sec:Mod}.
 }
\end{center}
\end{figure}
These expressions look complicated but they are actually simple to interpret.
We basically have two possibilities, depending on the exact value of $k$ and, consequently, the value of $\mathcal{F}_k$.
These two cases are summarized in fig.\,\ref{fig:ShiftTest}.
\begin{enumerate}
\item In the first case the value of $k$, say $k_1$, is such that $\mathcal{N}_0$ lies within the ultra-slow-roll phase. This means,  as already discussed, that the duration of the ultra-slow-roll phase is such that $e^{(2\eta_{\rm II} - 3)\Delta N}$ compensates the suppression given by $x^2$. In this case we can safely approximate
\begin{align}\label{eq:Fk1}
\mathcal{F}_{k_1}\simeq - \frac{x_1^2}{(2\eta_{\rm II} - 3)}e^{(2\eta_{\rm II} - 3)\Delta N}\,,
\end{align}
which as we can see is large and negative. As far as the other terms in eqs.\,(\ref{eq:Nice1},\,\ref{eq:Nice2}) are concerned, we can further approximate (neglecting the decaying mode and considering $e^{(2\eta_{\rm II} - 3)\Delta N}\gg 1$)
\begin{align}
\left.\frac{u_{k_1}}{z}\right|_{{\rm III}} & \simeq \frac{iH}{\sqrt{4\epsilon_{\rm I}k_1^3}}\bigg[
1 - \frac{x_1^2e^{(2\eta_{\rm II} - 3)\Delta N}}{(2\eta_{\rm II} - 3)}
\bigg] = {\rm constant}\,,\\
\left.zp_{k_1}\right|_{{\rm III}}  & \simeq \frac{\sqrt{\epsilon_{\rm I}k_1^3}}{H}\frac{1}{x_1}\bigg[
1 - \underbrace{\frac{x_1^2 e^{-2\Delta N}}{(2\eta_{\rm II} - 3)}e^{(1-2\eta_{\rm III})(N-N_{\rm end})}}_{\rm final\,exp\,growth}
\bigg]\,.\label{eq:siMpl}
\end{align}
We find that $|\mathcal{J}_{k_1}(N)|^2$ in region\,III stays constant until the exponential growth of $zp_{k_1}$ takes over. Before this happens, however, another dip is expected for values of $N$ such that 
the square bracket in eq.\,(\ref{eq:siMpl}) vanishes as the consequence of the negative sign that is acquired by $\mathcal{F}_{k_1}$ in eq.\,(\ref{eq:Fk1}).
This behavior is shown in the left panel of fig.\,\ref{fig:ShiftTest}.

\item In the second case the value of $k$, say $k_2$, is such that $\mathcal{N}_0 > N_{\rm end}$. 
This means that the duration of the ultra-slow-roll phase is such that $e^{(2\eta_{\rm II} - 3)\Delta N}$ is not enough to fully compensate the suppression given by $x^2$. In this case we can  approximate
\begin{align}\label{eq:Fk2}
\mathcal{F}_{k_2}\simeq 1\,.
\end{align}
Right after $N_{\rm end}$, $|\mathcal{J}_{k_2}(N)|^2$ still decreases (down to some value that depends on the specific choice of $k_2$) but it settles exponentially fast to a constant value which 
is maintained until the final exponential growth sets in. This is because in this case we can further approximate (using eq.\,(\ref{eq:Fk2}) and disregarding the exponentially decaying solution)
\begin{align}
\left.\frac{u_{k_2}}{z}\right|_{{\rm III}} & \simeq  \frac{iH}{\sqrt{4\epsilon_{\rm I}k_2^3}} \\
\left.zp_{k_2}\right|_{{\rm III}}  & \simeq  \frac{\sqrt{\epsilon_{\rm I}k_2^3}}{H}\frac{1}{x_2}\bigg[
1 + \underbrace{\frac{e^{(1-2\eta_{\rm II})\Delta N}}{(1-2\eta_{\rm III})}e^{(1-2\eta_{\rm III})(N-N_{\rm end})}}_{\rm final\,exp\,growth}
\bigg]\,.
\end{align}
Notice that in this case we do not expect the presence of a  dip in region\,III. This scenario is shown in the right panel of fig.\,\ref{fig:ShiftTest}.
\end{enumerate}

This explicit --even though simplified-- computation shows that the issue of quantum-to-classical transition in the presence of an ultra-slow-roll phase is anything but trivial. 
Let us try to draw some conclusions. 
If we consider the condition  $|\mathcal{J}_k(\tau)|^2 \gg 1$ as a test for classicality, 
it is evident from our analysis that for modes which are affected by the presence of the ultra-slow-roll phase the only robust statement 
that one can do is that only after the end of the ultra-slow-roll phase the condition $|\mathcal{J}_k(\tau)|^2 \gg 1$ can be unambiguously satisfied as a consequence of its final exponential growth. 
The details of what happens during the ultra-slow-roll phase depend --as discussed in depth by means of the analytical approximation exploited in this section-- on the specific value of the comoving wavenumber analyzed. 
At the qualitative level, this result remains valid if one considers the large occupation number condition  $n_k(\tau) \gg 1$ as a test for classicality. 
From the expression for $n_k(\tau)$ in eq.\,(\ref{eq:Explicitnk}) and the explicit form of the solutions for $u_k$ and $p_k$ in eqs.\,(\ref{eq:MSsuXHub},\,\ref{eq:ConjMom}) 
we find that it is only after the end of the ultra-slow-roll phase that the condition  $n_k(\tau) \gg 1$ remains fulfilled.



\end{document}